\documentclass[review]{elsarticle}

\makeatletter
\def\ps@pprintTitle{%
 \def\@oddfoot{}%
 \let \@evenfoot\@oddfoot}
\makeatother

\usepackage{lineno,hyperref}
\usepackage{lineno,hyperref}

\usepackage{amsmath,amssymb}

\usepackage{tabularx}
\usepackage[ruled]{algorithm2e} 
\usepackage{algorithmic}

\usepackage{mathtools}
\usepackage{amsmath} 
\usepackage{amsthm}
\newcommand{\Emptyset}{\text{\O}}
\usepackage{amsfonts}
\usepackage{tikz}
\usepackage[utf8]{inputenc}

\newtheorem{theorem}{Theorem}
\newtheorem{corollary}{Corollary}
\newtheorem{lemma}{Lemma}

\theoremstyle{definition}
\newtheorem{definition}{Definition}
\newtheorem{property}{Property}

\begin{document}

\begin{frontmatter}

\title{Invariants and Home Spaces in Transition Systems and Petri Nets}

\author{Gerard MEMMI}
\address{LTCI, Telecom-Paris, Institut polytechnique de Paris}



\address{19 place Marguerite Perey F-91120 Palaiseau, France}

\begin{abstract}
\textit{This lecture note focuses on comparing the notions of  invariance and home spaces in Transition Systems and more particularly, in Petri Nets. We also describe how linear algebra relates to these basic notions in Computer Science, how it can be used for extracting invariant properties from a parallel system described by a Labeled Transition System in general and a Petri Net in particular. 
We endeavor to regroup a number of algebraic results dispersed throughout the Petri Nets literature with the addition of new results around the notions of semiflows and generating sets.
\newline
Several extensive examples are given to illustrate how the notion of invariants and home spaces can be methodically utilized through basic arithmetic and algebra to prove behavioral properties of a Petri Net. 
Some additional thoughts on invariants and home spaces will conclude this note.}
\end{abstract}


\end{frontmatter}


\tableofcontents

\section{Introduction}


\subsection{Motivations}
Parallel programs, distributed systems, telecommunication networks, cyber-physical systems are complex entities to design, model, and verify. Sometimes, they can include analog components, however, this note will focus on their discrete behavior considering these entities as \textit{Discrete Digital Systems} (DDS). Using formal verification at different stages of the system development life cycle is a strong motivation that spread throughout this note and provide us with rationale for many concepts, definitions, and behavioral properties.
Discrete Digital Systems are often prototyped then simulated or emulated at great expense. Their behavior will be intensively tested and stressed over multiple kinds of inputs and their outputs carefully analyzed. 
Just to illustrate the magnitude of the importance of design verification, let us recall that integrated circuit design verification costs keep increasing and today, significantly surpass system design costs 
\footnote{https://blogs.sw.siemens.com/verificationhorizons/2021/01/06/part-8-the-2020-wilson-research-group-functional-verification-study/}. 

One problem is that DDS are often designed to operate during a long period of time or even indefinitely and continuously produce a stream of outputs.
Another problem is that they often scale up by integrating components designed and developed independently.
Last but not least of these problems, is that systems components are running concurrently, sharing common resources or objects, and generating  an enormous amount of possible behaviors, many of them unforeseen during the design and integration phases. 
One then speaks of combinatorial explosion.

DDS description must cope with concurrency and complexity as well as scalability. 
One solution is to recourse to system modeling before prototyping and use abstraction mechanisms while being very careful to minimize loss of accuracy and to preserve wanted properties. 
For instance, system designers may use genericity (via a kind of  polymorphism or more simply parameterization) in their models at the cost of making their verification much more difficult and sometimes, almost intractable (see \cite{M83}, \cite{BEISW20}, or section \ref{subsec: mutex-param} for specific cases).
System behavioral complexity depends upon the number of actions, the size, type, and structure of the data they can handle; however, this is the entanglement between concurrent or parallel actions that exacerbates their behavioral complexity and makes their design as well as their verification quite challenging. 

Following R. Descartes' second and third principles 
\footnote{``To divide each of the difficulties under examination into as many parts as possible, and as might be necessary for its adequate solution." then ``by commencing with objects the simplest and easiest to know, I might ascend by little and little, and, as it were, step by step, to the knowledge of the more complex" in \textit{Discours de la M{\'e}thode} (1637)}, a system observer (be a designer, a developer, or a tester) will want to cope with this situation by dividing and regrouping properties into subsets \footnote{These subsets designed for verification may differ from components ailing at an efficient architecture.} for which he will simplify and adapt the system model. Then, he will recompose the system step by step in a bottom up approach; detailing for each subset only relevant elements constituting what is sometimes called a system view or projection. 
Some of these behavioral properties will be called $invariants$ the study of which is at the core of this note. 

By the way, what is an invariant? We could paraphrase W. Bageholt 
\footnote{``we know what it is when you do not ask us, but we cannot very quickly explain or define it." in \textit{Physics and Politics} (1887) in: 

https://archive.org/ details/workslifeofwalte08bage/page/14/mode/2up} 
since on the one hand, so many papers use this notion without formal definition and on the other hand, so many papers introduce quite different definitions. 
How to best define and use invariants to analyze the behavior of a model? Can we not only prove that a formula is an invariant, but find a way in which they can be organized or concisely described, a way in which they can be discovered or computed? How can invariants be combined to represent a meaningful behavior? How can invariants be decomposed into simpler and verifiable properties? How to determine whether a given decomposition is more effective than another one with regard to formal verification?

In this note, we want to show how linear algebra or algebraic geometry can efficiently sustain invariant calculus as long as conceptual models can exhibit what we will call later in this note a structure allowing to deduce a system of equations with state variables as unknowns. 
In such a setting, linear algebra can also be applied and utilized to prove a large variety of behavioral properties.

One of our motivation was to go beyond regrouping a number of algebraic results dispersed throughout the Petri Nets literature and to compare and better position these results by considering semirings such as $\mathbb{N}$ or $\mathbb{Q^+}$ then over a field such as $\mathbb{Q}$ especially regarding generating sets of semiflows (\cite{M23}).  
In this regard, we were able to complete the current state of the art with a range of new results on minimal semiflows, minimal supports, and finite minimal generating sets for a given family of semiflows.

At the same time, we try as much as possible to link models, conceptual models, and behavioral properties with actual system examples to support our motivations. 

We sometimes looked at these conceptual models, basic properties, examples, and results from an historical perspective, which explains many of the citations from the 70's and 80's, two decades of intense research activities where the domain really took off.


\subsection{Outline and contributions}
After providing several basic notations Section \ref{sec: notations}, we organize in Section \ref{sec: models} various informal definitions and concepts that will provide some intuition with what we mean by terms such as observer, model, or system under study. The notion of state variable and the fundamental notions of transitions and states together with the notions of trajectories, traces, and computations are introduced Section \ref{sec: state-transition}. This will support the definition and properties of home space described Section \ref{subsec: HS} and later of the notion of invariant Section \ref{sec: invariant}.

The reader of this note is assumed to be somewhat familiar with Transition Systems described in Section \ref{subsec: transition systems}, more especially with Petri Nets. Basic concepts and definitions are provided in Section \ref{subsec: PN-plus} in order to produce a note as self-contained as possible. 
However, many more definitions, theoretical results, and examples can be found in \cite{K76} or \cite{FinkelS01} for Transition Systems and in many books such as \cite{BR82} or \cite{GV03} for Petri Nets.

Then, the notion of invariance is described in Section \ref{sec: invariant} trying to sort out various definitions found in the literature. Mutual exclusion is described and analyzed as a first example of invariant in Section \ref{sec:mutex}. 
Then, this note focuses on how linear algebra relates to this basic notion in Computer Science, and how it can be used for extracting invariant properties from a parallel or concurrent system described by specific Transition Systems or Petri Nets. 

In Section \ref{sec: semiflows}, semiflows, and the set $\mathcal{F^{+}}$ of all semiflows with non-negative coordinates of a Petri Net are introduced.
The notions of generating sets, minimal semiflows, and minimal supports are then defined in Section \ref{sec: generating sets}. 
In this section, semiflow theory is shown to be related to mathematical concepts from linear algebra, convex geometry, or discrete mathematics such as Sperner's theorem to provide with a bound for the number $c$ of minimal supports  or  Gordan's lemma about the existence of finite generating sets. This last connection has never before been attempted before. 

The three decomposition theorems of Section \ref{sec: 3theorems} have been first published in \cite{M78}. 
Here, two of them are extended to better characterize minimal semiflows and generating sets over $\mathbb{N}$ and better covering $\mathbb{N}$, $\mathbb{Q^+}$, and $\mathbb{Q}$.
They are at the core of this paper, since they have made it possible to consolidate results scattered in the literature in a meaningful way and are at the source of the new results presented in the following sections.
 
Succinct examples and counterexamples are provided to illustrate various results throughout Sections \ref{sec: 3theorems} to \ref{subsec: mgs}.  
Differences between minimal and canonical semiflows are made precise in the lemma \ref{lem: canonical-minimal} in Section \ref{sec: canonical}, which is dedicated to this notion; the section ends with a theorem which states under which condition the number of canonical semiflows is infinite (Section \ref{subsec: nb of canonical}).

The notion of minimal generating sets for semiflows described in Section \ref{subsec: mgs} can already be found in \cite{M77}, then by Colom, Silva, and Teruel in \cite{STC1998} p. 319 and later by the same authors in \cite{ColomTS2003}, p. 68 or more recently, in \cite{CMPW09}.
However, theorem \ref{th: lgs=mgs} is new, to the best of our knowledge, and proves the coincidence between a least generating set and a minimal generating set which is not always true in order theory. 
Uniqueness of particular generating sets is presented in Section \ref{subsec: uniqueness}.

In Section \ref{sec: summary}, a table is summarizing main results, in particular showing differences 
\footnote{A few inaccuracies found in the literature (e.g. in \cite{ColomS89,CMPW09}) stem from these differences.} when considering $\mathbb{N}$, $\mathbb{Q^+}$, or $\mathbb{Q}$ and highlighting our contribution.

Examples are given to illustrate how invariants can be used to prove behavioral properties of a Petri Net in Section \ref{sec: ex}. A concrete example is drawn from the telecommunication industry Section \ref{subsec: example-summary}; it illustrates  how the choice of minimal semiflows of minimal support is an important factor in simplifying the proof of safeness and liveness for a given Petri Net.

Section \ref{sec: concl} concludes and provides possible avenues of future research.
At last, some additional thoughts will conclude this note.

\subsection{Notations and abbreviations}
\label{sec: notations}

The following basic notations are also introduced where they are used for the first time.

\textbf{Numbers and sets}

$\mathbb{F}_{2} = \{0,1\}$ is the 2-element field.

$\mathbb{N}$ is the set of natural numbers including 0, $\mathbb{N_+}$ is the set of natural numbers excluding 0.

$\mathbb{Z}$ is the set of integer, $\mathbb{Q}$ the set of rational, and $\mathbb{R}$ the set of real numbers. $\mathbb{Q^+}$ and $\mathbb{R^+}$ are the semirings associated with $\mathbb{Q}$ and $\mathbb{R}$ respectively i.e. the sets of non-negative rationals and reals. 

If $E$ is a set, $|E|$ denotes the cardinal number of $E$.

$\omega$ denotes infinity and is absorbing for the addition ($\omega= \omega + 1$).
We abusively use the same symbol ‘0' to denote 
$(0,...,0)^\top$ of $\mathbb{N}{^n},\ \forall n \in \mathbb{N}_{\omega} = \mathbb{N} \cup \{\omega\}$. 

For any countable set $\mathbb{X}$,
resulting of an n-ary Cartesian product,  $\mathbb{X_\omega} = \mathbb{X} \cup \{(\omega,\ \omega,...,\ \omega)\}$ where $\omega$ is repeated n times, which is quite different from $\mathbb{X^\omega} = \mathbb{X} \times \mathbb{X} ...\ \omega$ times. 
With these conventions, it becomes clear that if $A$ and $B$ are two sets then $(A \times B)_\omega \neq A_\omega \times B_\omega$.

\textbf{Relations, functions, and vectors}

If $E$ and $F$ are two sets, and a \textit{relation} $r$ from $E$ to $F$ is a subset of the Cartesian product $E \times F$. Several notations can be used to express that $x$ of $E$ is in relation with $y$ of $F$ by $r$: we can use the infix notation $xry$ or $x\overset{r}{\rightarrow}y$, but also say that the pair $(x,y)$ belongs to $r$ and define $r(x) = \{y \in F | (x,y) \in r\}$ as the \textit{image} of $x$ under $r$.

The domain of $r$ is defined as: 
$\texttt{Dom}(r) = \{x\in E|r(x)\neq \varnothing \}$ 
and the image of $r$ (sometimes called codomain) is defined as:  $\texttt{Im}(r) = \{ y \in F | \exists x \in E,\ (x,y) \in r \}$.

Given a relation $f$ from $E \times F$ to $G$ and $x \in E$, $f(x, \cdot)$ denotes the relation from $F$ to $G$ such that: 
$ \forall y \in F, f(x,\cdot)(y) = f(x,y)$. 
Similarly, given $y\in E$, $f(\cdot,y)$ denotes the relation from $E$ to $G$ such that: 
$ \forall x \in F, f(\cdot,y)(x) = f(x,y)$; 
and can be seen as a projection of $f$ on $E$.

A \textit{partial function} $f$ from $E$ to $F$ is a relation such that $\forall x \in E, \  |f(x)| \leq 1$, then $f$ is a \textit{function} (sometimes also called \textit{mapping}) when it is a partial function and $\texttt{Dom}(r)=E$.

If $f$ is a partial function from $E$ to $\mathbb{N}$ or $\mathbb{F}_{2}$, then the \textit{support of $f$} is denoted by $\left \| f \right \|$ and is defined by 
$\left \| f \right \| = \{x\in \texttt{Dom}(f)\ |\ f(x) \neq 0\}$.

With $x_i$ denoting the $i^{th}$ coordinate of a vector $x$, two vectors $f$ and $g$ can be compared and $f \leq g$ if and only if $f_i \leq g_i$ for all coordinate of the two vectors. $f^T$ denotes the transpose of $f$ and $f^Tg$ denotes the scalar product of $f$ and $g$, we have $f^Tg = \underset{i}\sum f_i g_i$. 

\textbf{Words}

Let $T$ be an alphabet, $T^*$ denotes the free monoid generated by $T$ that is to say the set of finite words over $T$. $\lambda$ is the empty word. 
If $T$ is ordered such that $T=\{t_1,...,t_m\}$ and $r$ is a word over $T^*$, $\left|  r \right|$ is its length and $\overrightarrow{r}$ (sometimes called the Parikh vector of $r$) is such that $\overrightarrow{r_i}$ is the number of occurrences of the letter $t_i$ in the word $r$ and
$\overrightarrow{r} \in \mathbb{N}^{|T|}$.
A word has a notion of support similar to that of function: $\left \| r \right \| = \left \| \overrightarrow{r} \right \|= \{t_i \in T\  \|\ \overrightarrow{r_i} > 0\}$

$T^\omega$ is the set of infinite words and $\left|  r \right| = \omega$ if and only if $r \in T^\omega$; and $T^{\infty} = T^* \cup T^\omega$ is the set of finite and infinite words.

\textbf{Various abbreviations}

$DDS$ stands for Discrete Digital System, $SuS$ for System under Study, $M$ for model, $CM$ for conceptual model, $TS$ for Transition System, $LTS$ for Labeled Transition System, $PN$ for Petri Net, $VAS$ for Vector Addition System. $\langle M, Init \rangle$ denotes a model $M$ paired with a set of initial states $Init$.

$\mathcal{SV}$ is denoting the set of \textit{state variables} and $|\mathcal{SV}| = d$ denoting the number of state variables. $\mathcal{SV}$ can be ordered and we have $\mathcal{SV}=\{x_1,...,x_d\}$.

$T$ is denoting the set of named or labeled transitions and $|T| = m$.

$\emph{Q(M)}$, $\emph{$RS(M,q_0)$}$, $\emph{$RG(M,q_0)$}$, $\emph{$LRG(M,q_0)$}$ denote the \textit{state space}, the set of \textit{reachable states} (i.e. Reachability Set), the Reachability Graph, the Labeled Reachability Graph of $\langle M, q_0 \rangle$ respectively. 
These sets will more simply be denoted $\emph{Q}$, $\emph{RS}$, $\emph{RG}$, and $LRG$ respectively when there is no ambiguity.

\newpage
\section{Observer, systems under study, models, and conceptual models}
\label{sec: models}

We cannot address the notions of invariant and home space without first describing even succinctly the basic objects and concepts motivating our study. 
A model can support in various different ways, the description and the analysis of a program or of all kind of systems. In our case, a model must describe its \textit{behavior} that is to say how a DDS evolves over time from one state to the next one as it is running and performing various operations or actions.

\subsection{Observer}
\label{subsec: observer}
The notion of model cannot in our opinion, be separated from the notion of observer. 
An observer may have different additional roles which can influence the kind of information he needs to observe. He can be a system designer, an architect, a developer, a test or a maintenance engineer, or even a user.  
An \textit{observer} is interacting with the model before and after its progression during which he is only observing the model evolution. 
His interaction consists in selecting the level of information he can or wants to extract from the model.
To do so, he defines a model environment, instrumenting the model and defining output data, preparing input data, tuning parameters, selecting an observer ``clock frequency" (i.e. the level of granularity or accuracy with which he will be able to extract or sample information). 
In essence, the observer wants to draw sufficient historical information about the dynamic functioning of the model to answer the following typical questions: 

$\bullet$ under which circumstances will such system element change the value of such variable, or such statement becomes true? Is this evolution accurately predictable?

$\bullet$ Is the observed evolution of the model in accordance with the observer's expectations expressed under the form of behavioral properties or in accordance with the behavior of the actual system under study? 

As a designer or an architect, the observer may want to proceed by reification with a new version of the model, by refining the model environment with more inputs, or by increasing of his own clock frequency and be able to observe the model with more details. 
He will be satisfied if the two levels of observation are somewhat equivalent, preserving behavioral properties ; and he will look for an improved degree of determinism, in other words that the past and present has some predictive value 
\footnote{This is not without reminding the stoic doctrine that Seneca expresses in \cite{Sen10} book2, 32, 4: ``Everything that happens
is a sign of some future event. Chance events and purposeless, chaotic ones do not admit of divination; where there is order, there is also predictive force."
}.
  
As a maintenance engineer he may want to observe alignment between  what he observed on the model and the behavior of the actual system under study, increasing confidence in the model looking for avoiding costly system emulation.

The notions of system under study, model, and conceptual model are now informally introduced. Models and conceptual models will be introduced later in Section \ref{sec: state-transition} with formal definitions  together with a few behavioral properties.

\subsection{Systems under study (SuS)}
We will call \textit{system under study} the original entity be a program or a digital system to be described then verified. We are interested in systems evolving or progressing most of the time indefinitely (as in UNITY \cite{ChMi88}). 
They will be performing statements, actions, activities, or reacting to events, sharing, consulting, producing, or consuming \textit{resources}. 
They will be handling resources which are consumed, tested, modified, shared, or produced in discrete quantities while statements, actions or activities are being performed. 
Some of these actions could be running in parallel when they are independent of each other or in concurrency when they race for a shared resource.

In this note, we will consider that a system under study is able to perform a finite number of different elementary or \textit{indivisible} actions (i.e. actions that are not a composition of observable actions) over a finite set of different types of resources. 
This does not prevent systems to present infinite behaviors or being able to produce a resource of a given type in an infinite quantity. 
The dynamic \textit{progression} of an SuS is modeled by a notion of \textit{computation} i.e. a sequence finite or infinite of actions altering a set of resources described by a notion of \textit{state}.

We are interested in the study of the system \textit{behavioral properties} such as reachability (section \ref{sec: reachability}), liveness, boundedness, or fairness (defined in section \ref{sec: behav}). 
That is to say, we are interested by the analysis of these properties along a set of possible progressions of the SuS characterized by the first time they start to be observed (i.e. their initial state). 
In this regard, we will need a model representing a simplified version of our SuS.

At last, we will not consider the notion of time in terms of dates or duration but rather in terms of sequencing: something is happening before, after, or at the same time than something else (independently from a notion of causality).
Therefore, in this note, the words \textit{action} (which often bears a meaning of duration) or \textit{event} (which often is understood as instantaneous) will be synonymous (see for instance \cite{Kuz18}, for an illustration of the difference between the two notions where time (as a duration) is taken into account).

\subsection{Models (M)}
\label{subsec: models}
We will call $model$ the formal description of a discrete digital system by an observer. 
This description is discrete and must support the design, the analysis, or the verification (or at least the observation) of the behavior (in other words, the dynamic evolution or progression of the system under study over time).

Model and observer are linked together. 
The model is defined by the observer in accordance to the level of observation he wants or can reach. Of course, a model is never \textit{fully} equivalent to the system under study it describes. It is sometimes considered as an ideal description to be matched by the actual system seen as an approximation. 
A model is developed with the intent of describing and verifying a specific set of properties not any  property the system under study may or must satisfy. As such, many "irrelevant details" can be abstracted away.

A ``good" model will retain just enough details and will be as small as possible in order to conduct as an effective verification as possible and at the same time, it must be large enough to convincingly behave like the system under study in regard to the specific set of properties to be proven. 
It must be recalled here, that most of the time, a verification algorithm complexity depends among other parameters on the size of the model as well as the level of interactions between its different operations. 
In this regard, one can speak of an \textit{art of system modeling} especially when considering techniques such as model  projections, reductions (see \cite{BR82,ColomTS2003}), or properties rewriting (see \cite{JouDer90}) that can be used to better organize or simplify the model while preserving its desired properties. 
However, describing in details these very interesting techniques are out of the scope of this note.

\subsection{Conceptual models (CM)}
Like programs are written in a programming language, models are   described in a \textit{conceptual model}.
Sometimes the terms abstract machine (as in CAML \footnote{https://caml.inria.fr/about/history.fr.html}), computational model, or formal model are used instead (\cite{FinkelS01} cites the two later terms indifferently, for instance). 
With this in mind, Turing machines are a conceptual model as well as automata \cite{E74} or Kripke structures \cite{Kri63}.

A specific automaton which would represent the addition of two bounded integers would be a model of an adder and we use to say that such a model is an automaton in a kind of metonymy where we use the same name for a specific model and the underlying conceptual model. 

In this note, we will be interested in CMs expressive enough to support the description and the analysis of the dynamic progression (also called behavior or evolution) of a system under study. 
In the digital world we are interested in, this progression will be discrete and characterized by a notion of state taking its values in countable sets. 
States will be characterized by a set of state variables which in turn will allow speaking of invariants. 
CMs will also include a notion of transition (or action) that will support describing how the model progresses from one state to the next. 
A survey on various CMs can be found in \cite{NieSas98}.

\section{State Variables, States, and Transitions}
\label{sec: state-transition}
\subsection{State variables, states, and trajectories}

\subsubsection{State variables}
\label{subsubsec: SV}
As said in section \ref{subsec: observer}, the observer is central in selecting state variables. He is deciding which variables are required (sometimes involving the need to reify the model), useful, or accessible (directly or indirectly for instance, by means of instrumentation in the SuS), which ones can influence the progression of the system and alter the behavioral properties he wants to analyze and prove. Let us point out that a state variable can be useful during the design phase and optimized away later in the SuS. 

A state variable can represent any kind of information even outside of the system itself. In some of his own examples, Hoare \cite{CARH15} includes what he calls the interaction with the system environment to take such external elements into account.
For instance, an ambient temperature, an altitude, 
or a number of users can qualify as state variables despite the fact that they very likely are outside the system under study, but interact with it to the point where they can change its outputs or behavior and at the same time, are of interest to the observer.

A state variable can be a simple Boolean variable or a complex data structure controlling all sort of synchronization mechanism: queues, shared resources, clocks, watchdogs,... 

Concerning DDS, authors often distinguish two kinds of state variables:

$\bullet$ the memory, the registers, data variables witnessing (such as outputs) or influencing (such as inputs) the system progression, or data variables which can support monitoring resource variation pertinent to properties to be proven;  

$\bullet$ the control variables such as Boolean variables or guards of a guarded commands language (\cite{Dijkstra75}) to determine a possible choice between two or more courses of action (e.g. pieces of code), semaphore variables to control access to shared resources, locations, labels, or ordinal counters to point towards the next possible statements that can be executed by the SuS. 

\begin{definition} 
$\mathcal{SV}$ will denote the \textit{set of state variables} of a given model and unless specified otherwise, $\mathcal{SV}$ will be finite with $|\mathcal{SV}|=d$. For convenience, we may want to order $\mathcal{SV}$ such that $\mathcal{SV}=\{x_1,...,x_d\}$.
A \textit{state variable} $x_i$ is an element of $\mathcal{SV}$. It has a \textit{type} that determines the range $D_i$ of its potential values, i.e. the values it can possibly take during any evolution of the system.
\end{definition}
State variable potential values may be in $\mathbb{Q}$ as in \cite{BlondinFHH17}, more often in $\mathbb{R}$ as in\cite{ChMi88} or \cite{Tiwari01}; however in this note, most of the time, we will limit ourselves to values in countable sets particularly $\mathbb{N}$, $\mathbb{F}_{2}$, or any finite enumerated set.  

\subsubsection{States and state spaces}
Most conceptual model designed to analyze system behaviors and address concurrency will manipulate explicitly or implicitly a notion of \textit{state}. A state must contain enough information to monitor or even predict the system progression at a level of observation sufficient to analyze and determine whether a set of desired properties is met or not. 
\begin{definition}
The set of all \textit{potential states} is called the \textit{state space} and is denoted by $\emph{Q}$ when there is no ambiguity, and \emph{Q(M)} when it is necessary to relate it to a particular model \emph{M}.
\end{definition}
In this note, we will assume $Q(M)$ countable as in many publications, for instance in \cite{QSif83}.

Two subsets of $\emph{Q}$ are usually distinguished: $Init$ denoting the set of \textit{initial states} i.e. the states from which the model begins its progression, $Term$ denoting the set of \textit{terminal states} i.e. the states where the model progression is expected to halt (or at least, where the observer stops his observation). These two subsets have no particular properties in terms of values or connectivity with other states.
In the next sections, models will often be paired  with a single initial state $q_0\  (Init = \{q_0\})$; and unless specified otherwise, we will have: $Term = \varnothing$ since we are interested by models with infinite behaviors. 

Some conceptual models such as automata or transition systems do not go further in defining states and states spaces looking for being as general as possible. 
However, understanding and analyzing behavioral properties can often require providing more details about states, especially when dealing with invariant analysis. 
Which values can they? Within which domain? 
A state can be defined with the help of the set $\mathcal{SV}$ of \textit{state variables} determined by the observer.

In \cite{ASHCROFT75} or in \cite{K76} and in many other theoretical computer science papers, a state will be structured as a couple of vectors, separating the two kinds of state variables we just mentioned (i.e. observation or memory and control variables or locations as in \cite{ColonSS03}). 
This distinction will not be made in this note. 

\begin{definition}
Given a conceptual model equipped with a set of state variables $\mathcal{SV}$, a \textit{state} $q$ as an element of the state space $Q$ is a mapping from $\mathcal{SV}$ to the set of potential values taken by the elements of $\mathcal{SV}$. 
\end{definition}

More precisely, if $x_i \in \mathcal{SV}$ varies in the set $D_i$ then $q(x_i) \in D_i\ \text{and}\  Q = \prod_{i=1}^{i=d} D_i$ by construction. 
A state can also be seen as a vector of dimension $d$ since $\mathcal{SV}$ can be ordered as seen in section \ref{subsubsec: SV}. 
The most convenient notation between $q_i ,\ q(x_i)$, or  $x_i$ will be chosen consistently with the notation adopted for the model description and its behavioral properties (see for instance notations used section \ref{sec: ex}). 

Regarding a system under study, it is often sufficient to consider amounts of $d$ different  \textit{state variables} (including Boolean variables) modeling the number of resources or the truth values of control elements the system under study disposes of. 
Henceforth, we may assume $\emph{Q} = \mathbb{N}^{d}$ by default and a state $q$ can be then seen as a vector of $\mathbb{N}^{d}$ unless more information is made available. For instance, if we know that there are $d_1$ boolean state variables and $d_2$ non-negative integer state variables such that $d=d_1+d_2$, we have $\emph{Q} = {\mathbb{F}_2}^{d_1}\times\mathbb{N}^{d_2}$. 

 
We say that $\mathcal{SV}$ is \textit{complete} if and only if the model progression including its desired behavioral properties depends uniquely on the values of the current state. 
As an example, a model can progress according to its past history rather than its sole current state. 
In that case, historical data (i.e. model progression) will have to be encoded in a specific state variable in order to make $\mathcal{SV}$ complete.
\subsubsection{trajectories}
\label{subsubsec: trajectories}
Two states $q$ and $q'$ are \textit{consecutive} if there exists an execution of the model from $q$ to $q'$ during which no intermediary state could not be observed. We also say that $q'$ \textit{directly succeed} to $q$ or that $q$ \textit{directly reach} $q'$.

This notion is associated to a concept of indivisibility of the progression of the model: there is a way to move from state $q$ to state $q'$ in one single indivisible step. 
However, it must be pointed out that it may exist several different ways to reach $q'$ from $q$; and even if $q$ and $q'$ are consecutive, $q'$ can be reached from $q$ after having reached intermediary states: $q_1, q_2, ...q_n$ forming a sequence of states.
\begin{definition}
A sequence of consecutive states is called a \textit{trajectory}.
\end{definition}

Trajectories can be infinite when there exists a cycle (a sub-sequence of the said trajectory that begins and ends with the same state) repeated an infinity of times or when there exists an infinity of different states (constituting a ray in term of graph theory \cite{Diestel10}). 

\subsection{Transitions and traces}
\label{subsec: TTL}
A transition is a key element for any conceptual model. 
It is modeling a specific indivisible event, statement, or action of the SuS describing how it is possible to move from one state to another.

\begin{definition}
\label{def: transition}
A \textit{transition} $t$ is a relation in $Q \times Q$. If $(q,q') \in t$, then we say that $q'$ is \textit{directly reachable} from $q$ consistently with section \ref{subsubsec: trajectories}. 
We then write $q \overset{t}{\rightarrow}q'$. If $q \in \texttt{Dom}(t)$, then we write $q\overset{t}{\rightarrow}$. 
When the name or label of the transition is unknown or omitted, we write $q \rightarrow q'$ and $q \rightarrow$ respectively.
T denotes the set of transitions (sometimes transitions names or labels) for a given model.
\end{definition}

Some CMs such as transition systems, are only interested by the possibility for a state to directly or indirectly (through a more complex progression) reach another one, ignoring by which kind of computation or sequence of actions, it can happen. 
In that case, $\rightarrow$ ($|T|=1$) may suffice to describe the SuS. 
However, there is often a need for distinguishing various actions. In that case, transitions are given a name or a label in a set $T$ of transition labels or names. In this note, $T$ will be finite and by convention $\left| T \right| = m$, where $m \ \in \mathbb{N}$.
Associating labels and transitions, we have: $(x,y) \in \rightarrow$ if and only if  $\exists t \in T$ such that $(x,y) \in t$. In other words $\rightarrow =\bigcup_{t \in T} t$.

A \textit{sequence of transitions} is a word of $T^{*} \cup T^{\omega} = T^{\infty}$.

Only a subset of $T^{\infty}$ can represent actual evolution of a given model. 
In this regard, a notion of trace is introduced to make a difference between a word and an actual system behavior.
The word ``trace" can already be found in many articles such as \cite{CARH15} to define a sequence of events. In \cite{Kuz18}, we can find a very similar notion with the concept of ``history".
\begin{definition}
\label{def: trace}
A \textit{trace} $\sigma_n$ where $n \in \mathbb{N}_\omega$ is a relation in $Q \times Q$ which is a composition of transitions say $t_1, \ t_2,...t_n$ such that $(q,q') \in \sigma_n$ if and only if $\exists\ q_1, q_2,...q_{n+1} \in Q$ such that $
\forall i \in \{1,...n\},\ (q_i \overset{t_i}{\rightarrow}q_{i+1})$ 
 with $q_1=q$  and  $q_{n+1}=q'$.
 We then write $q \overset{\sigma_n}{\rightarrow}q'$.
If $q \in \texttt{Dom}(\sigma_n)$ then we write $q \overset{\sigma_n}{\rightarrow}$. 

When the names or labels of transitions are unkown or omitted, we write $q \overset{*}{\rightarrow}q'$ and $q \overset{*}{\rightarrow}$ respectively;
$ \overset{*}{\rightarrow}$ denoting the reflexive transitive closure of ${\rightarrow}$.
 \end{definition}
 
We sometimes say that $\sigma_n$ begins at $q$ and ends at $q'$; we then write $q \overset{\sigma_n}{\rightarrow}q'$. 

Any sub-sequence of a trace is a trace. 
Similarly to trajectories, traces can be infinite; in that case, a trace $\sigma_n$ can be indefinitely extended by a transition: $\sigma_{n+1}=\sigma_n t$ and $\sigma_\omega = \displaystyle \lim_{ n\to \infty } \sigma_n $. 

\begin{figure}[ht]
\centering
\includegraphics[width=0.8
\textwidth]{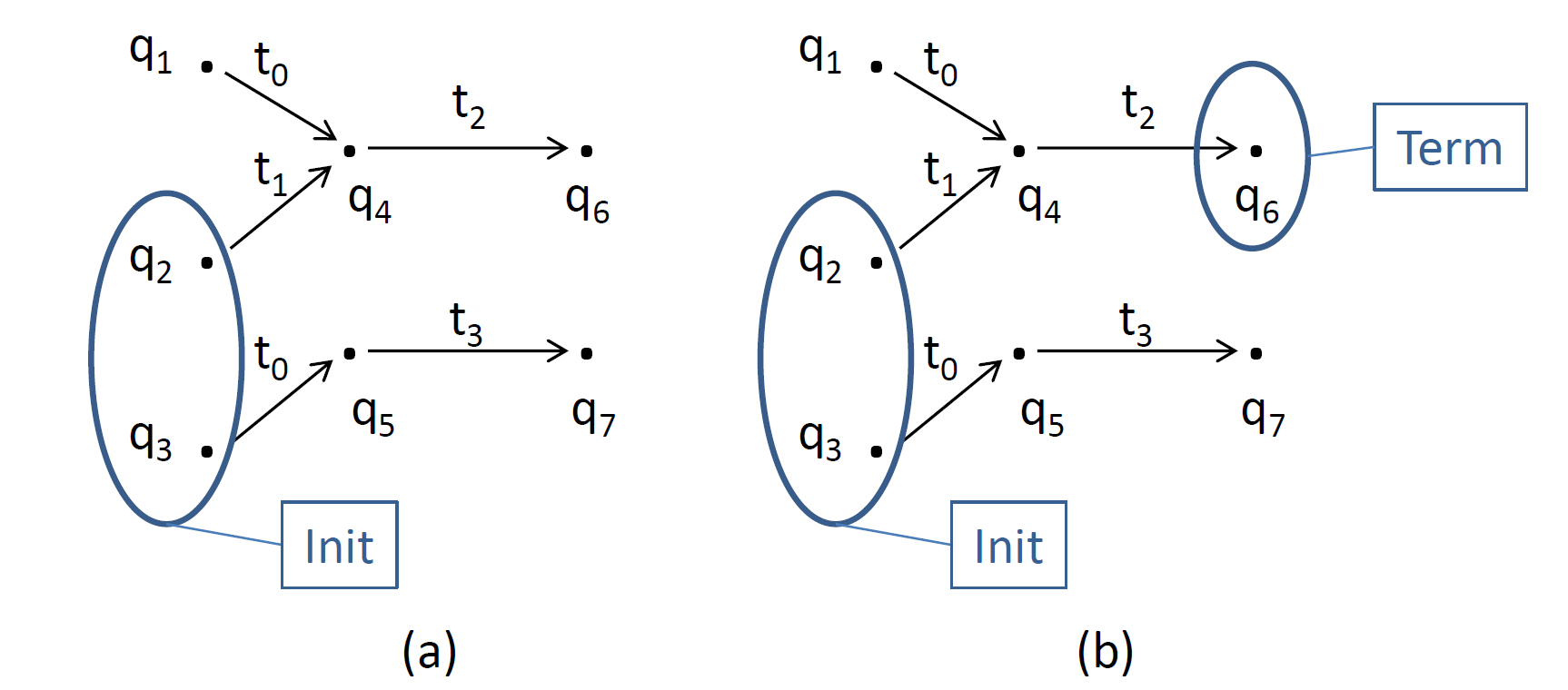}
\caption{$Q = \{q_1, q_2, q_3, q_4, q_5, q_6, q_7\}, T =\{t_0, t_1, t_2, t_3\}$. 
The sequence of transitions $t_1t_3$ can never be a possible trace whatever is $Init$.
\newline
In (a), $\textit{Term} = \varnothing$ and the sequence of transitions $t_0t_2$ is a possible trace beginning at $q_1$; however, it is not a trace from $\textit{Init} = \{q_2,q_3\}$.
\newline
In (b), $q_2 \overset{t_1}{\rightarrow} q_4\overset{t_2}{\rightarrow}q_6$ is the only computation given $Init$ and $Term$.}
\label{fig: trace}
\end{figure}

A trace can be seen as a particular sequence of transitions, we then write $\sigma_n = t_0t_1...t_n$ as a word of $T^{\infty}$. 
However, a sequence of transitions $s$ may or may not be a trace. 
It may just be impossible to compose the transitions in the order specified by the sequence of transitions. 
This is the case when $\exists s' \in T^*\setminus \{\lambda\}, t \in T, s" \in T^{\infty}$ such that $s=s'ts"$ and $\texttt{Im}(s') \cap \texttt{Dom}(t) = \varnothing$.
Other cases may happen once the set \textit{Init} of initial states paired with a model under consideration is chosen. 
Figure \ref{fig: trace} is showing two elementary sequences of transitions the first one can be or cannot be a trace according to the considered initial states; the second one can never be a trace.

\subsection{Languages and computations}
\label{subsec: L&C}
\begin{definition}
Given a model $M$ and a subset $Init$ of states, the set $\mathcal{L}(M,Init)$ of all possible traces beginning at $q \in Init$ defines the \textit{language} of $\langle M,Init \rangle$.
\end{definition}
As long as $Term = \varnothing$, a language $\mathcal{L}(M,Init)$ is closed for the prefix relation $pre$ defined by: $\forall w \in \mathcal{L}(M,Init),\ pre(w)= \{v \ |\ \exists u \in T^\infty, w = vu \} \subseteq \mathcal{L}(M,Init) $. This property is valid for Petri Nets \cite{BR82}.

\section{Reachability}
\label{sec: reachability}

Reachability or accessibility is a central question when studying concurrent system behavior. 
Most behavioral properties if not all can be described using the notion of reachability: is a given property true when reaching a specific subset of states or is it true for any state that can be reached from a subset of states?
Can the same state be reached several times, or even after any evolution of the model?

Reachability can be studied through a number of theoretical tools: from graph theory such as the reachability graph, or the covering graph (see \cite{KM69,F93}) or from algebra such as semiflows (see section \ref{sec: generating sets}).

\subsection{Reachability and reachability sets}
\label{subsec: RS}
As we learn how a model progresses over time, a relation of reachability can be introduced between states and we will say that a state $b$ can be \textit{reached} from a state $a$ (sometimes, $b$ is said \textit{accessible} from $a$ as in automata theory \cite{Saka09} or \cite{E74}) if there a mean for the model to progress from $a$ to $b$ and we write $a \overset{*}{\rightarrow}b$ to follow the notation introduced in Section \ref{subsec: TTL}.
\begin{definition}
The set of all reachable states from an initial state  $q_0$ is traditionally called the \textit{reachability set} \footnote{It includes $q_0$ since $\overset{*}{\rightarrow}$ is defined as the reflexive transitive closure of ${\rightarrow}$ in Section \ref{subsec: TTL}.} and is denoted by \emph{RS}; when there is no ambiguity and \emph{$RS(M,q_0)$} when we consider the specific model \emph{M} paired with the initial state $q_0$.
Formally, the reachability set $RS(M,q_0)$ is defined by: 

$RS(M,q_0) = \{q' \in Q\ | \ q_0 \overset{*}{\rightarrow} q'\}$.
\end{definition}

Similarly to the definition of the reachability set, we can be interested by the set of states reaching a given state $q$ denoted by $RS^{-1}(M,q)$:  
$RS^{-1}(M,q)= \left\{ q' \in Q\ |\ q' \overset{*}{\rightarrow} q \right\}$.

From this definition, it can be immediately deduced the following property:
\begin{property}
\label{prop: RS succession}
Given two states $q$ and $q'$, 
$\ q \overset{*}{\rightarrow} q'$ if and only if $RS(M, q') \subseteq \emph{RS(M,q)}$ or equivalently, if and only if 
$RS^{-1}(M, q) \subseteq RS^{-1}(M,q')$.
\end{property}

The notion of reachability set immediately raises the question of knowing which state is in $\emph{RS}(M, q_0)$ which state is not. 
At the level of conceptual models, this question is better known as the reachability problem:

\textit{the reachability problem for a conceptual model $CM$ is decidable if there exists an algorithm taking as inputs a model $M$ described in $CM$, a state $a$, and a state $b$, able to determine whether or not $b$ is reachable from $a$ in $M$.}

This problem has been solved and its proof simplified several times for a particular CM called Petri Nets, for instance in \cite{Mayr84, L09}.


If $Init$ is a nonempty subset of $Q$, we can extend the definition of the reachability set and $RS(M,Init)$ will denote 
$\bigcup_{q \in Init} RS(M,q)$; 
similarly:

$RS^{-1}(M,Init)= \bigcup_{q \in Init} RS^{-1}(M,q)$. 
A state $q$ is said \textit{isolated} if and only if $RS(M,q) = RS^{-1}(M,q) = \{q\}$ and \texttt{Isol} denotes the set of isolated states. This means that the only possible successor or predecessor of an isolated state is itself.

\begin{property}
\label{prop: Qdef}
Let's $Q$ be the state space of the model $M$ with $\mathcal{SV}$ its set of state variables, each state variable $x_i$ varying in a domain $D_i$, we have:

$Q = \prod_{i=1}^{i=d} D_i = \bigcup_{q \in Q} RS(M,q) = \bigcup_{q \in Q} RS^{-1}(M,q) = \bigcup_{t \in T} \texttt{Dom}(t) \cup \bigcup_{t \in T} \texttt{Im}(t) \cup \texttt{Isol}$.

\end{property}

\subsection{Reachability graph and labeled reachability graph}
\label{subsec: RG}
To explore and analyze the reachability set in a concise manner and support the resolution of the reachability problem, a \textit{reachability graph} $RG(M,q_0)$ ($RG$ when there is no ambiguity) can be constructed where the set of vertices is in bijection with \emph{$RS(M,q_0)$} that is to say that each vertex $n$ is labeled by exactly one element $q$ of \emph{$RS(M,q_0)$} and that any reachable state $q$ is the label of exactly one vertex. 
We will not make distinction between a vertex and its label any longer. 
\begin{definition}
Given a model $M$ and a subset of states $Init$, 
its associate reachability graph is a directed multigraph defined by:

$RG(M,Init)= \langle RS(M,Init),E \rangle$ where $(a,b) \in E$ if and only if
state $b$ is directly reachable from state $a$ in $\langle M,Init \rangle$.
When $M$ is equipped with a transition relation, we equivalently write $(a,b) \in E$ or $a \rightarrow b$.
When $M$ is equipped with a set of transitions $T$, we can defined a \textit{labeled reachability graph} as a directed multigraph $LRG(M,Init)= \langle RS(M,Init),E \rangle$ where an edge $(a,b)$ of $E$ is labeled by a transition $t$ if and only if $a \overset{t}{\rightarrow} b$.
\end{definition}
$LRG(M,Init)$ is genuinely a multigraph when there exists several different direct progression from $a$ to $b$ \footnote{Actually, the observer could be able to observe multiple different progressions from $a$ to $b$. 
That would define a subgraph of RG or more precisely, a sub-model with $a$ as its initial state and $b$ as its terminal state containing strictly all paths from $a$ to $b$. 
This could be seen as a kind of weak non-determinism since the system would have the choice to pick one path or another among a determined set of paths, to progress from $a$ to $b$.}.

A directed path will join a state $a$ to a state $b$ if and only if $b$ is reachable from $a$ in $\langle M,q_0 \rangle$. 
This directed path can be seen as a sequence of consecutive states called a \textit{trajectory} section \ref{subsubsec: trajectories}.

The reachability graph can have an infinite number of edges for two reasons: there is an infinite multiplicity of edges from or to a vertex, or there is an infinite number of vertices. 
In this note, we will limit ourselves in considering reachability graphs where vertices have a finite degree called \textit{locally finite} as in \cite{Diestel10}. 
In \cite{FinkelS01}, vertices just have their outdegree finite (\textit{finitely branching}) \footnote{The interested reader can look at \cite{BlondinFM18} for reachability graphs where vertices may have an infinite degree.}. 
We then may have an infinity of vertices and this is precisely why the reachability problem is such an interesting decidability problem.  

Let us recall some basic graph theory concepts that can be found in \cite{Diestel10} or \cite{Berge70} and are related to the notion of reachability and later to the notion of home space.

\subsubsection{Sources, sinks, and strongly connected components}
\label{subsubsec: ssc-sk}
We consider a model $M$, its reachability set $RS(M,Init)$ given a subset of states $Init$, its reachability graph $RG(M,Init)$.

\begin{definition}
\label{def: source & sink}
A subset of states $S$ is a \textit{source} in $RG(M,Init)$ if and only if $RS^{-1}(M,S)= S$.
Similarly, a subset of states $S$ is a \textit{sink} in $RG(M,Init)$ if and only if $RS(M,S)= S$.
\end{definition}

From these definitions, we directly deduce that the intersection and the union are stable over the sets of sources or the set of sinks.
\begin{property}
If $S_1 \text{and} S_2$ are two sources then $S_1 \cap S_2$ and $S_1 \cup S_2$ are two sources.
If $S_1 \text{and} S_2$ are two sinks then $S_1 \cap S_2$ and $S_1 \cup S_2$ are two sinks.
\end{property}
If $S_1$ and $S_2$ are sinks ans $x \in S_1 \cup S_2$ then $RS(M,x) \subseteq S_1 \cup S_2$ directly from the definitions of RS and sink. Hence, $S_1 \cup S_2$ is a sink. A similar argument can be used for the other statements of the property.
\hfill
$\square$

Usually, a source or a sink or are single vertices with no predecessor or successor respectively (see \cite{Boll02} for instance). 
Here, we expended their definition to a subset of vertices to take into consideration the possibility for a reachability graph to have an infinity of vertices.

If it is clear that a source can only be associated with an element of the set of initial states, it is worth pointing out that 
first, $Init$ is not necessarily a source (not even included in a set of sources); 
second, a sink is not necessarily associated with a \textit{terminal state} and in the case where a sink is finite, a sink can be interpreted as an unexpected system halt (sometimes called a deadlock); 
third, a terminal state is not necessarily a sink.

The reachability graph as any directed graph has its vertices partitioned into \textit{strongly connected components}. 
We recall here, their definition that can be found in \cite{Berge70} and holds for infinite graphs. 
Two vertices $x$ and $y$ are strongly connected if and only if there is a path from $x$ to $y$ and from $y$ to $x$. The reflexive closure of the relation \textit{strong connectivity} is clearly an equivalence relation, and a class defined by this relation is called a \textit{strongly connected component} (scc in short). 
If $x$ is a vertex then $S_x$ will denotes the scc containing $x$.
We immediately deduce from the definitions:
\begin{property}
\label{prop: sink, source, scc}
If $A \subseteq RS$  is a source then $\forall x \in A, S_x \subseteq RS^{-1}(M,x) \subseteq A$. Moreover, $RS^{-1}(M,x)$ is a source.

If $A \subseteq RS$  is a sink then $\forall x \in A, S_x \subseteq RS(M,x) \subseteq A$. Moreover, $RS(M,x)$ is a sink.
\end{property}
Let's consider a sink $A$ and $x \in A$ , then 
$R(M,x) \subseteq RS(M,A)$.
As a sink, we have $RS(M,A)=A$.

Since $S_x$ denotes the strongly connected component containing $x$, we have:
$S_x = RS(M,x) \cap RS^{-1}(M,x)$.
Therefore, $S_x \subseteq RS(M,x) \subseteq A$. 
From property \ref{prop: RS succession}, we have
$RS(M,(RS(M,x)=RS(M,x)$  therefore, $RS(M,x)$ is a sink.
\hfill
$\square$

\begin{definition}
An scc $S$ of $RG(M,q_0)$ is an \textit{scc source} (scc-s in short) if and only if $\nexists\ x \in RS(M,q_0) \setminus S$  such that $\exists\ y \in S$ and $x \rightarrow y$.

Similarly, we say that an scc $S$ is an \textit{scc sink} (scc-sk in short) if and only if $\nexists\ y \in RS(M,q_0) \setminus S$ such that $\exists\ x \in S$ and $x \rightarrow y$.
\end{definition}

Sources and sinks are interesting when they are minimal.
\begin{definition}
A source is \textit{minimal} if and only if it is not empty and contains no other sources but itself and the empty set.
Similarly, a sink is \textit{minimal} if and only if it is not empty and contains no other sources but itself and the empty set.
\end{definition}
Of course, if $x$ is a vertex then $RS(M,x)$ is a sink (property \ref{prop: sink, source, scc} but not necessarily a minimal sink.

\begin{theorem}
\label{th: sink, source min}
A source is minimal if and only if it is an scc-s.
Similarly, a sink is minimal if and only if it is an scc-sk.
If $S$ is an scc and $A \subseteq S$ is a sink (source respectively) then $S$ is a minimal sink (source respectively).
\end{theorem}

Let's $S$ be a minimal sink and $x_1, x_2 \in S$. $RS(M,x_1)$ and $RS(M,x_2)$ are sinks (by property \ref{prop: sink, source, scc}. 
Since $S$ is minimal, we have : $S = RS(M,x_1) = RS(M,x_2)$. 
Therefore, there exists a path from $x_1$ to $x_2$ since $x_2 \in RS(M,x_1)$ and a path from $x_2$ to $x_1$ since $x_1 \in RS(M,x_2)$. 
Hence $S$ is an scc-sk.
Let's consider $S$ an scc-sk and $A \subseteq S$ a minimal sink then $\forall x \in S, a \in A$ there is a path from $a$ to $x$ since S is strongly connected. Therefore, $x \in RS(M,a) \subseteq A$. Hence, $S = A$ and $S$ is a minimal sink.
If $S$ is an scc and there exists a sink $A \subseteq S$ then $\forall x \in S$ there is a path from an element of $A$ to $x$; therefore, $x \in RS(M,A)$ and $S = RS(M,A)$. Since $A$ is a sink, $S = RS(M,A) = A$ and $S$ is a minimal sink.
\hfill
$\square$

At last, let's consider models paired with $q_0$ as a single initial state. 
The reachability graph of $\langle M,q_0 \rangle$ is then a directed multigraph with exactly one strongly connected component source (scc-s) containing the vertex labeled by $q_0$ which is often called \textit{root}.
\begin{definition}
A \textit{root} $q_0$ is a distinguished vertex of the reachability graph $RG$ such that $\forall q \in RS,\ q_0 \overset{*}\rightarrow q$.
\end{definition}
Indeed, only vertices of the unique scc-s can be distinguished as root of RG, a root exists only if there is a unique scc-s. 

\subsection{Structures and topologies}
\label{subsec: struc}
In order to more precisely figure out how to move from one state to another, the notions of transition and Transition System will be introduced in Section \ref{subsec: transition systems} where in addition, the definition of reachability graph will be refined further.

As soon as a conceptual model allows organizing states we will say that it possesses a \textit{structure}, notion that we will keep informal for now and which is referring to the information detained about states, transitions, or state variables. Therefore, a reachability graph is an example of a structure. 
A structure can hold more or less information about the states of the SuS and is key to support accurate and efficient behavioral analysis. 
In particular, the reachability graph is a powerful tool to study all kind of properties such as reachability questions or to solve the reachability problem.

\section{Other behavioral properties}
\label{sec: behav}
In this section, we consider a pair $\langle LTS, q_0 \rangle$ where $q_0$ is the initial state of a Labeled Transition System (LTS) equipped with $\leq$ an ordering relation in $Q$.

\subsection{Monotonicity and repetitivity}
A first property that we can already find in \cite{BR82} p.43 or in \cite{V81} relatively to PNs is the property of monotonicity. It was expanded in \cite{Fin20, FinkelS01} for LTS with the notion of strong monotonicity. 

\begin{definition}
A transition $t$ is \textit{strongly monotone} if and only if $\forall (q,q') \in Q^{2}$ such that $q \leq q'$ and $\texttt{Dom}(t)$, we have $q' \in \texttt{Dom}(t)$ and $\forall p \in t(q) \exists p' \in t(q')$ such that $p \leq p'$. We say that LTS is \textit{strongly monotone} if and only if every transition of $T$ is strongly monotone.
\end{definition}

We also find in \cite{Fin20} a notion more relaxed of monotonicity. 
We say that an LTS is \textit{monotone}
if and only if $\leq$ is upward compatible. 
Strong monotonicity preserves traces, that is to say that a trace in $\langle LTS, q \rangle$ still exists in any $\langle LTS, q' \rangle$ such that $q \leq q'$.
\begin{definition}
A trace $\sigma$ is \textit{repetitive} if and only $\forall q \in \texttt{Dom}(\sigma), \forall n \in \mathbb{N}, \sigma^n$ is a trace in the LTS paired with $q$.
\end{definition}

\begin{lemma}
Let's LTS be strictly monotone and $\sigma$ be a trace, if $\exists\ q,\ q' \in Q$ such that $q \leq q'$ and $q\overset{\sigma}{\rightarrow}q'$ then $\forall n \in \mathbb{N},\ \sigma^n$ is a trace of $\langle LTS,q \rangle$ and $\sigma$ is repetitive.
\end{lemma}

\subsection{Liveness, Starvation, and Fairness}
\label{subsec: liveness}
\begin{definition}
\label{def: live}
A transition $t$ is \textit{live} in $\langle LTS, Init \rangle$ if and only if 

$\forall q \in RS(M,Init),\  \exists \ q' \in RS(M,q)$ such that $q' \overset{t}{\rightarrow}$, 

or equivalently, if and only if $\forall q \in RS(M,Init),\ 
q \overset{* t}{\rightarrow}$.
\end{definition}

We say that the pair $\langle LTS, q_0 \rangle$ is live if and only if every transitions of $T$ are live.

Liveness is an important property for a transition $t$. It says that whatever is the progression of the model, it will always be possible to execute $t$. 
However, this property does not prevent the existence of infinite progression where $t$ never occurs. At any step of such a progression, a derivation that allow to execute $t$ must be possible for $t$ to be live. 
Such situations lead to the notions of starvation and fairness.

The term \textit{starvation} comes from a famous case study called the ``five dining philosophers" by Dijkstra described in \cite{Dijkstra71} then in \cite{CARH15} or in \cite{BEISW20} among many other authors. 
In this use case, two philosophers have the ability to eat and at the same time starve their their common neighbour. Despite this possible situation, the system is live, that is to say that at any moment, a philosopher has the ability to eat.

Later, came the term \textit{fairness} for which we found the following informal definitions of : 
 
 ``every statement is selected infinitely often" in \cite{ChMi88} p9,

``every event which becomes possible infinitely often is realized infinitely often" in \cite{QSif83}.
\subsection{Home spaces and home states}
\label{subsec: HS}

The notion of home space was first defined in \cite{M83} for Petri Nets relatively to a single initial state. 
Here, we effortlessly extend its definition relatively to a nonempty subset of states and beyond Petri Nets to any CM equipped with a reachability graph. 

Home spaces are extremely useful to analyze various behavioral properties such as liveness, or resilience (see \cite{FinHil24} for a formal definition of resilience).
This notion can also be used to support the proof of behavioral properties that require to eventually become satisfied (for a subset of states) after executing a specific sequence of transitions. 

Verifying that a set is a home space is not always easy. 
In this regard, the knowledge of the reachability sets and associated reachability graph is often very useful. 
A corpus of decidability properties can be found in \cite{ValkJ85,EJ89} or more recently in \cite{FinHil24,JALE22,JLr2024}. 
For instance, in \cite{JALE22}, the fact that a given set is an A-home space is proven decidable for home state when the conceptual model is a Petri Net but is still open in the general case.

\subsubsection{Definitions and basic properties}
Given a model $M$, its associated state space $Q$ and a subset $Init$ of $Q$, we say that a set \emph{HS} is an \textit {Init-home space} if and only if for any progression of the model $M$ from any element of $Init$, there exists a way of  prolonging this progression and reach an element of \emph{HS}. In other words:

\begin{definition}
\label{def: homespace}
Given a nonempty subset $Init$ of $Q$, a set $\emph{HS}$ is an \textit{Init-home space} if and only if $\forall q\in RS(M,Init)$, $\exists h \in \emph{HS}$ such that $h$ is reachable from $q$ (i.e. such that $q \overset{*}{\rightarrow}h$).

$\emph{HS}$ is a \textit{well-structured home space} if and only if: 
$\forall q\in Q$, $\exists  h \in \emph{HS}$ such that $q \overset{*}{\rightarrow}h$.
\end{definition} 
In \cite{JALE22}, we can find for Petri Nets, an equivalent definition : $\emph{HS}$ is an \textit{Init-home space} if and only if $RS(M,Init) \subseteq RS^{-1}(M,HS \cap Q)$.

We must point out that $HS$ is not necessarily included in $Q$ and we may have $h \in HS$ such that for some state variables $x_i, \ h(x_i) \notin D_i$ ($D_i$, being the domain in which $x_i$ varies in $Q$); see examples section \ref{sec: ex}.
Of course, we immediately have $HS \cap RS(M,Init) \neq \varnothing$.
 
\begin{property}
\label{prop: struct-home}
If $HS$ is a well-structured home space then $\forall Init \in 2^Q \setminus \{\Emptyset\}$, $HS$ is an \textit{Init}-home space.

Also, if $A$ is such that $Q=RS(M,A)$ and $\emph{HS}$ is an A-home space then $\emph{HS}$ is a well-structured home space. 
\end{property} 
\begin{definition}
\label{def: home state}
Given a nonempty subset $Init$ of $Q$, a state $s$ is an \textit{Init-home state} if and only if 
$\{s\}$ is an \textit{Init-home space}. 
\end{definition}
If $s$ is an Init-home state, then it is straightforwardly an $\{s\}$-home state and we simply say that $s$ is a home state when there is no ambiguity. 
This is the usual definition that can be found in \cite{BR82} p.59 or in \cite{ColomTS2003} p. 63 for Petri Nets. 
When $Init = s$, it is said that $\langle M, s \rangle$ is \textit{reversible} \cite{ColomTS2003,Ler13}. 
A Petri Net can have a home state without being reversible. 
In \cite{Ler13}, this notion is defined as a relation between two states which will be in a \textit{reversible relation} if and only if they are reachable one from each other.

\begin{property}
\label{home-state}
    Considering a model $M$ paired with a single initial state $q_0$, the three following statements are equivalent: 
    \begin{itemize}
    \item [(i)] the initial state is a home state (i.e. $\langle M,q_0 \rangle$ is reversible), 
    \item [(ii)] any reachable state is a home state, 
    \item [(iii)] the reachability graph is strongly connected.
    \end{itemize}
\end{property} 

If $q_0$ is the initial state, then $\forall x, y \in RS(M, q_0)$, there exist a path from $q_0$ to $x$ and a path from $q_0$ to $y$ and since $q_0$ is a home state there also exist a path from $x$ to $q_0$ and from $y$ to $q_0$ in the reachability graph. Hence, $q_0, x, y$ belong to the same strongly connected component. We easily conclude that the reachability graph is strongly connected. The other elements of the property become obvious.
\hfill
$\square$

In many systems, the initial state $q_0$ represents an \textit{idle} state from which the various capabilities of the system can be enabled. In this case, it is important for $q_0$ to be a home state.
This property is usually guarantied by a reset function which can be modeled in a simplistic way by adding a transition $r$ such that $\forall q \in RS(M,q_0),\ q_0 \in r(q)$ (which means that $r$ is executable from any reachable state and that its execution reaches $q_0$). 
However, requiring to add too much complexity to $RG$ (one edge per node), this function is most of the time abstracted away when describing the corresponding system model.

\begin{figure}[ht]
\centering
\includegraphics[width=0.4\textwidth]{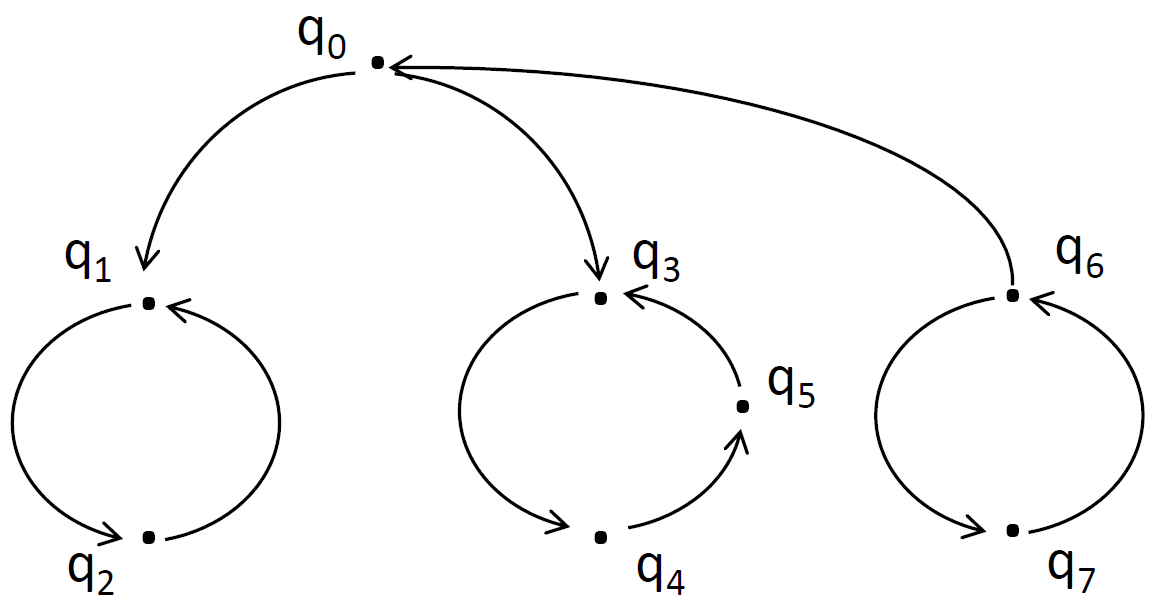}
\caption{$Q=\{q_0,q_1,q_2,q_3,q_4,q_5,q_6,q_7\},\ HS_1 = \{q_1,q_3,q_4\},\ HS_2= \{q_1, q_5\}, \ HS_3=\{q_1,q_3,q_5\}$ are three $\{q_0\}$-home space. 
$HS_4=\{q_1, q_4,q_7\}$ is a $\{q_6\}$-home space as well as a $\{q_0\}$-home space.
}
\label{fig: inter-hs}
\end{figure}

In this regard, it may be worth mentioning the straightforward following properties, given a subset $A$ of states: 
\begin{property}
\label{prop: inter-home}
    Any set containing an A-home space is also an A-home space. If $HS$ is an A-home space, it is a B-home space for any nonempty subset $B$ of $A$. If $HS_1$ is an A1-home space and $HS_2$ is an A2-home space then $HS_1 \cup HS_2$ is an ($A1 \cup A2$)-home space.  
\end{property} 
However, the intersection of two home spaces is not necessarily a home space. Figure \ref{fig: inter-hs} is representing the reachability graph of a model with eight states, $HS_1, HS_2, HS_3$ as defined Figure \ref{fig: inter-hs} are three $\{q_0\}$-home spaces. 
While $HS_1 \cap HS_3 = \{q_1, q_3\}$ is a $\{q_0\}$-home space, $HS_1 \cap HS_2 = \{q_1\}$ is not a $\{q_0\}$-home space (even if it is a $\{q_1\}$-home state).
    
\subsubsection{Sinks and Home spaces}

\label{subsubsec: sink+HS}
\begin{property}
\label{prop: sink & home spaces}
If there exists a sink $q_f \in RG(M,Init)$ then it belongs to any Init-home space.
More generally, any home space has at least one element in each strongly connected component sink (scc-sk) of the reachability graph.
At last, if $S$ is the only scc-sk of $RG(M,Init)$ then $S$ is an Init-home space. Moreover, $q \in S$ if and only if $q$ is an Init-home state.
\end{property}

It is easy to prove that this property holds even when the reachability graph is infinite considering that the definitions of sources, sinks, or strongly connected components are the same as in the case where the directed reachability graph is finite. 

\subsection{Boundedness}
\label{secsec: boundedness}

As soon as $\mathcal{SV}$ and $Q$, its associated state space are defined, the observer of the corresponding model $M$ will want to understand and if possible, simplify and reduce the domains in which state variables vary. 
Doing so will also reduce the state space $Q$ (see property \ref{prop: Qdef}) and accelerate analysis based on state exploration. 
Furthermore, he will want to see if their progression follow simple rules or equations, or if there exists an upper bound for a subset of $\mathcal{SV}$ during the potentially infinite progression of $M$.

\begin{definition}
\label{def: boundedness}
Given the sets $X$ and $A$ such that $X \subseteq  \mathcal{SV},\ A \subseteq Q$, we say that $X$ is \textit{A-bounded} (or bounded when there is no ambiguity) 
if and only if: 

$\exists k \in \mathbb{N}$ such that $\forall x \in X ,\ \forall q \in RS(M,A),\ q(x) \leq k$.

If $X=\mathcal{SV}$ then we say that $M$ is A-bounded.

If $\exists k \in \mathbb{N}$ such that $\forall x \in X ,\ \forall q \in Q,\ q(x) \leq k$ then $X$ is \textit{structurally bounded}.

If $X=\mathcal{SV}$ and $RS(M,A)=Q$ then $M$ is structurally bounded.

\end{definition}

\begin{figure}[ht]
\centering
\includegraphics[width=0.6\textwidth]{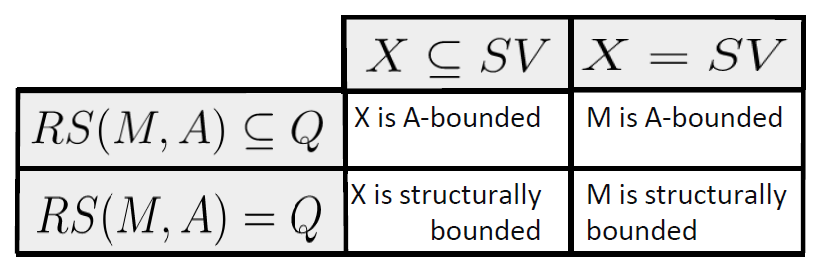}
\caption{Different definitions of boundedness according to the fact that every state variable is bounded or not for every possible state or not.}
\label{Boundedness}
\end{figure}

These somewhat overlapping definitions are positioned in relation to each other in Figure \ref{Boundedness}.

The fact that a state variable variation follows a rule or an equation along the model progression will be illustrated in examples in Section \ref{sec: ex}. Structural boundedness  will be discussed again in particular in Section \ref{subsubsec: semi}.

\section{Conceptual Models: from transition systems to Petri Nets}
\subsection{Transition Systems (TS) and Labeled Transition Systems (LTS)}
\label{subsec: transition systems}

Many papers use the notion of Transition Systems (sometimes State Transition Systems as in \cite{ChMi88}) without even providing a formal definition. 
Hereunder, we use the definitions introduced in \cite{Kel72,K76}, then definitions by \cite{Sankar03} and \cite{FinkelS01} will be introduced. 
We point out that the definition of Transition Systems given in \cite{L76a} is in fact equivalent to the definition that will be provided for Petri Nets in Section \ref{subsub: PN}. 

\begin{definition}
\label{def: TS}
A \textit{Transition System} is a pair $TS=\langle\emph{Q}, \rightarrow \rangle$ where: $\emph{Q}$ is a non empty \textit{state space}, 
$\rightarrow \subseteq \emph{Q} \times \emph{Q}$ is a binary relation on $\emph{Q}$ and is called the \textit{set of transitions}. $(q,q') \in \rightarrow $ is denoted by $q \rightarrow q'$.
\end{definition}

$\emph{Q}$ as well as $\rightarrow$ can be countably  infinite. Neither states nor transitions are precisely defined. This is intentional to make the conceptual model as general as possible. No wonder the reachability problem is not decidable for $TS$.
However, a large class of Transition Systems called Well-Structured Transition Systems (WSTS) were introduced in \cite{FinkelS01} (see Section \ref{secsec: WSTS}) for which questions close to the reachability problem are decidable thanks to a particular ordering over Q which considerably enrich the conceptual model's structure.

At this stage, Transition Systems are a conceptual model with no topology. States are just elements of $Q$ with no details (no information is provided on state variables in the definition of a Transition System). 
A model is simply described by successive states it can take during its progression (i.e. traces, see Figure \ref{fig: trace}).
Each pair of states of the binary relation $\rightarrow$ is an anonymous transition: they have no names to help reporting about possible properties.
Moreover, as systems under study are able to perform a finite set of possible indivisible actions, we will want to handle a \textit{finite} set of \textit{identifiable} transitions.

In this regard, we consider the notion of Labeled or named Transition System (\emph{LTS}) that we borrow from \cite{K76} but that we can find in many other publications such as \cite{JS20}.

\begin{definition}
\label{def: LTS}
A \textit{Labeled (or Named) Transition System} is a triple $LTS=\langle \emph{Q}, T, \rightarrow \rangle$ where: 
$\emph{Q}$ is a non empty set of states, 
$T$ is a non empty set of labels and
$\rightarrow \subseteq \emph{Q} \times T \times \emph{Q}$ relates labels with states.
\end{definition}

As for Transition Systems, $\emph{Q}$ can be infinite as well as $T$. 
The reader will have recognized the key elements of the definition of an automaton (as defined p.51 of \cite{Saka09}) where $T$ is an alphabet and where only initial and terminal states are missing to the definition. 
Actually, one element $q_0$ of $\emph{Q}$ is often distinguished and called \textit{initial state}, however, terminal states are scarcely used since most of the time, the system under study is supposed to run forever. 

Henceforth, we will call \textit{transition} a label $t$ of $T$; $t$ will also denote a relation in $\emph{Q} \times \emph{Q}$ such that $(q,q') \in t$ if and only if $(q,t,q') \in \rightarrow$, in that case, we will use the notations $q\overset{t}{\rightarrow}q'$ or $q' \in t(q)$. 
We can compose transitions to form computations (see \cite{Saka09} p.54) and sequences of transitions : if $(q_{i-1},q_i) \in t_i$ for $i \in \{1,...,n\}$ we then form $(q_0,t_1,q_1),(q_1,t_2q_2),...(q_{n-1},t_n,q_n)$ defined as a \textit{computation} (or run) and denoted by: $q_0 \overset{t_1}{\rightarrow}q_1\overset{t_2}{\rightarrow}q_3...\overset{t_n}{\rightarrow}q_n$. 
We then say that $\sigma = t_1t_2...t_n$ is a trace as defined in Section \ref{subsec: TTL} and $(q_0,q_1,...q_n)$ is a trajectory as defined in Section \ref{sec: reachability}. We write $q_0 \xrightarrow[]{t_0t_1...t_n}q_n$ or $q_0 \xrightarrow[]{\sigma}q_n$, or sometimes $q_n \in \sigma(q_0)$. 
We also say that $q_n$ is reachable from $q_0$
\footnote{In \cite{FinkelS01}, a computation is a maximal sequence of transitions. This extends the definition in \cite{Saka09} where a successful computation starts from the set of initial states to reach a state in the set of terminal states.}. 

We must keep in mind that a transition is a relation and in general, cannot be seen as a function or even a partial function from $\emph{Q}$ to $\emph{Q}$ (we will see however, that Petri Nets assume that transitions are partial function). 

\subsection {Labeled Reachability Graph (LRG) and behavioral properties}
\label{subsec: LRG}

Now that we have names for identifying transitions, we can move to the notion of Labeled Reachability Graph from the Reachability Graph defined in Section \ref{sec: reachability}. Moreover, we can speak of behavioral properties focusing on frequencies or patterns of sequences of transitions such as liveness.

\subsubsection{Labeled Reachability graph (LRG)}

Given a Labeled Transition System $M = \langle Q,T,\rightarrow \rangle$ paired with an initial state $q_0$, $LRG (M, q_0)$ denotes the reachability graph of $\langle M,q_0 \rangle$. $LRG (M, q_0)$ is a directed connected multigraph $( RS(M,q_0),E )$ where $RS(M,q_0)$ is the set of vertices and E the set of edges such that $(x,y) \in E$ if and only if $\exists t \in T$ such that $x\overset{t}{\rightarrow}y$. The edge (x,y) is then labeled by $t$ and we then say that t \textit{appears as a label} in $LRG (M,q_0)$.

$LRG(M,q_0)$ differs from $RG(M,q_0)$ only by the addition of labels on its edges. $LRG(M,q_0)$ (LRG for short) can be an infinite graph with an infinite number of ends even if $T$ is finite and LRG locally finite (see \cite{Diestel10} chapter 8 for a comprehensive study on infinite graphs).
Of course, 
there is a path from $q_0$ to any vertex $x$ of LRG (which makes $LRG$ a connected directed graph with at most one root).

The set of \textit{computations} of $\langle M, q_0 \rangle$ is exactly the set of labeled paths of LRG. 
Let $c = x, q_1,...q_n,y$ be such a path from $x$ to $y$ and $t_0,t_1,...,t_n$ be the sequence of labels on the edges of this path. The \textit{sequence of states} $x, q_1,...q_n,y$ is said to be a \textit{trajectory} while the \textit{sequence of transitions} $t_0,t_1,...,t_n$ constitutes a \textit{trace} also denoted by the word $t_0t_1...t_n$ of $T^{*}$. 

\subsubsection{Determinism as a property between traces and trajectories}
A trace $t_1t_2...t_n$ \textit{corresponds} to a trajectory $a_0,a_1,...a_n$ or reversely, a trajectory $a_0,a_1,...a_n$ corresponds to a trace $t_1t_2...t_n$ if and only if $\exists \ q \in Q$ 
such that we have the labeled path $a_0 \overset{t_1}{\rightarrow}a_1\overset{t_2}{\rightarrow}a_3...\overset{t_n}{\rightarrow}a_n$ in $LRG(M,q)$.

Let us point out that traces and trajectories are not in general in a one-to-one correspondence. Several different trajectories may correspond to the same trace even if they start from the same state and several different traces may correspond the same trajectory. 
For a given model, a one-to-one correspondence would at least involve the two following properties:

\begin{definition}
\label{prop: deterministic}
We say that an LTS is \textit{deterministic} if and only if the two following properties are satisfied:
\item[P1]For any given sequence of states $\rho$ there exists at most one sequence of transitions to form a computation with $\rho$.
\item[P2]for any given state $q$, a given sequence of transitions $\sigma$ there exists at most one sequence of states starting at $q$ to form a computation with $\sigma$.
\end{definition}

To obtain such a one-to-one correspondence would require to put some restrictions on the set of transitions such as the ones of the following lemma which makes the properties P1 and P2 easily verifiable.

\begin{lemma} 
\label{lem: deterministic}
In an LTS, we have the two following properties:
\item[L1] Any sequence of states has at most one corresponding trace if and only if:
$\forall t_1, t_2 \in T,\ 
\forall q \in Q, \
t_1(q) \cap  t_2(q) = \varnothing $

\item[L2] Any sequence of transitions has at most one corresponding trajectory starting at a given state $q$ if and only if any transition is a partial function.
\end{lemma}
The proofs are somewhat similar for L1 and L2, so they are jointly performed. 
If we consider a sequence of only two states $q_i$ and $q_f$, according to P1, there is at most one trace to form a unique computation. This trace can only be a single transition. So, there is at most one transition between any couple of states therefore $\forall t_1, t_2 \in T,\ 
\forall q \in Q, \
t_1(q) \cap  t_2(q) = \varnothing $.
Similarly, if we consider a sequence of one transition $t$ and one state $q_i$, from P2 we can form at most one computation, therefore from $q_i$ $t$ can reach at most one state. Hence, $t$ is a partial function.

Reversely, 
we consider a sequence of states $ \rho =a_0, a_1,...a_k$, two traces $\sigma=t_1,...t_k$ and $\sigma'=t'_1,...t'_k$ forming a computation with $\rho$, and $i$ the first index such that for $j<i, t_j=t'_j$ and $t_i \neq t'_i$, then we would have a contradiction with $t_i(q_i) \cap  t'_i(q_i) = \varnothing$.

At last, we consider a sequence of transitions $\sigma = t_1...t_k$ and two trajectories $\rho=q_i,...q_k$ and $\rho'=q_i,q'_1,...q'_k$ starting at the same state $q_i$ and forming a computation with $\sigma$ and $i$ the first index such that for $j<i, q_j=q'_j$ and $q_i \neq q'_i$, then we have $\{q_i,q'_i\} \subseteq t_i(q_{i-1})$ which would be a contradiction with the fact that $t_i$ is a partial function.
\hfill 
$\square$ 

It is worth pointing out that the property P2 means in particular that given an initial state $q_i$ and a sequence of transitions $\sigma$ there is at most one state $q_f$ such that $q_i\overset{t}{\rightarrow}q_f$. This is consistent with the intuitive notion of determinism \footnote{For instance in \cite{ChMi88}, the 'N' in UNITY stands for non-determinism with the following: ``Different runs of the same program may execute statements in different orders, consume different amounts of resources, and even produce different results."}.
In \cite{E74}, deterministic machines are defined with conditions similar to the two properties L1 and L2 of lemma \ref{lem: deterministic}. 
We also found in \cite{Saka09} p.77 a  property slightly different from L2 of lemma \ref{lem: deterministic} stating that any deterministic automata (briefly meaning that any transition is a partial function) is an unambiguous automata (briefly meaning that any sequence of transitions starting at an initial state and finishing at a terminal state forms at most one computation). 

Determinism is a good property especially when it is about to guarantee that a test is reproducible. However, many authors define non-determinism in their conceptual model to take into account statements that are independent and can run in parallel \cite{ChMi88} or to represent a possible arbitrary choice (between guarded commands \cite{Dijkstra75} or between sequential processes \cite{CARH15}).

\subsection{Well Structured Transition Systems (WSTS)}
\label{secsec: WSTS}

Motivated by extending the domain of CMs for which boundedness and other behavioral properties would be decidable, Finkel (see \cite{FinkelS01} in particular) introduced the conceptual model of well structured transition systems as a subclass of TS or LTS. In this regard, the author enriched a TS or an LTS with an order relation denoted $\leq$ in $Q \times Q$. Here we give the definition with an LTS notation:

A labeled transition system is a \textit{well structured transition system} (WSTS) $M = \langle Q,T,\rightarrow,\leq  \rangle$  where $ \langle  Q,T,\rightarrow  \rangle$ is an LTS and:

$\leq$ is reflexive, transitive, and such that for any infinite sequence of states $\rho = q_1,...q_k,...$ there exist two indices i, j such that $i < j$ and $q_i \leq q_j$. 
We then say that $\leq$ is a \textit{well-quasi-ordering} (wqo),

$\forall (q,q',s) \in Q^{3}$ such that $q \leq q'$ and $t \in T$ such that $s \in t(q)$ then there exist a sequence $\sigma \in T^{*}$ and a state $s'$ such that $s' \in \sigma (q')$ and  $s \leq s' $. We then say that $\leq$ is \textit{upward compatible} with $\rightarrow$.

The definition holds with using a TS instead of an LTS (i.e. with no name for the set of transitions).

\subsection {Petri Nets and a handful of derived models}
\label{subsec: PN-plus}
In order to support an algebraic invariant calculus, dependencies between transitions and state variables need to be more precise than with TS or LTS. 
We call \textit{topology} the set of information referring to dependencies between transitions and state variables in the same informal fashion as the notion of structure. A topology must present enough information for building $RS$ up. We find this topology in Petri Nets. In \cite{PETRI96}, Petri extended much further the notion of topology, however, we will limit ourselves to Petri Nets as defined in this section.

Petri Nets are among the very first conceptual model to present a topology under the format of a bipartite graph between transitions and state variables.

\subsubsection{Petri Nets (PN)}
\label{subsub: PN}
Petri Nets have been introduced by C. A. Petri circa 1962 \footnote{Peruse the book cited in \cite{M19} for understanding the influence of C. A. Petri in this research community.}. Here, we give definitions that are similar to definitions in \cite{BR82} or in \cite{GV03}. It can be  pointed out that \textit{semi-transition systems} defined in \cite{L76a} are equivalent to PNs.

\begin{definition}
\label{def: PN}
A \textit{Petri Net} is a tuple $PN = \langle P, T, Pre, Post \rangle$, where $P$ is a finite set of \textit{places}, $T$ a finite set of \textit{transitions} such that $P \cap T = \varnothing$. 
A transition $t$ of $T$ is defined by its $Pre(\cdot,t)$ and $Post(\cdot,t)$ \textit{conditions} 
\footnote{We use here the usual notations: $Pre(\cdot,t)(p) = Pre(p,\cdot)(t) = Pre(p,t)$ and  $Post(\cdot,t)(p) = Post(p, \cdot)(t) =Post(p,t)$.}, i.e. its pre and post dependencies with P:
$Pre: P \times T \rightarrow \mathbb{N}$ is a function providing a weight for pairs ordered from places to transitions, $Post: P \times T \rightarrow \mathbb{N}$ is a function providing a weight for pairs ordered from transitions to places.
\end{definition}

Extensive definitions, properties, and case studies can be found in particular in \cite{BR82,GV03}.

In \cite{PETRI96}, $P$ and $T$ can be infinite; however in this note, we will consider $P$ and $T$ finite which is the common convention. $d$ will denote the cardinal number of $P$, $m$ the cardinal number of $T$.

The dynamic behavior of Petri Nets is modeled via markings. A  \textit{marking} $q$ is a function from $P$ to $\mathbb{N}$.
When $q(p) = k$, it is often said that the place $p$ contains $k$ \textit{tokens}.
A \textit{marked Petri Net} is a pair $ \langle PN, q_0 \rangle$ where $PN$ is a Petri Net and $q_0$ its initial marking.

We dynamically progress from one state $q$ to the next $q{'}$ by \textit{executing} \footnote{The words 'firing' or 'occurring' are often used in the literature.}  a transition $t$ from $q$ if and only if its preconditions ($Pre$) are satisfied i.e.:
\begin{equation}
\label{eq: precondition-t}
q \geq Pre(\cdot,t)
\end{equation}
In other words, in a PN we have: $\texttt{Dom(t)} = 
\left\{ q \in Q \ |\ q \geq Pre(\cdot,t)\right\}$.

We let the reader verify that regarding property \ref{prop: deterministic} and for any PN, P2 holds but not P1.

We say that $t$ is \textit{enabled} in q.
Then, $q{'}$ can be computed from $q$ with the following equation:
\begin{equation}
\label{eq: state-by-t}
q{'}= Post(\cdot,t)-Pre(\cdot,t)+q
\end{equation}
A marked Petri Net $ \langle PN, q_0 \rangle$ 
can be often associated with a particular class of Labeled Transition Systems $\langle S, T_{TS}, \rightarrow \rangle$ paired with the same initial state $q_0$ such that: 

\begin{lemma}
\label{lemma: PN-LTS}
For any marked PN $\langle \langle P,T_{PN},Pre,Post \rangle, q_0 \rangle$ we can associate an LTS $\langle S, T_{TS}, \rightarrow \rangle$ with the same initial state $q_0$ such that:
 $T_{TS}=T_{PN}$, $S = RS(PN,q_0)$,  and $\rightarrow$ is such that: $(s,t,s') \in \rightarrow$ if and only if  $ s \geq Pre(\cdot,t)$ and  $s'=Post(\cdot,t))-Pre(\cdot,t )+s$. 
\end{lemma}
This association is non injective: several marked Petri Nets may be associated with the same LTS (since loops can be added to a Petri Nets without modifying its behavior). A similar association can be found in \cite{Sankar03}. 
In this note, in order to follow the idea of the lemma, places will model state variables and transitions will model atomic actions described by equations over these variables (see equation \ref{eq: state-by-t}).

From this analogy, we can transfer the definitions of reachability set and home space to Petri Nets and other related conceptual models, a state being called a \textit{marking}, state variables being called \textit{places}. 

A Petri Net is often represented by an \textit{incidence matrix} over $\mathbb{Z}$ directly derived from $Pre$ and $Post$ considered as matrices of dimension $P \times T$ with entries in $\mathbb{N}$:
\begin{equation}
	C = Post-Pre 
\end{equation}
It must be pointed out that $C$ does not take into account loops between a place $p$ and a transition $t$ (we have a loop between $p$ and $t$ if $Pre(p,t) \times Post(p,t) > 0$, in particular If $Pre(p,t) = Post(p,t)$ then $C(p,t)=0$ as if there would be no connection (i.e. dependency) between $p$ and $t$). In PNs, a loop between $p$ and $t$ is nevertheless useful since it can model a test on a state variable $p$ constituting a pre-condition to execute $t$ without modifying the value of $p$. 

As for LTS, a trace or a sequence of transitions $r$ is associated with a vector $\overrightarrow{r}$ such that $\overrightarrow{r_i}$ is the number of occurrences of the transition $t_i$ in the sequence $r$. $\overrightarrow{r}$ is sometimes called the Parikh vector of $r$ since $r$ can be seen as a word of $T^{*}$ (by lemma \ref{lemma: PN-LTS} a Petri Net can be seen as a Labeled Transition System which in turn can be seen as an automata!).
If from $q_0$, we reach $q$ by executing the sequence $r$ then we can compute $q$ with the equation:
\begin{equation}
 q = C\overrightarrow{r} + q_0
 \label{eq: state}
\end{equation}
This equation is known to as the \textit{state equation} in reference to control theory. 

First, let's recall that the matrix $C$ lost possible loops, second moving from $r$ to $\overrightarrow{r}$ looses all information about the order in which the transitions are executed.
Therefore, one should not expect to deduce everything about the behavior of the PN from this single equation. 
It is nevertheless legitimate to start any NP analysis by studying the equation of state for at least three reasons.
First, one can still deduce many properties (as we will see in the examples of the following sections)
Second, the complexity of other algorithms that would bring a more complete analysis is often exponential. 
Finally, the information provided by the study of the state equation can allow simplifying more complex algorithms by pruning away states that can be proven unreachable because they would contradict the state equation.

Since $P$ and $T$ are disjoint, the functions $Pre$ and $Post$ define a bipartite graph between $P$ and $T$.
Petri Nets have inherited a pictorial representation which has not been a minor factor in their success compared to other similar models such as Vector Addition Systems (VAS) (introduced in \cite{KM69} and used in \cite{L09} to solve the reachability problem). 
Transitions are represented by rectangle (or bars) and places by circles (see Figure \ref{fig: tiny} or \ref{Mame} for examples). If $Pre(p,t) = k>0$ then the graph will have an edge from $p$ to $t$ labeled by $k$. If $Post(p,t) = k>0$ then the graph will have an edge from $t$ to $p$ labeled by $k$. We thus formed a bipartite graph: there is no edge neither between any two transitions nor between two places.

If for a place $p$, $q(p)=n$ we will say that $p$ contains $n$ \textit{tokens} represented by the letter $n$ (by the value of $n$ when this value is known or by $n$ dots when $n$ is sufficiently small) inside the circle representing the place $p$ (see Figure~\ref{mutex} for a first example, Figures~\ref{fig: tiny} and ~\ref{mutex-param} for examples with parameters). As the Petri Net evolves from marking to marking, the number of tokens in places evolves dynamically, this is often denoted by the \textit{token game} of a given Petri Net.

\subsubsection{About other models derived from Petri Nets}

Often, researchers or engineers would find Petri Nets too low level to describe complex systems and would  find Petri Nets almost as complicated to handle as Turing machines in terms level of abstraction if it was not for their handy graphical aspect. 
Indeed, we know since the reachability problem has been solved for Petri Nets with finite sets of places and transitions (see \cite{L09} for an elegant proof using VAS (i.e. PN with no loops \footnote{This slight difference is already noticed in \cite{L76} later in \cite{V81}})) that Petri Nets are strictly less powerful in terms of algorithmic description than Turing machines.

Stressing the need for more concise conceptual models providing a higher level of abstraction, many (maybe too many) proposals were introduced to enrich the token game. These conceptual models are built from PNs using two methods.

First, they are making the enabling rule more elaborate, using for instance, inhibitor arcs (to be able to test that the marking of a place is null), priorities between transitions, time (introduced in \cite{Ram73}), or probabilities. 

Second, they are making token types more complex. 
One of the most famous extension is called \textit{color} \cite{JR91, GV03}: a place $p$ can take its values in $\mathbb{N}^{\mathcal{C}}$ where $\mathcal{C}$ is an enumeration of colors potentially infinite. When $\mathcal{C}$ is finite, it is possible to ``unfold" $p$ to get a traditional Petri Net. 
Other authors transform tokens into letters (introduced in \cite{M83, FinkelM83}), predicates \cite{Genrich1987}, or abstract data types \cite{VautherinM84}. 
At last, let us mention the self-modifying Petri Nets \cite{Valk78} where the enabling rule itself dynamically evolves depending on a linear combination of the current state (expanded with a constant). 
Self-modifying Petri Nets include a number of derived models; they allow very concisely modeling entangled enabling rules supporting the description of complex and flexible systems such as collaborative processes \cite{GinM93}.

As long as they retain a bipartite topology between places and transitions and define a token game of some sort, these extensions keep contributing to forming the wide family of Petri Nets variants.

\newpage
\section{What exactly is an invariant?}
\label{sec: invariant}

\subsection{Towards a definition}
\label{subsec: invariant-definition}


Intuitively, each time you start a sentence by a phrase such as \textit{{'}is constantly{'}}, \textit{{'}from now on{'}}, \textit{{'}must always{'}}, \textit{{'}must never{'}}, \textit{{'}for all possible situation{'}}, \textit{{'}is impossible under any circumstances{'}}, or \textit{{'}never will{'}} you can suspect that your sentence can be associated with an invariant. However, the use of words such as \textit{{'}almost{'}}, \textit{{'}sometimes{'}},
\textit{{'}likely{'}}, or
\textit{{'}probable{'}} can change everything and let you suspect that your sentence will almost surely not be considered as an invariant. Actually, if you start to interleave these phrases, a reader may quickly have some hard time to determine what exactly the resulting statement is meant. For instance, \textit{{'}is almost constantly{'}} could sound like an oxymoron and certainly does not suggest an invariant property. This may explain why it is so uneasy to capture the notion of invariant in a consensual manner.

``In mathematics, an invariant is a property of a mathematical object (or a class of mathematical objects) 
which remains unchanged after operations or transformations of a certain type are applied to the objects." 
\footnote{https://en.wikipedia.org/wiki/Invariant\_(mathematics)}
In our case, an invariant would be a property over state variables, states, or transitions which remain unchanged after transitions of a certain type are applied to the model under consideration.

In computer science or in software engineering, an invariant is a system behavioral property or relationship the value of which does not vary over time, that is over the evolution of the said system. It can be a property stable over the entire state space of the system under study; it can also be a sequence of actions that happens regularly in a predictable manner within a context defined by a set of initial sets.

In system modeling, invariants will be used to describe behavioral requirements or specifications as for instance, in \cite{CARH15}. 
In programming, invariants will be used to take care of system exceptions which are often used to support the detection of a possible physical failure or a software bug and provide valuable guards (preventing from an unwanted event leading to an erroneous state) or alerts (detecting an unwanted event requiring the execution of an exception). Often, the concepts of constraint, property, strong requirement (which must always happen), assertion, or exception (which must never happen) will be described by using an invariant, especially in the context of the specification of a parallel system.

Invariants have a constant truth value that is always satisfied for any evolution of the Transition System (see \ref{def: TS} or \cite{K76} for a general definition) modeling a system or a concurrent program under study.
An \textit{invariant} is a property over the state variables true for all elements of the reachability set (in $RS$). This is indeed different from a property true in $Q$ i.e. for all states (reachable or not) which would be close to be a tautology in regard to the model behavior.

Invariants can be described and verified in a modal logic, particularly in a temporal logic (see the seminal work by A. Pnueli in \cite{P77} or the definition of a Computation Tree Logic (CTL) in \cite{HT87} or CTL* in \cite{EH86}). 

Concerning Petri Nets, these properties or relationships are expressed or deduced by using markings, places, and transitions.  
The bipartite graph which supports the topology of the system under study, connecting places and transitions can be seen as a graphical view of a system of Diophantine equations. Together with the\textit{``token game"}, which in turn models the dynamic evolution of the system under study, we have two distinctive elements which make Petri Nets so unique and popular. 
As long this topology remains in any derivation of a Petri Nets, we believe that the notion of invariant will also remains.


In a way, considering markings and tokens, a vast family of invariants can also be seen as constant functions over the distribution of tokens in a Petri Net independently of the sophistication of its enabling rule or the complexity of the structure of its tokens.


\subsection{Mutual exclusion: a classic example of invariant}
\label{sec:mutex}
A classic example of invariant is related to the mutual exclusion mechanism  (see \cite{R86} for many elaborated discussions and solutions on the topic) between activities known as critical sections of two or more concurrent programs. 
A \textit{critical section} is a piece of code handling a set of (shares) resources in a way that must be unambiguous \cite{Dijkstra71}; this can be achieved by excluding other pieces of code from simultaneously modifying or manipulating the considered shared resources. 
In other words, the concurrent programs must never\footnote{The reader would have recognized one of the phrases of Section \ref{subsec: invariant-definition} leading to an invariant definition} be active in their respective critical section at the same time. 

\subsubsection{Two programs in mutual exclusion}

\begin{figure}[ht]
\centering
\includegraphics[width=0.6\textwidth]{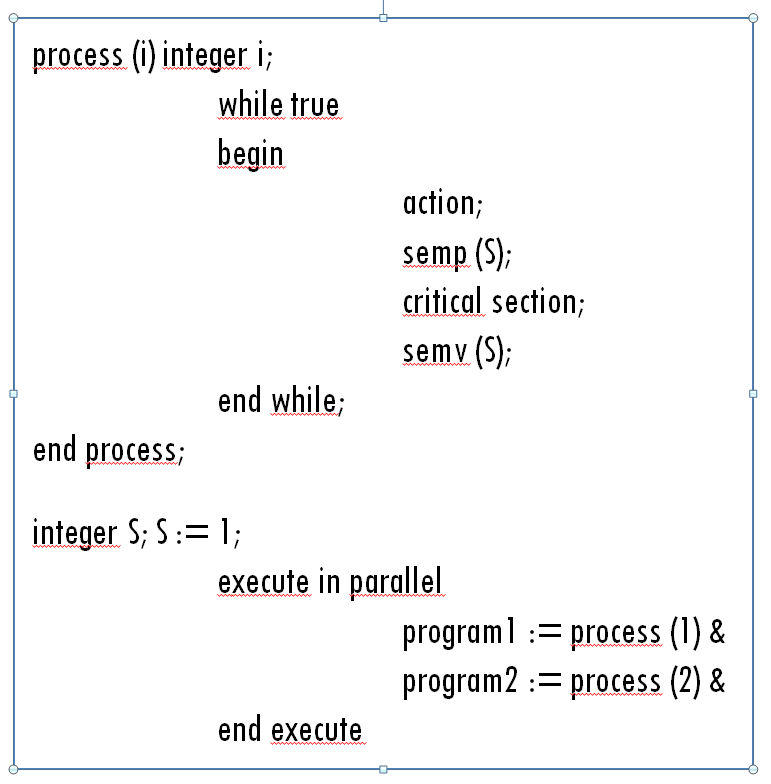}
\caption{\texttt{Program1} and \texttt{Program2} are two instances of the process(i) and are concurrently running sharing a common resource guarded by a semaphore $\texttt{S}$. $\texttt{S}$ is initialized to $1$ to guarantee the mutual exclusion between the \texttt{critical section} of each program. We assume that there are no dependencies (i.e. shared resources) between an initial piece of code called \texttt{action} and the \texttt{critical section}, in particular that only the \texttt{critical section} handles the common resource protected by the semaphore.}
\label{fig: mutex-prog}
\end{figure}

In the example Figure \ref{fig: mutex-prog}, we consider two programs sharing a common resource. The mutual exclusion between the \texttt{critical sections} is ensured by a Dijkstra semaphore $\texttt{S}$ \cite{Dijkstra71}.
The shared resource is only handled during the execution of the \texttt{critical sections} and is abstracted away.

After running an \texttt{action} which has no use of the common resource, a program can enter in its \texttt{critical section} only by executing the indivisible primitive \texttt{semp} of the semaphore $\texttt{S}$. Initializing $\texttt{S}$ to $z$ will not authorize more than $z$ execution of \texttt{semp} in a row. To ensure the mutual exclusion between the two programs, $\texttt{S}$ must be initialized to $1$. 

A program executing \texttt{semp(S)} will be put on hold (in a sleeping mode in a queue for instance) if $\texttt{S} \leq 0$ else if $S>0$ then within an indivisible sequence, the program will execute $\texttt{S} := \texttt{S}-1$ and will access its \texttt{critical section} where it safely can handle the shared resource. 
This must be permitted by the semaphore only if the other program is not itself in its \texttt{critical section}, for this reason, the semaphore is initialized to one. 

The program exits from its \texttt{critical section} by executing \texttt{semv(S)}. 
A program executing \texttt{semv(S)} will increment $\texttt{S}$ by $1$ and will wake up any other program put on hold by a \texttt{semp(S)}, granting to only one of them an access right to enter in their own \texttt{critical section} by successfully executing again \texttt{semp(S)}. 
To guarantee the mutual exclusion, we must be sure that $\texttt{S} < 2$ for all reachable state. 

Let us model the two programs of figure \ref{fig: mutex-prog} by an LTS and build its associated labeled reachability graph (Figure \ref{mutex-reachability}). 
We will model the same example by A Petri Net in the section \ref{subsubsec: modelization in a PN}. 
We want to prove that the following property: ``for all reachable state, if one of the two programs is its critical section then the other one is not." is an invariant.

\subsubsection{Modelization with an LTS}
\label{subsec: modelization with LTS}

Let us model this example by a labeled transition system $LTS=\langle \emph{Q}, T, \rightarrow \rangle$. First, each Process(i) is associated with one state variable: its ordinal counter $OCi$ (sometimes called label as in \cite{ASHCROFT75}). Usually, an ordinal counter is a register that points to the memory address of the next statement to be  executed. In this example, our ordinal counter $OCi$ varies in $Instr= \{1,2,3,4\}$ with the following meaning:

1 : when process(i) is ready to execute \texttt{action},

2 : when process(i) is ready to execute \texttt{semp},

3 : when process(i) is ready to execute \texttt{critical section},

4 : when process(i) is ready to execute \texttt{semv}.

The set $\mathcal{SV}$ is composed of 3 state variables: $\mathcal{SV} = \{OC1, OC2, S \}$ one state variable per ordinal counter and one state variable for the semaphore. The statement ``$S=S-1$" being guarded by testing if  ``$S>0$" we can conclude that the state space $Q = Instr\times Instr\times \mathbb{N}$; the initial state is $q_0 = (1,1,1)$. 

\begin{figure}[htbp!]
\centering
\includegraphics[width=0.9\textwidth]{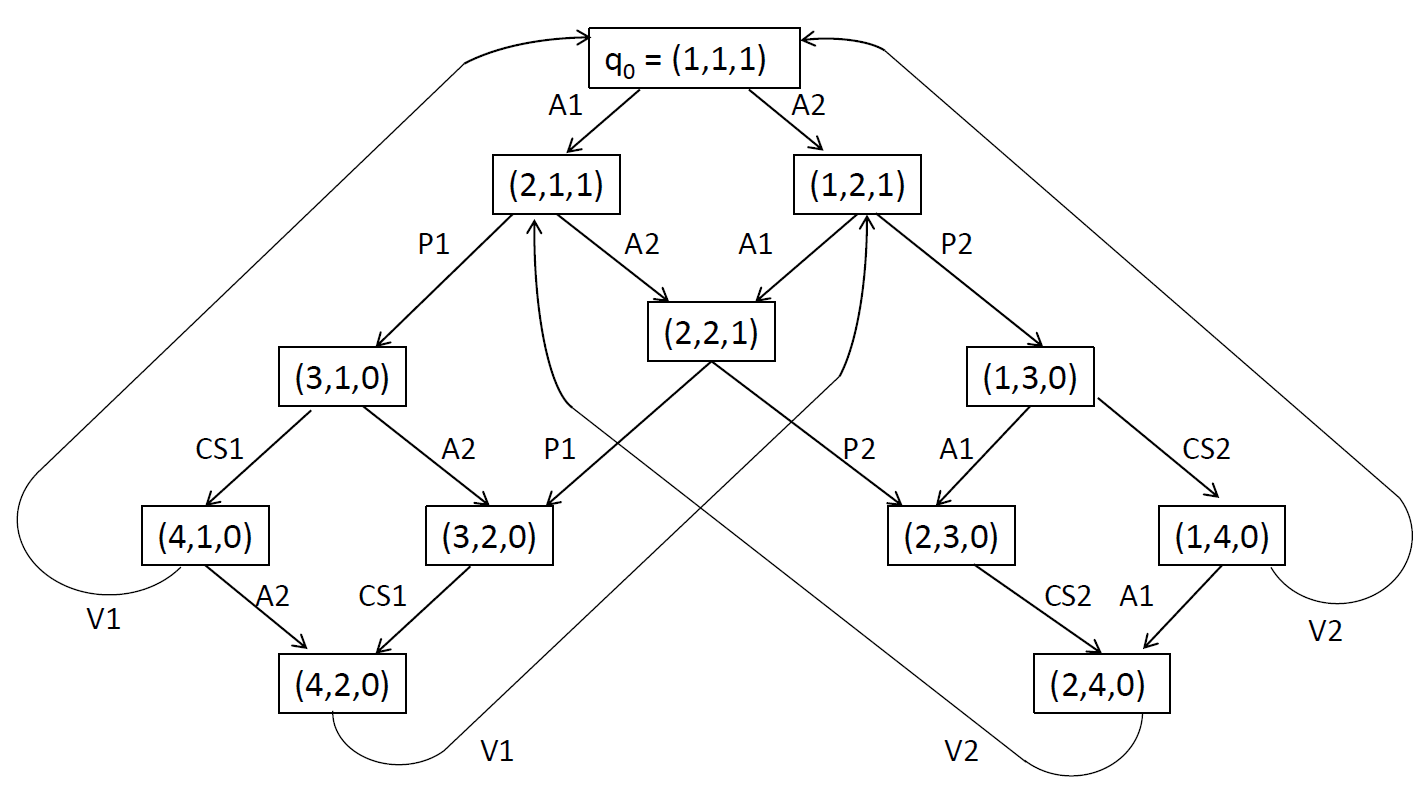}
\caption{Labeled Reachability Graph (LRG) of the LTS modeling the two concurrent programs in mutual exclusion of figure \ref{fig: mutex-prog}. Each vertex is labeled by a state of $RS$, each edge is labeled by a transition of $T$. We can observe that $q(\texttt{S}) < 2$ for all reachable state: the semaphore is accomplishing its mission in preventing the execution of \texttt{semp} twice in a row.}
\label{mutex-reachability}
\end{figure}

The set of transitions $T$ is matching the set of statements of the two programs: 

$T=\{A1, Semp1, CS1, Semv1, A2, Semp2, CS2, Semv2\}$ 

where each program is represented by 4 transitions with the following definitions for the relation $\rightarrow$: 

$\bullet$ $Ai$ for \texttt{action}, with $\forall x, y \in Instr, \forall z\in \mathbb{N}$,   

$A1: (1,y,z) \rightarrow (2,y,z)$ and $A2: (x,1,z) \rightarrow (x,2,z)$,

$\bullet$ $Sempi$ for \texttt{semp(S)} with
$\forall x, y \in Instr, \forall z\in \mathbb{N^+}$,   

$Semp1: (2,y,z) \rightarrow (3,y,z-1)$
and $Semp2: (x,2,z) \rightarrow (x,3,z-1)$,

$\bullet$ $CSi$ for \texttt{critical section} with $\forall x, y \in Instr, \forall z\in \mathbb{N}$,   

$CS1: (3,y,z) \rightarrow (4,y,z)$ and $CS2: (x,3,z) \rightarrow (x,4,z)$,

$\bullet$ $Semvi$ for \texttt{semv(S)} with $\forall x, y \in Instr, \forall z\in \mathbb{N}$,   

$Semv1: (4,y,z) \rightarrow (1,y,z+1)$ and $Semv2: (x,4,z) \rightarrow (x,1,z+1)$,

The knowledge of the overall program and the definition of $\mathcal{SV}$ make possible the construction of the reachability set $RS$ of this LTS. Actually, starting from $q_0$, its labeled reachability graph (LRG) can easily be constructed as shown in Figure \ref{mutex-reachability}.
From $q_0$ we can only execute the transitions $Ai$ to reach the states (2,1,1) (by executing $A1$) and (1,2,1) (by executing $A2$) and so on. We can observe that $RS$ is finite and that there is no state $q \in RS$ such that $q = (3,3,z)$ where $z \in \mathbb{N}$. 
Therefore, the two programs are never executing their \texttt{critical section} at the same time and our invariant is proven by simple exploration of LRG.

\subsubsection{Modelization with a PN}
\label{subsubsec: modelization in a PN}

\begin{figure}[ht]
\centering
\includegraphics[width=0.8\textwidth]{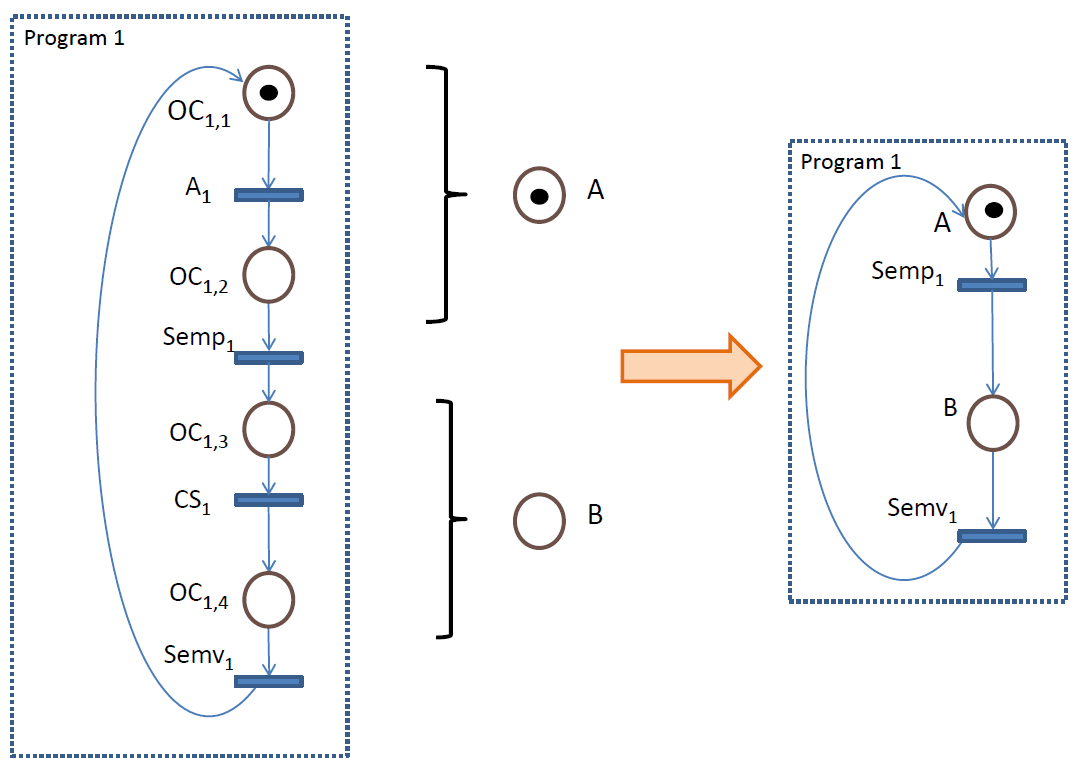}
\caption{On the left, \texttt{program1} is modeled by a Petri Net where all the elements of the LTS of Section \ref{subsec: modelization with LTS} are represented. 
Then, it is reduced to a smaller one behaving similarly in regard to mutual exclusion. A token in the place $A$ means that the program 1 is not in its \texttt{critical section} and may be executing another piece of code (i.e. \texttt{action}). 
A token in the place $B$ means that the program 1 is in its \texttt{critical section}. 
We must not simplify the Petri Net any further since we do  want to observe the mutual exclusion with \texttt{program2} through the connections via the transitions ${Semp_1}$, $Semp_2$, $Semv_1$, $Semv_2$, and the place $\textbf{S}$ which initially contains one token to comply with the process described in Figure \ref{fig: mutex-prog}.}
\label{mutex-reduction}
\end{figure}

\begin{figure}[htbp!]
\centering
\includegraphics[width=0.7\textwidth]{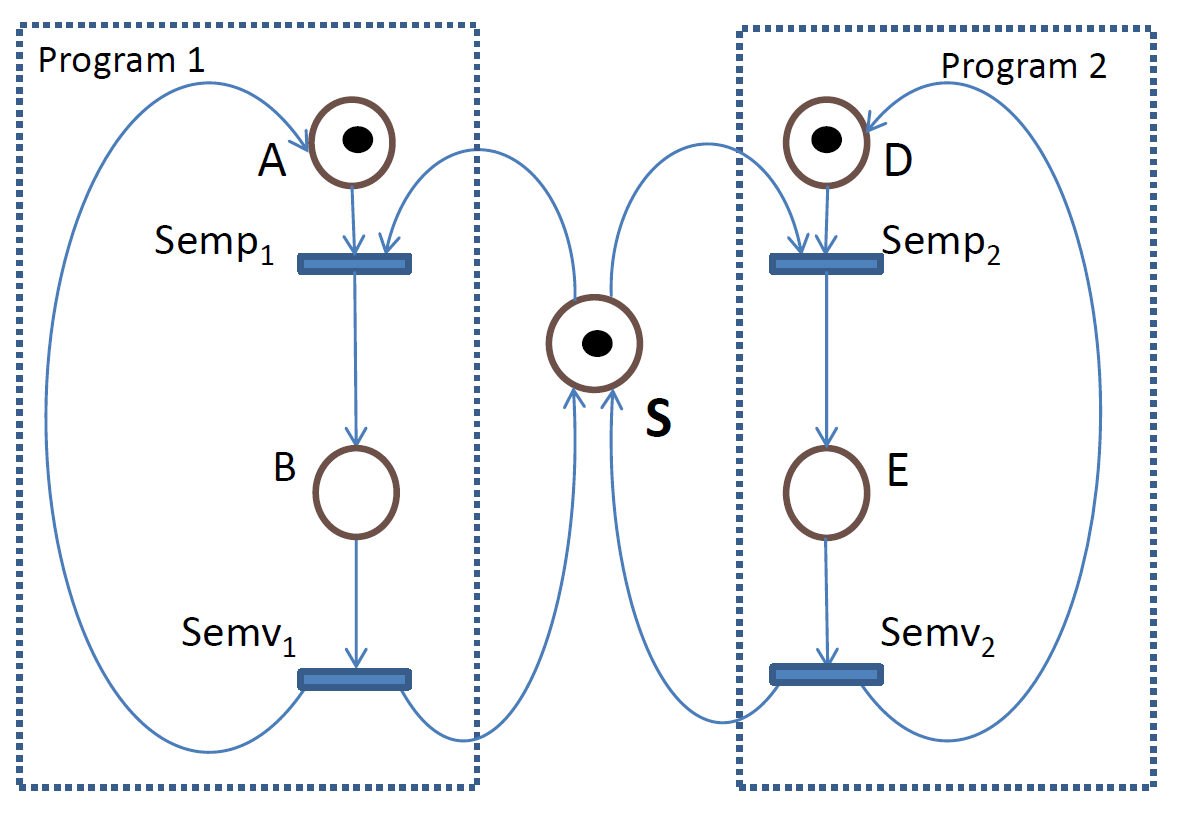}
\caption{Program 1 and Program 2 must be in mutual exclusion. For instance, as they are sharing a common resource guarded by a semaphore $\textbf{S}$, they must not access (and modify) this shared resource at the same time to ensure its consistency. The state of the shared resource is being consistent with the state of $\textbf{S}$ and does not need to be described.}
\label{mutex}
\end{figure}

We can also model our example with a Petri Net directly borrowing its representation from the LTS we just defined: with the same set $T$ of transitions and since we know $\mathcal{SV}$ and since $Instr$ is finite, we will have one place for each value of the ordinal counter of each program. The left hand side of Figure \ref{mutex-reduction} is showing how \texttt{program1} can be modeled by a Petri Net. 
Moreover, noticing that the transitions $A1$ and $CS1$ have only their ordinal counter for input and the ordinal counter of their respective next statement for output, we can reduce the places $OC_{1,1}$ and $OC_{1,2}$ with the transition $A1$ into a single place $A$ and similarly reduce the places $OC_{1,3}$ and $OC_{1,4}$ with the transition $CS1$ into a single place $B$. 
Places $A$ and $B$ are sometimes called \textit{macroplace} (see chapter 15 in \cite{GV03, Berth86}, or chapter 4 in \cite{BR82} for more details on reduction rules). 
We get the reduced Petri Net of \texttt{program1} on the right hand side of Figure \ref{mutex-reduction}. 
The semaphore is now modeled by a single place $\text {S}$ to obtain the Petri Net Figure \ref{mutex}.

In the Petri Net described in Figure~\ref{mutex}, the two concurrent programs are represented by the two subgraphs within dotted lines. The semaphore primitives \texttt{semp} and \texttt{semv} by the transitions ${Semp_1}$, $Semp_2$ and $Semv_1$, $Semv_2$ respectively. 
The initial marking $q_0$ has one token in $A$ (\texttt{program1} is not in its \texttt{critical section}), one token in $D$ (\texttt{program2} is not in its \texttt{critical section}), 
and one token in $\textbf{S}$ (the semaphore is ready to let pass one and only one program in its \texttt{critical section}: either ${Semp_1}$ or $Semp_2$ will be executed). 
The fact that a $ \texttt{program}_i$ 
is active in its \texttt{critical section} is modeled by the presence of a token in the places $B$ or $E$ respectively. 

The mutual exclusion between the two programs can be expressed in various different ways, and the corresponding invariant can be rewritten under at least the four following equivalent forms \footnote{where $\oplus$ stands for the xor Boolean operator.}: for any marking $q$ reachable from the initial marking $q_0$:

$(ME)\ \ q(B)\times q(E)=0 $

$(ME')\ \ (q(B) = 0) \oplus (q(E) = 0) $ 

$(ME'')\ \ q(B) + q(E) \leq 1$

$(ME''')\ \ q(S) < 2$

It is possible to use a theorem prover triggering rewriting rules (for instance, rules described in \cite{JouDer90} already hinted in section \ref{subsec: models}) in order to automate the proof of the equivalency of these four statements given the Petri Net paired with the initial marking of Figure \ref{mutex}. 
However, it is not the goal of this note to utilize rewriting systems to analyze Transition Systems (see for instance, \cite{StehrMO01}). 
Here, we just want to raise the attention of the reader in considering rewriting his property or his model in different equivalent ways especially when using a verification tool such as a model checker or a prover which can be very sensitive to the way a formula is written (as a trivial example, the order in which commutative terms of a Boolean disjunctive normal form are written and evaluated) and can very well develop a proof with extremely different performance.

We let the reader build the labeled reachability graph (LRG) of this Petri Net and compare it with the $LRG$ of Figure \ref{mutex-reachability} which is obtained before reduction. 
It is easy to prove the mutual exclusion simply by exploring each node of the LRG and check that the property is satisfied. 
However, this technique relies on the size of the LRG which can be infinite. 
Other methods have been developed; in particular, inductive proofs (as in \cite{Kel72} and many other authors). Up to now, these methods require human intervention and are not always converging \cite{Tiwari01}.

This classic example will be parameterized and analyzed Section \ref{subsec: mutex-param} using algebraic method developed in this note.
 
Other synchronization mechanisms for ensuring mutual exclusion between critical sections of concurrent processes can be found in the literature, see for instance, \cite{Raynal19} for many different examples with various degrees of sophistication or \cite{MVD2003} for three interesting case studies using Petri Nets.

\section{Invariants, conservative components, and semiflows}
\label{sec: inv, conserv, semiflows}
One straightforward method to prove that a proposition is an invariant would be by exploring all reachable state of the model, especially by using a model checker, a powerful tool that automatizes and optimizes such an exploration (see \cite{MOSS99} for a tutorial on model checking, see \cite{Wolf2019} for Petri Net model checking). Symbolic state-space exploration can also be used in the presence of parameters in the model. Other methods would be for instance, by induction (see \cite{K76} for an example of an induction proof in a Transition System). However, an important question is not only to prove that a given proposition is an invariant (which is no small undertaking), but also to design algorithms able to discover as many of them as possible with as little guidance as possible and ultimately, understand which ones are the most potent. 

We first associate to any property or relationship $PR$, the set of states $supp(PR) $ for which $PR$ is satisfied. $supp(PR)$ is said to be the \textit{support} of $PR$. Likewise, an invariant $I$ is associated with a set $supp(I)$ of all states satisfying the invariant property or relationship (i.e., for which the invariant is true): $I$ will be an invariant if and only if $RS \subseteq supp(I)$ 
\footnote{Here, it is worth pointing out that it is possible to have $supp(I) \setminus  Q \neq  \varnothing$ depending on the definition of $Q$, and the restrictions on the state variables.}; and $I$ will be a structural invariant if and only if $Q \subseteq supp(I)$. 
In other words, an invariant can be used to describe (most of the time concisely) a superset of the reachable states of a model. The smaller the support the more potent the corresponding invariant will be since it can be seen as pruning away a larger set of unreachable states (which will not satisfy the invariant).

The set of states $supp(I)$ associated with an invariant $I$ is a home space since it includes the entire reachability set. 

\subsection{Invariant calculus}
\label{subsec: inv calculus}
If $\mathcal{I}$ and $\mathcal{J}$ are two invariants associated with their respective supports $supp(I)$ and $supp(J)$, then it is easy to define two invariants $K_u$ and $K_i$ such that $supp(K_u) = supp(I) \cup supp(J)$ and $supp(K_i)= supp(I) \cap supp(J)$. 
A similar observation can be found in \cite{ChMi88}.
The intersection is stable over the set of supports of invariants because the reachability set is included in any support of invariant; this property is important since we are looking at manipulating sets as small as possible. 
Let's recall that the intersection is not stable for home spaces (property \ref{prop: inter-home} section \ref{subsec: HS}).
It is also worth pointing out that it is easy to combine invariants in various ways and to generate many of them. It is therefore important to understand whether invariants are organized and how to develop some computational rules over invariants relative to a given Transition System or a given Petri Net.

These observations led to look for the smallest set of states associated with invariants, or at least to look for the smallest ones that can be easily handled from an algebraic point of view. 

Until now, all these definitions have been pretty general and can be applied to any Labeled Transition System model equipped with  state equation such as equation (\ref{eq: state-by-t}) or (\ref{eq: state}), in particular, any extension of the Petri Net conceptual model such as the numerous ones that can be found in \cite{V81}, \cite{JR91}, \cite{FinkelS01}, or \cite{GV03}.

If a property $\mathcal{I}$ can be associated with an evaluation function $eval_I$ over Q, then from equation (\ref{eq: state-by-t}) in section \ref{subsub: PN} we have: 
\begin{equation}
eval_I(q{'})=eval_I(Post(\cdot,t)-Pre(\cdot,t)+q) \ \ \ \  \forall t \in T
\ such\ that \ q\overset{t}{\rightarrow}q' 
\end{equation}

Moreover, if $eval_I$ is defined over $Q \bigcup_{t \in T}Pre(\cdot,t) \bigcup_{t \in T}Post(\cdot,t) $ and if $eval_I$ has the good taste to be a morphism that preserves addition, then we may rewrite our system of equations to obtain:
\begin{equation}
eval_I(q{'})=eval_I(Post(\cdot,t))-eval_I(Pre(\cdot,t))+eval_I(q) \ \ \ \  \forall t \in T
\end{equation}

If $\mathcal{I}$ is an invariant, $eval_I(q{'})=eval_I(q)$ therefore, $eval_I$ must verify the following system of equations:
\begin{equation}
\label{eq: topo-invariant}
	eval_I(Post(\cdot,t))=eval_I(Pre(\cdot,t)) \ \ \ \  \forall t \in T
\end{equation} 

\section{Semiflows Calculus}
\label{sec: semiflows}

Semiflows calculus is possible as soon as states and transitions are described with state variables.
Any subset of transitions allows deducing a system of equations with state variables which must not be modified outside the considered subset of transitions. Then, a solution to this system of equations will be called a semiflow.

In many Petri Nets papers, a specific class of invariants is considered. They are sometimes unduly called \textit{linear invariants} because they are associated with an integer weight function $f$ over the set $P$ of places. With regard to a Petri Net PN, these invariants have the following form: 
\begin{equation}
\label{eq: invariant}
    eval_I(q) = f^\top q = eval_I(q_0) = f^\top q_0 \ \ \ \ \forall q \in RS(PN,q_0) 
\end{equation}
where $f^\top q = \sum_{p \in P} f(p)q(p)$ is the scalar product of $f$ and $q$ seen as vectors in $\mathbb{Z}^d$. In that case $supp(I) = \{q \in Q\ | \ f^\top q = f^\top q_0\}$. The first definitions of semiflows were given in $\mathbb{N}^d$.

Y.E. Lien \cite{L73} \footnote{Where a Petri Net is said conservative if and only if $\exists f \in \mathcal{F}$ such that $f > 0$.}and then independently, K. Lautenbach, and H. A. Schmid \cite{LS74} \footnote{Where minimal semiflows are called simple invariants.}were among the first ones, if not the first ones, to describe such a concept for Petri Nets. The algebraic calculus underneath this concept was first described in \cite{M77} followed by many other publications such as \cite{Si78, STC96, M23}.
The vast corpus of linear algebra results and algorithms can be applied to analyze Petri Nets and compute semiflows. 

These weight functions can be computed by solving the following homogeneous system of $m$ Diophantine equations directly derived from equation (\ref{eq: topo-invariant}):
\begin{equation}
\label{eq: inv-semiflow}
    f^\top Post(\cdot,t) = f^\top Pre(\cdot,t) \ \ \ \  \forall t \in T
\end{equation} 
\begin{definition}
A \textit{semiflow} is a solution of the system of equations (\ref{eq: inv-semiflow}) in $\mathbb{Z}$.     
\end{definition}
The system of equations (\ref{eq: inv-semiflow}) is well-known in the literature \cite{BR82, STC96, GV03}. 
It can be directly derived from the more general system of equations (\ref{eq: topo-invariant}) that can already be found in \cite{M19}.
This system of equations is fundamental; it describes some constraints over the topology of the Petri Net independently of any initial state and expresses a law of conservation over the net and its evolution. 
It says that during the dynamic evolution of the Petri Net, a function of token distribution remains constant: 
what is consumed and needed to enable $t$ is equal to what is produced after executing $t$ modulo the weight function $f$. 
This is a law similar to Kirchhoff's first law on the conservation of current flows but only applied on transition nodes i.e. half of the nodes of the bipartite graph representing $PN$. 
This (with the fact that $\mathbb{N}$ is a semiring) explains why such functions over $P$ (or equivalently vectors in $\mathbb{Z}^d$) have been baptised \textit{semiflow} in \cite{M77}.

Several algorithms were independently developed to compute a generating set of semiflows \cite{T81} or later in \cite{martinez1982,AM82}. 
All of them can be considered as variations of Farkas or Fourier algorithms related to integer linear programming and convex geometry 
(see \cite{ColomS89} for a comparative study or \cite{Schrijver87} for the underlying mathematical theory).

Of course, any linear combination of semiflows is still a semiflow. 
The set $\mathcal{F}$ of semiflows with integer coefficients is a module since $\mathbb{Z}$ is a ring. 
In subsection ~\ref{subsubsec: semi}, we will examine the subset 
$\mathcal{F}^+ = \{f \in \mathcal{F}\ |\ \forall p \in P, f(p) \geq 0 \}$ 
of $\mathcal{F}$ and see why this subset which is a monoid (as soon as we accept $0$ as a semiflow of $\mathcal{F}^+$) is of particular interest when it comes to analyze behavioral properties. 
However, it is still possible to characterize this set with a generating set by considering the notion of support for semiflows. 
\begin{definition}
\label{def: support-f}
The support of semiflows $f$ is a non-injective function from $\mathcal{F}$ to $2^P$ such that 
$\left \| f \right \|$  denotes the \textit{support of $f$} and $\left \| f \right \| = \{x \in P\ |\ f(x) \neq 0\}$.
\end{definition}

The support of semiflows as a function from $\mathcal{F}$ to $2^P$ is not a morphism for the addition in $\mathcal{F}$ to the union in $2^P$. 
We have $\left\|f+g \right\| \subseteq\ \left\|f \right\|\cup \left\| g\right\|$, however the equality between $\left\|f+g \right\|$ and $\left\|f \right\|\cup \left\| g\right\|$ can be reached only under conditions such as the condition (a) in following property:
\begin{property}
\label{prop: support-plus-union}
Let $f$ and $g$ be two semiflows:
\begin{itemize}
    \item [(a)]
$\forall i \in \{1,...d\}, \ f(x_i)*g(x_i) \geq 0 $,
\item[(b)] 
$\nexists x_i \in \left\|f \right\|\cup \left\| g\right\|, f(x_i)+g(x_i) = 0$
\item[(c)] 
$\left\|f+g \right\|=\left\|f \right\|\cup \left\| g\right\|$.
\end{itemize}
We have: if (a) then (b); and (b) if and only if (c).

Moreover, $\left\|0 \right\|= \varnothing$ and
if $\alpha$ is a non-null scalar then 
$\left\|\alpha f \right\|=\left\|f \right\|$.
\end{property}

Then, we can define the \textit{positive and negative supports} of a semiflow $f \in \mathcal{F}$ as:

$\left\| f\right\|_+ = \left\{ p\in P\ |\ f(p)> 0\right\}$,

$\left\| f\right\|_- = \left\{ p\in P\ |\ f(p)< 0\right\}$,

with $\left\| f\right\| = \left\| f\right\|_- \cup\   \left\| f\right\|_+$.

We can then rewrite the system of equations (\ref{eq: invariant}):

\begin{equation}
\label{eq: differential}
f^\top q=\left| \sum_{p \in \left\| f\right\|_+}f(p)q(p)\right| - \left| \sum_{p \in \left\| f\right\|_-}f(p)q(p)\right|=f^\top q_0
\ \ \ \ \forall q \in RS(PN,q_0)
\end{equation}

As we can see, the formulation of the system of equations (\ref{eq: differential}) is a differential between the weighted number of tokens in the places of the positive support and the weighted number of tokens in the places of the negative support of $f$.
This differential allows deducing an invariant since by equations (\ref{eq: differential}) it remains constant during the evolution of the Petri Net. 
A first general property can be immediately deduced recalling that the initial state $q_0$ belongs to $\mathbb{N}^d$.

\begin{property}
\label{prop: boundedness in F}
    For any semiflow $f \in \mathcal{F}$,
    $\exists p \in \left\| f\right\|_+$ not bounded if and only if $\exists p \in \left\| f\right\|_-$ not bounded.
\end{property}
When in $\mathcal{F}^+$, $\left \| f \right \|$ is also called \textit{conservative component} \cite{L76} or \cite{BR82}.

\subsection{Semiflows and Invariants in the mutex example}
\label{subsec: semiflows in simple mutex}

The Petri Net of Figure~\ref{mutex} has 3 semiflows $f_1$, $f_2$, and $sem$:

$f_1$ such that $f_1(A) = f_1(B) = 1$, $f_1(p) = 0$ for any other place $p$.

$f_2$ such that $f_2(D) = f_2(E) = 1$, $f_2(p) = 0$ for any other place $p$.

$sem$ such that $sem(B) = sem(E) = sem($\textbf{S}$) = 1$, $sem(p) = 0$ for any other place $p$.

It is easy to verify that $f_1$, $f_2$, and $sem$ are minimal of minimal support.
$f_1$ is associated with Program 1. Its associated invariant means that there is always only one token in the support $\{A,B\}$ of $f_1$ i.e. either in $A$ or in $B$. Similarly, $f_2$ is associated with Program 2. Its associated invariant means that there is always only one token between D and E. The token in $\{A,B\}$ can be interpreted as modeling the ordinal counter of Program1; similarly, the token in $\{D,E\}$ can be interpreted as modeling the ordinal counter of Program2. The semiflow $sem$ is associated with the semaphore $\textbf{S}$ and its associated invariant is: 

$sem^\top q = sem^\top q_0$ in other words: $sem(B)+sem(E)+sem($\textbf{S}$) = 1$ for any reachable marking $q$ from $q_0$.

It allows to straightforwardly deduce that (ME) defined in Section \ref{subsubsec: modelization in a PN} is an invariant since there is at most one token in $\{B,E\}$.
\hfill 
$\square$ 

\subsection{Invariants pave the way to boundedness, liveness, or fairness analysis}

In regard to Petri Nets, invariants are interesting not only because they express a property that is looked for itself (for instance, to comply with the SuS specification) but also because they straightforwardly allow the exclusion of a large number of markings that cannot be reached from the initial marking without violating the invariant. In other words, if a marking does not satisfy an invariant then it is unreachable. This important aspect is at the source of pruning techniques and is critical to analyze and prove many important behavioral Petri Net properties such as liveness, boundedness, or even fairness which otherwise must be analyzed by developing the reachability or one coverability graph of the Petri Net under consideration (see \cite{KM69} and \cite{F93} for a minimal coverability graph). Many of these results are described and proven in \cite{BR82}.


\subsubsection{Home spaces, semiflows, and liveness}
\label{subsubsec: hs-semiflow-liveness}
Semiflows are intimately associated with home spaces and invariants and can greatly simplify the proof of fundamental properties of Petri Nets (especially when the description includes parameters as in \cite{BEISW20}) such as safeness, boundedness, or more complex behavioral properties such as liveness.  
Let us provide three properties supporting this idea.  
\begin{property}
A transition $t$ is live if and only if \texttt{Dom(t)} is a home space.

Moreover, if \texttt{Dom(t)} is a home space then \texttt{Im(t)} is also a home space.
\end{property} 
Directly deduced from the definition \ref{def: live} of liveness and definition \ref{def: homespace} of home space.
\hfill
$\square$

We consider a model $\langle M,q_0 \rangle$ with its associated reachability set $RS$ labeled reachability graph $LRG$, a home space $HS$ and $H=HS \cap RS$.
\begin{lemma}
If $H$ induces a strongly connected subgraph of LRG then 

a transition $t$ is live if and only if $\exists h_t \in H, \exists \sigma \in T^*$ such that $h_t \xrightarrow {\overset {\sigma t }{ }}$.
\end{lemma}
If $HS$ is a home state then $H$ is also a home state and $\forall q \in RS, \ \exists s_1 \in T^*,\ \exists h \in H$ such that $q \overset{s_1\ }{\rightarrow}h$.

The subgraph induced by $H$ (see \cite{Diestel10} for the notion of induced subgraph) being strongly connected, there exists a path from $h$ to $h_t$, in other words, $\exists s_2 \in T^*$ such that $h \overset{s_2\ }{\rightarrow} h_t$. We can construct a sequence $s=s_1s_2\sigma$ such that $\forall q \in RS, q\overset{st }{\rightarrow}$. Hence $t$ is live in $RS(M,q_0)$. 
The reverse is obvious.
\hfill
$\square$

From this lemma, we can easily deduce the following property regarding home states:

\begin{property}
\label{prop: home-state-liveness}
Let $M$ be a model and $q_0$ be a home state then:

any transition that is enabled in $q_0$ is live,

more generally,

a transition is live if and only if it appears as a label in $LRG(M,q_0)$.
\end{property} 

This can easily be proven directly from the definition of liveness section \ref{subsec: liveness} and the property \ref{home-state} about home states.
\hfill 
$\square$

Given an initial state $q_0$, any semiflow $f$ can be associated with an invariant $I_f$ defined by equation \ref{eq: differential}. $I_f$ can in turn can be associated with a home space. In other words, if $f \in \mathcal{F}$, then $supp(I_f) = HS(f,q_0) = \{q \in Q \ |\ f^\top q = f^\top q_0\}$.

$HS(f,q_0)$ is a $\{q_0\}$-home space since $RS(M,q_0) \subseteq HS(f,q_0)$. Indeed, we have $HS(0,q_0)=Q$. 
We define by $\mathbb{HS}$ and $\mathbb{HS}^+$ the set of home spaces defined by semiflows in $\mathcal{F}$, $\mathcal{F}^+$ respectively; then $\mathcal{HS}$, $\mathcal{HS}^+$ their respective closure for the intersection. The following property says in particular that any element of $\mathcal{HS}$ is still a home space.

\begin{property}
\label{prop: inter-home-spaces}
    If $f \in \mathcal{F}$ then:
    $\forall \alpha \in \mathbb{Q} \setminus \{0\}, HS(\alpha f,q_0)= HS(f,q_0)$,

    $\forall f,g \in \mathcal{F}, \forall \alpha, \beta \in \mathbb{Q}, HS(f,q_0) \cap HS(g, q_0) \subseteq HS(\alpha f+ \beta g,q_0)$. 
    
    Moreover, $HS(f,q_0) \cap HS(g, q_0)$ is a $\{q_0\}$-home space.
\end{property}
$HS(f,q_0)\ \cap \ HS(g, q_0)$ is straightforwardly a $\{q_0\}$-home space since they both contain $RS(M,q_0)$. 
Let us recall that in general, the intersection of home spaces is not a home space (see Figure \ref{fig: inter-hs}). 
If $q \in HS(f,q_0) \cap HS(g,q_0)$, then $\forall \alpha , \beta \in \mathbb{Q},\ \ \alpha (f^\top q) = \alpha (f^\top q_0)$ and $\beta (g^\top q) = \beta (g^\top q_0)$, 

so $(\alpha f+\beta g)^\top q = (\alpha f+ \beta g)^\top q_0$, therefore, $q \in HS(\alpha f+\beta g,q_0)$
\hfill 
$\square$

These three properties provide us with a methodology to analyze and prove that a subset of transitions are live.
From a set of invariants, we can define a first home space $HS$ that concisely describe how tokens are distributed over places. 
From this token distribution, we can analyze what transition are enabled in order to prove that a specific given marking $q$ ($q_0$ being the usual case) is always reachable from any element of $HS$. 
When this is possible, it can easily be deduced that $q$ is a home state. 
Then, it may be possible using property \ref{prop: home-state-liveness} to prove which transition are live and whether the Petri Net is live or not. 
This will be illustrated later with a few examples in section \ref{sec: ex}.

\subsubsection{Semiflow basic behavioral properties}
\label{subsubsec: semi}

Solutions of the system of equations (\ref{eq: inv-semiflow}) have their coefficients in $\mathbb{Z}$ in general. 
However, the most interesting semiflows from a behavioral analysis point of view are defined over natural numbers instead of integers. This can be seen through the three following properties. 

First, if a given semiflow is such that $\left\| f\right\|_- = \varnothing$ then $f \in \mathcal{F}^+$ and by property \ref{prop: boundedness in F}, $\left\| f\right\|$ is necessarily structurally bounded. 
More generally, considering a weighting function $f$ over $P$  being defined over non-negative integers, the following property and its corollary can be easily proven \cite{M78}:  
\begin{property}
\label{prop: boundedness-alg}
If $f \geq 0$ is such that $f^{T} Pre(\cdot ,t) \geq  f^{T} Post(\cdot ,t)\ \forall t \in T$,
then the set of places of $\left\|f \right\|$ is structurally bounded \footnote{following the general definitions given in Section \ref{secsec: boundedness}}.
Moreover, the marking of any place $p$ of $\left \| f \right \|$ has an upper bound:

$q(p) \leq \frac{f^{T}q_0}{f(p)}, \  \ \forall q \in RS(PN,q_0)$.
\end{property}

This bound can be reached only if every tokens of $\left \| f \right \|$ can be moved to $p$ which is indeed not always the case as in the example section \ref{subsubsec: euclidean}.

If $f > 0$ then $\left\|f \right\|_+ = \left\|f \right\| = P$ and the Petri Net is also structurally bounded. The reverse is also true: if the Petri Net is structurally bounded, then there exists a strictly positive solution for the system of inequalities above (see \cite{Si78} or \cite{BR82}). 
This property is indeed false for a semiflow that verifies the system but would have at least one negative element and constitutes a first reason for particularly considering weight functions $f$ over $P$  being defined over non-negative integers including $\mathcal{F}^+$. 

The following corollary can directly be deduced from the fact that any semiflow in $\mathcal{F}^+$ satisfies property \ref{prop: boundedness-alg}:

\begin{corollary}
\label{prop: bound-for-support}
    For any place $p$ belonging to at least one support of a semiflow of $\mathcal{F}^+$, an upper bound $\mu$ can be defined for the marking of $p$ such that:   

    $\forall q \in RS(PN,q_0), \ \ q(p) \leq \mu(p,q_0) =  \min_{\{f\in \mathcal{F}^+\ |\ f(p) \neq 0\}} \frac{f^\top q_0}{f(p)} \ $
\end{corollary}
A second reason for particularly considering a semiflow $f$
as being defined over non-negative integers is that the system of inequalities: 

\begin{equation}
 f^{T} q_0  \geq f^{T} Pre(\cdot,t), \ \ \ \ 
 \forall t \in T
\label{eq:threshold}
\end{equation}
  
becomes a necessary condition for any transition $t$ to stand a chance to be enabled in any reachable marking from $q_0$, then to be live. In \cite{BR82}, $f^{T} Pre(\cdot,t)$ is called the \textit{enabling threshold} of $t$. 
\begin{property}
\label{prop: enabling-threshold}
If $t$ is a transition and $\exists f \in \mathcal{F}^+ \setminus \{0\}$ such that:

 $f^{T} q_0  < f^{T} Pre(\cdot,t)$, then $t$ cannot be executed from $\langle M,q_0 \rangle$.
\end{property}
This property can be used when looking for a frugal management of resources (i.e. a marking as small as possible) and still fully functioning (i.e. live). Another application is when the model is parameterized, then the equation (\ref{eq:threshold}) can help to easily discard some values of these parameters for which the model is not live (see example of figure \ref{fig: tinyi}).

The  property \ref{prop: support-plus-union} can already be found in \cite{M78} or \cite{MR79, BR82} for semiflows in $\mathcal{F}^+$. We just give its mathematical formulation (see for instance, \cite{Lan02} for definitions). 

\begin{corollary}
\label{cor: support-morphism}
The support of semiflows as a function from $\mathcal{F}^+$ to $2^P$ is a (monoid) homomorphism from $\langle \mathcal{F}^+ , + \rangle$ to $\langle 2^P , \cup \rangle$. In other words:

If f and g are two semiflows in $\mathcal{F}^+$ then

$\left\|f+g \right\|=\left\|f \right\|\cup \left\| g\right\|$.
Moreover, $\left\|0 \right\|= \varnothing$
\end{corollary}

These results have been cited and utilized many times in various applications going beyond computer science, electrical engineering, or software engineering. 
For instance, they have recently been used in the domain of biomolecular chemistry relatively to chemical reaction networks \cite{JACB18} which brings us back to the original vision of C. A. Petri when he highlighted that his nets could be used in chemistry.

\section{Generating sets and minimality}
\label{sec: generating sets}

\subsection{Generating sets}

The notion of generating sets for semiflows is well known and efficiently supports the handling of an important class of invariants sometimes unduly called linear invariants in the literature. 
Several results have been published starting from the initial definition and structure of semiflows  \cite{M77} to a large array of applications used especially to analyze Petri Nets \cite{Colom2003,DworLo16,JACB18,Wol2019}.

Minimality of semiflows and minimality of their supports are critical to understand how to best decompose semiflows and manage analysis of behavioral properties.
Invariants directly deduced from minimal semiflows relate to smaller quantities of resources.
Furthermore, the smaller the support of semiflows, the more local their footprint.
In the end, these two notions of minimality will foster analysis optimization.

\subsection{Basic definitions and results}
\label{subsec: gs-basic-results}

\begin{definition}
\label{def: FG}
A subset $\mathcal{G}$ of $\mathcal{F^{+}}$ is a \textit{generating set over a set $\mathbb{S}$} if and only if $\forall f \in \mathcal{F}^+$ 
we have $f = \sum_{g_i \in \mathcal{G}} \alpha_ig_i $ where $\alpha_i \in \mathbb{S}$ where $\mathbb{S} \in \{\mathbb{N}, \mathbb{Q^{+}}, \mathbb{Q}\}$ with $\mathbb{Q^{+}}$ denoting the set of non-negative rational numbers.
\end{definition}

Since $\mathbb{N} \subset \mathbb{Q^{+}} \subset \mathbb{Q}$, \footnote{ Where $\subset$ denotes the strict inclusion between sets.} a generating set over $\mathbb{N}$ is also a generating set over $\mathbb{Q^{+}}$, and a generating set over $\mathbb{Q^{+}}$ is also a generating set over $\mathbb{Q}$. However, the reverse is not true and, in our opinion, one source of some of inaccuracies that can be found in the literature.
Therefore, it is important to precise over which set the coordinates (used for the decomposition of a semiflow) belong.

The notion of the generating set is strongly related to algebraic concepts especially when it is finite. 
Let's consider $\mathcal{G}$ a finite generating set of $\mathcal{F^{+}}$ such that $\mathcal{G} = \{g_1,...g_k\}$, the following mathematical definitions can be recalled and contribute to exemplify the care that must be taken with this notion.

\textbf{-}If $\mathcal{G}$ is a generating set over $\mathbb{N}$ then 

$\mathcal{S(G)} = \{ f \in \mathbb{N}{^d}\ |\ f = \sum_{i=1}^{i=k} \alpha_ig_i \ \text{and}\ \alpha_i \in \mathbb{N}\}$ is a \textit{monoid} and $\mathcal{F^{+}} = \mathcal{S(G)}$.

\textbf{-}If $\mathcal{G}$ is a generating set over $\mathbb{Q}^+$ then 

$\mathcal{C(G)} = \{ f \in (\mathbb{{Q}^+}){^d}\ |\ f = \sum_{i=1}^{i=k} \alpha_ig_i\ \text{and}\ \alpha_i \in \mathbb{Q}^+\}$ is a \textit{convex polyhedral cone} and $\mathcal{F^{+}} = \mathcal{C(G)} \cap \mathbb{N}{^d}$. 
It is interesting to recall a result from \cite{Lasserre89} stating that $\mathcal{F^{+}} \neq \{0\}$ if and only if $\mathcal{C(G)} \neq \{0\}$.

\textbf{-}If $\mathcal{G}$ is a generating set over $\mathbb{Q}$ then 

$\mathcal{V(G)} = \{ f \in \mathbb{Q}{^d}\ |\ f = \sum_{i=1}^{i=k} \alpha_ig_i\ \text{and}\ \alpha_i \in \mathbb{Q}\}$ is a \textit{vector space} and $\mathcal{F^{+}} = \mathcal{V(G)} \cap \mathbb{N}{^d}$. We can extract from $\mathcal{G}$ a basis of $\mathcal{V(G)}$ (see, for instance, \cite{Lan02} p. 85) which also is a generating set of $\mathcal{F^{+}}$ over $\mathbb{Q}$ since the elements of this basis are in $\mathcal{F^{+}}$.

\subsection{Minimal supports and minimal semiflows}
The fact that there exists a finite generating set over $\mathbb{N}$ is non trivial. This result was proven by Gordan circa 1885 then Dickson circa 1913. Here, we directly rewrite Gordan's lemma \cite{AB86,Oda12} by adapting it to our notations.

\begin{lemma} (\textbf{Gordan circa 1885})
\label{lem: Gordan}
Let  $\mathcal{F^+}$ be the set of non-negative integer solutions of the system of equations (\ref{eq: inv-semiflow}). Then, there exists a finite generating set of vectors in $\mathcal{F^+}$ such that every element of $\mathcal{F^+}$ is a linear combination of these vectors with non-negative integer coefficients.
\end{lemma}

The question of the existence of a finite generating set being solved for $\mathbb{N}$, it is necessarily solved for $\mathbb{Q^+}$ and $\mathbb{Q}$. 

Several definitions of the notion of minimal semiflow were introduced in \cite{STC1998} p. 319, in \cite{ColomTS2003} p. 68, \cite{Kruck86}, \cite{CMPW09}, or in \cite{M78,M83}. It can be confusing to look into these in details. 
In light of this, we propose to consider only two basic notions in order theory: minimality of support with respect to set inclusion and minimality of semiflow with respect to the componentwise partial order on $\mathbb{N}{^d}$ since the various definitions we found in the literature as well as the results of this note can be described in terms of these sole two classic notions.

\begin{definition}[minimal support]
A non-empty support $ \left \| f \right \| $ 
of a semiflow $f$ is \textit{minimal} with respect to set inclusion if and only if $\nexists \ g \in \mathcal{F}^+\setminus\{0\}$ such that $\left \| g \right \| \subset \left \| f \right \| $. 
\end{definition}

Since $P$ is finite, the set $\mathcal{MS}$ of minimal supports in $P$ is a \textit{Sperner family} (i.e., a family of subsets such that none of them contains another one) and we can apply Sperner's theorem \cite{Sper28} over $c=|\mathcal{MS}|$ which states that:
\begin{equation}
\label{eq: Sperner}
    c \leq  \binom{d}{\left \lfloor d/2 \right \rfloor}.
\end{equation}
This result, which was already mentioned in \cite{M83,STC1998}, provides us with a general upper bound for the number $c$ of minimal supports in a given Petri Net. 
To the best of our knowledge, this bound can be reached only by three families of degenerate Petri Nets containing isolated elements: only one, two, or three places and as many isolated transitions as desired. 
We conjecture that this bound cannot be reached for Petri Nets with more than three places and should be refined on a case-by-case basis by exploiting connectivity between places and transitions as hinted in the example Section \ref{subsec: example-summary}.

\begin{definition}[minimal semiflow]
A non-null semiflow $f$ is \textit{minimal} with respect to $\leq$ if and only if $\nexists \ g \in \mathcal{F^{+}}\setminus\{0,f\}$ such that $g \leq f$.
\end{definition}

In other words, a minimal semiflow cannot be decomposed  as the sum of another semiflow and a non-null non-negative vector. 
This remark yields initial insight into the foundational role of minimality regarding the decomposition of semiflows.
We are looking for characterizing generating sets such that they allow analyzing various behavioral properties as efficiently as possible.

First, if we consider a generating set over $\mathbb{N}$, then we may have to explore every minimal semiflow. Although finite, the number of minimal semiflows can be quite large. Second, considering a basis over $\mathbb{Q}$ may not capture behavioral constraints quite easily (see the example of Section \ref{subsec: example-summary}).

\subsection{Three decomposition theorems}
\label{sec: 3theorems}

Generating sets can be characterized thanks to a set of three decomposition theorems that can be found in  \cite{M78} with their proofs. 
Here, theorem \ref{th: over N}  
is extended to fully characterize minimal semiflows and generating sets over $\mathbb{N}$, and is provided with a new proof using Gordan's lemma \ref{lem: Gordan}.
Theorem \ref{th: min support} was only valid over $\mathbb{N}$ and is now extended to include $\mathbb{Q^+}$ and $\mathbb{Q}$.
Theorem \ref{th: decomp} is only valid over $\mathbb{Q^+}$ and is unchanged from \cite{MR79}.

\subsubsection{Decomposition over non-negative integers}

\begin{theorem} (\textbf{Decomposition over $\mathbb{N}$})
\label{th: over N}

A semiflow is minimal if and only if it belongs to any generating set over $\mathbb{N}$.

The set of minimal semiflows of $\mathcal{F^{+}}$ is a finite generating set over $\mathbb{N}$.
\end{theorem}
Let's consider a semiflow $f \in \mathcal{F^{+}}\setminus\{0\}$ and its decomposition over any family of $k$ non-null semiflows $f_i, 1 \leq i \leq k$. Then, $\exists a_1,...,a_k \in \mathbb{N}$ 
such that $f = \sum_{i=1}^{i=k}a_if_i$. 
Since $f \neq 0$ and all coefficients $a_i$ are in $\mathbb{N}$, $\exists j \leq k$ such that $0< f_j \leq a_jf_j \leq f$. If $f$ is minimal, then $a_j=1$ and $f_j=f$.
Hence, if a semiflow is minimal, then it belongs to any generating set over $\mathbb{N}$. The reverse will become clear once the second statement of the theorem is proven.

Applying Gordan's lemma, there exists a finite generating set, $\mathcal{G}$ \footnote{This point is taken for granted in \cite{M78} as well as the rest of literature on semiflows.}. Since any minimal semiflow is in $\mathcal{G}$, the subset of all minimal semiflows is included in $\mathcal{G}$ and therefore finite. Let $\mathcal{E} = \{e_1,...e_n\}$ be this subset and prove by construction that $\mathcal{E}$ is a generating set. 

For any semiflow $f \in \mathcal{F^+}$, we 
build the following sequence leading to the decomposition of $f$:

i) $r_0 = f$

ii)  $r_i = r_{i-1} - k_ie_i$ such that $ r_i \in  \mathcal{F^+}$
and $r_{i-1} - (k_i+1)e_i \notin \mathcal{F^+}$

By construction of the non-negative integers $k_i$ , we have $r_n \in \mathcal{F^+}$ and $\nexists e_i \in \mathcal{E}$ such that $e_i \leq r_n$. 
This means that $r$ is either minimal or null. Since $\mathcal{E}$ includes all minimal semiflows, therefore, 
$r=0$, and any semiflow can be decomposed as a linear combinations of minimal semiflows, in other words, $\mathcal{E}$ is a finite generating set.
\footnote{If $\mathcal{E}$ was to be  infinite, the construction could still be used since the monotonically decreasing sequence $r_i$ is bounded by 0 and $\mathbb{N}$ is nowhere dense, so we would have:
\newline
$ \lim \limits_{n \to \infty} f- \sum_{j=1}^{j=n}k_je_j = 0$ with the same definition of the coefficients $k_j$ as in ii).}
It is now clear that if a semiflow $f$ belongs to any generating set, then it belongs in particular to $\mathcal{E}$, therefore, $f$ is a minimal semiflow. \hfill
$\square$

\begin{figure}[ht]
\centering
\includegraphics[width=0.35
\textwidth]{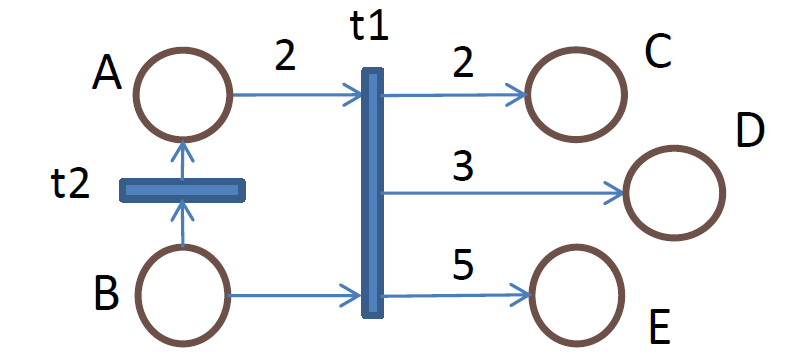}
\caption{$f_1^\top=(3,3,2,0,1),\ f_2^\top=(4,4,1,0,2),\ g_1^\top=(2,2,3,0,0),\ g_2^\top=(1,1,0,1,0),\ g_3^\top=(5,5,0,0,3)$ are five canonical and minimal semiflows.
$\left \| f_1 \right \|$ or $\left \| f_2 \right \|$ are not minimal. 
$f_1$ and $f_2$ are linear combinations of $g_1, g_2, g_3$ over $\mathbb{Q^+}$: $f_1= \frac{1}{3}(2g_1+g_3)$ and $f_2= \frac{1}{3}(g _1+2g_3)$. Therefore, the decomposition of a semiflow on $\{f_1, f_2, g_1, g_2, g_3\}$ is not unique. 
Moreover, $\mathcal{G}_1=\{g_1, g_2, g_3\}$ is a generating set over $\mathbb{Q^+}$ or over $\mathbb{Q}$.} 
\label{STC3}
\end{figure}

Let's point out that since $\mathcal{E}$ is not necessarily a basis, the decomposition is not be unique in general and depends on the order in which the minimal semiflows of $\mathcal{E}$ are considered as shown in Figure \ref{STC3} where the semiflow $h^\top= (9,9,6,0,3)$ can be decomposed in three different ways: $h=f_2+g_1+f_1=3f_1=2g1+g3$ just by changing the order in which the 5 minimal semiflows are considered.

\begin{figure}[ht]
\centering
\includegraphics[width=0.3\textwidth]{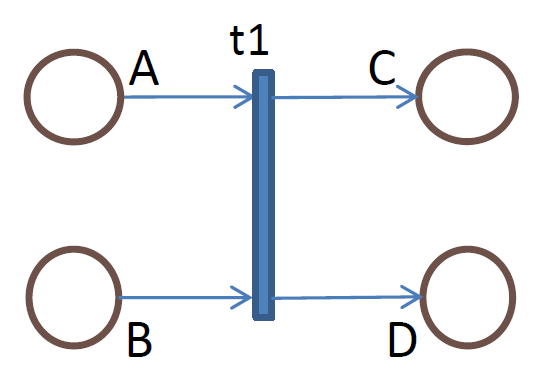}
\caption{$f^\top=(1,1,1,1),\ g_1^\top=(0,1,1,0),\ g_2^\top=(0,1,0,1),\ g_3^\top=(1,0,1,0),\ g_4^\top=(1,0,0,1)$ are five canonical semiflows. $f$ is not minimal and $\left \| f \right \|$ is not minimal. $\mathcal{G}_2=\{g_1,g_2,g_3,g_4\}$ is the unique generating set over $\mathbb{N}$ and $f= g_1+g_4=g_2+g_3$ has exactly two different decompositions in $\mathcal{G}_2$. $\mathcal{G}_3 = \{g_1, g_2, g_3\}$ is a generating set over $\mathbb{Q}$.}
\label{STC2}
\end{figure}

However, a minimal semiflow does not necessarily belong to a generating set over $\mathbb{Q^{+}}$ or $\mathbb{Q}$. In Figure \ref{STC3}, $\mathcal{G}_1$ does not include $f_1$, which is minimal or in Figure \ref{STC2} where $\mathcal{G}_3$ does not include $g_4$ 
which is of minimal support. 

\subsubsection{Decomposition over semiflows of minimal support}
Only the first part of theorem \ref{th: min support} can be found in \cite{MR79}.
\begin{theorem} (\textbf{Minimal support})
\label{th: min support}
If $I$ is a minimal support then 

i) there exists a unique minimal semiflow $f$ such that $I = \left \| f \right \|$ and $\forall g \in \mathcal{F^{+}}$ such that $\left \| g \right \| = I, \exists k \in \mathbb{N}$ such that $g = kf$,

ii) any non-null semiflow $g$ such that $\left \| g \right \| = I$ constitutes a generating set over $\mathbb{Q^{+}}$ or $\mathbb{Q}$ for $\mathcal{F}_I^+ = \left\{ g \in \mathcal{F^+} |\ \left\| g\right\|= I\right\}$.
\end{theorem}
In other words, $\{f\}$ is a unique generating set over $\mathbb{N}$ for $\mathcal{F}_I^+ =\{g \in \mathcal{F^{+}} \ | \ \left \| g \right \| = I\}$. 
However, this uniqueness property is indeed lost in $\mathbb{Q^{+}}$ or in $\mathbb{Q}$, since any element of $\mathcal{F}_I^+$ is a generating set of $\mathcal{F}_I^+$ over $\mathbb{Q^{+}}$ or $\mathbb{Q}$.

From Sperner's theorem, any support $I$ of a semiflow contains a finite number $c_I$ of minimal supports of semiflows. The following theorem states that these $c_I$ supports cover $I$, and provide a generating set deduced from these $c_I$ supports.
\begin{theorem} (\textbf{Decomposition over $\mathbb{Q^+}$})
\label{th: decomp}
Any support $I$ of semiflows is covered by the finite subset $\{I_1, I_2, \dots, I_N\}$ of minimal supports of semiflows included in $I$:

$I = \bigcup_{i=1}^{i=N} I_i$.

Moreover,
$\forall f \in \mathcal{F^{+}}$ such that $\left \| f \right \| \subseteq  I$, one has $f=\sum_{i=1}^{i=N} \alpha_ig_i$ where $ \forall i \in \{1,2,...N\},\ \alpha_i \in \mathbb{Q^{+}}$ and the semiflows $g_i$ are such that $\left \| g_i \right \| = I_i$.
\end{theorem}

A sketch of proof of theorem \ref{th: decomp} can be found in \cite{MR79, BR82}, a complete proof in \cite{M78}.

\subsection{Three extremums directly drawn from the notion of semiflow}
\label{subsec: bounds}
The knowledge of any finite generating set allows a practical computation of three extremums directly inspired from property \ref{prop: inter-home-spaces} Section \ref{subsubsec: hs-semiflow-liveness}, corollaries \ref{prop: bound-for-support} and \ref{cor: support-morphism} of Section \ref{subsubsec: semi}. Let us define them:
\begin{definition}
Given an initial state $q_0$ and the set of semiflows $\mathcal{F^+}$, the three following extremums can be defined:
\begin{itemize}
    
\item [\textbf{-}] $\iota = min(\mathcal{HS})$ is the least element of $\mathcal{HS}$,
\item [\textbf{-}] $\mu(p,q_0) = \min_{\{f\in \mathcal{F}^+\ |\ f(p) \neq 0\}} \left\lfloor \frac{f^\top q_0 }{f(p)} \right\rfloor$ is the least upper bound for the marking of $p$ that can be built directly from a semiflow the support of which contains the given place $p$ in $P$,
\item [\textbf{-}] $\rho = max\{S \subseteq \mathcal{SV}\ |\ \exists f \in \mathcal{F}^, S = \left\| f \right\|\}$ 
is the largest support of any semiflow in $\mathcal{F^+}$.
\end{itemize}
\end{definition}

Theorem \ref{th: bounds} expresses the fact that these bounds are independent from the chosen generating set utilized to compute them: 
\begin{theorem}
\label{th: bounds}
Let's $\mathcal{E} = \{e_1,...e_N\}$ be any finite generating set of $\mathcal{F}^+$, and $q_0 \in Q$ an initial state:
\begin{itemize}
  \item [\textbf{-}]
If $\mathcal{E}$ is over $\mathbb{S}$ then we have:
  
  $\iota = \bigcap_{f \in \mathcal{F^+}} HS(f,q_0) = \bigcap_{e_i \in \mathcal{E}} HS(e_i,q_0)$,
  \item [\textbf{-}]
If $\mathcal{E}$ is over $\mathbb{Q}^+$ or $\mathbb{N}$ then for any place $p$ belonging to at least one support of a semiflow of $\mathcal{F}^+$, $\forall q \in RS(PN,q_0)$, we have :   

$q(p) \leq \mu(p,q_0) = \min_{\{f\in \mathcal{F}^+\ |\ f(p) \neq 0\}} \left\lfloor \frac{f^\top q_0 }{f(p)} \right\rfloor = \min_{\{e_i\in \mathcal{E}\ |\ e_i(p) \neq 0\}} \left\lfloor \frac{e_i^\top q_0 }{e_i(p)} \right\rfloor $,

  \item [\textbf{-}]
  If $\mathcal{E}$ is over $\mathbb{S}$ then we have:

$\rho = \left\| \sum_{f\in \mathcal{F}^+} f\right\|= \bigcup_{f\in \mathcal{F}^+}\left\|f\right\| = \bigcup_{e_i \in \mathcal{E}}\left\|e_i\right\|$
 
\end{itemize}
\end{theorem}

First, the three bounds are computable since by Gordan's lemmma of Section \ref{subsec: gs-basic-results}, there exists a finite generating set such as $\mathcal{E}$.
\begin{itemize}
\item [\textbf{-}] 

For the first item of the theorem, let's consider $f \in \mathcal{F^+}$ with $f = \sum_{i=1}^{i=N} \alpha_ie_i$ 

and $q \in \bigcap_{e_i \in \mathcal{E}} HS(e_i,q_0)$, then:

$\alpha_i(e_i^\top q) = \alpha_i(e_i^\top q_0) \ \forall i\in \{1,...N\}$, hence:  

$ \sum_{i = 1}^{i = N} \alpha_i(e_i^\top q) = \sum_{i = 1}^{i = N} \alpha_i(e_i^\top q_0)$, then:

$f^\top q = f^\top q_0$, 
and $q \in HS(f,q_0)\ \ \forall f \in \mathcal{F^+}$ therefore, 

(since $\mathcal{E} \subseteq \mathcal{F^+}$ directly involves $(\bigcap_{f \in \mathcal{F^+}} HS(f,q_0))  \subseteq \bigcap_{e_i \in \mathcal{E}} HS(e_i,q_0)$) we have: 

$\bigcap_{e_i \in \mathcal{E}} HS(e_i,q_0) = \bigcap_{f \in \mathcal{F^+}} HS(f,q_0) = \iota$.

This first item means that $\iota$ is contained in any $q_0$-home space directly built from a semiflow $f \in \mathcal{F^+}$.

\item [\textbf{-}]
For the second item of the theorem, let's consider a state $q_0$, a place $p$, and a semiflow $f$ of $\mathcal{F}^+$ such that $f(p) > 0$ and $f = \sum_{i=1}^{i=N}\alpha_i e_i$ where $\alpha_i \geq 0\ \forall i \in \{1,...N\}$.

Let's define $\mu_\mathcal{E}$ such that:
$\mu_\mathcal{E} =  
\min_{\{e_i\in \mathcal{E}\ |\ e_i(p) \neq 0\}} \frac{{e_i}^\top q_0}{e_i(p)}$.

Then $\exists j$ such that $1 \leq j \leq N$, and $\mu_\mathcal{E} = \frac{{e_j}^\top q_0}{e_j(p)}$.

Therefore, $\forall i \leq N$, such that $e_i(p) \neq 0$,  
$\exists \delta_i \in \mathbb{Q}^+$ such that:

$\frac{{e_j}^\top q_0}{e_j(p)} = \frac{{e_i}^\top q_0 - \delta_i}{e_i(p)}$. It can then be deduced:

$\mu_\mathcal{E} = \frac{\alpha_j{e_j}^\top q_0}{\alpha_j e_j(p)} = \frac{\alpha_i({e_i}^\top q_0 - \delta_i)}{\alpha_i e_i(p)}\ \forall i$ such that $e_i(p) \neq 0$, therefore:

$\mu_\mathcal{E} = \frac{\sum_{\{i\ |\ e_i(p) > 0\}} \alpha_i({e_i}^\top q_0 - \delta_i)}{\sum_{\{i\ |\ e_i(p) > 0\}}\alpha_i e_i(p)}$

$= \frac{\sum_{\{i\ |\ e_i(p) > 0\}} \alpha_i({e_i}^\top q_0 - \delta_i) + \sum_{\{i\ | e_i(p) = 0\}}(\alpha_i{e_i}^\top q_0 - \alpha_i{e_i}^\top q_0)} {\sum_{\{i\ |\ e_i(p) > 0\}}\alpha_i e_i(p)} $

$= \frac{\sum_{\{i | e_i(p) > 0\}} \alpha_i{e_i}^\top q_0 + \sum_{\{i | e_i(p) = 0\}}\alpha_i{e_i}^\top q_0 - \sum_{\{i | e_i(p)>0\}}\alpha_i\delta_i - \sum_{\{i\ |\  e_i(p) = 0\}}\alpha_i{e_i}^\top q_0} {\sum_{\{i\ |\ e_i(p) > 0\}}\alpha_i e_i(p) + \sum_{\{i\ |\ e_i(p) = 0\}}\alpha_i e_i(p)}$

since $\sum_{\{i\ |\ e_i(p) = 0\}}\alpha_i e_i(p) = 0$. Then, since $\delta_i \geq 0$ and $\alpha_i \geq 0\  \forall i$ such that $1\leq i \leq N$

$\mu_\mathcal{E} = \frac{\sum_{i=1}^{i=N}\alpha_i{e_i}^\top q_0 -\sum_{\{i | e_i(p)>0\}}\alpha_i\delta_i - \sum_{\{i\ |\  e_i(p) = 0\}}\alpha_i{e_i}^\top q_0}{\sum_{i=1}^{i=N}\alpha_i e_i(p)}$

$\mu_\mathcal{E} = \frac{f^\top q_0 -\sum_{\{i | e_i(p)>0\}}\alpha_i\delta_i - \sum_{\{i\ |\  e_i(p) = 0\}}\alpha_i{e_i}^\top q_0}{f(p)} \leq  \frac{f^\top q_0 }{f(p)}$ 

This being verified for any semiflow of $\mathcal{F}^+$, we have: $\mu(p,q_0) = \left\lfloor \mu_\mathcal{E} \right\rfloor$.

\item [\textbf{-}]
For the third item of the theorem, let's consider $\mathcal{E}$ a generating set over $\mathbb{S}$ then, any semiflow $f$ can be decomposed as follows:

$f = \sum_{\alpha_i > 0} \alpha_ie_i + \sum_{\alpha_i < 0}\alpha_ie_i$ where $\alpha_i \in \mathbb{S}$.
$f \in \mathcal{F}^+$ means that at least one coefficient $\alpha_i$ is strictly positive and 

$\sum_{\alpha_i < 0}|\alpha_i|e_i + f = \sum_{\alpha_i > 0}\alpha_ie_i \neq 0$.

Therefore, applying corollary \ref{cor: support-morphism}: 

$\left\|f\right\| \subseteq \left\|\sum_{\alpha_i < 0}|\alpha_i|e_i + f\right\| = \left\|\sum_{\alpha_i > 0} \alpha_ie_i\right\| = \bigcup_{\alpha_i > 0} \left\|\alpha_ie_i\right\| \subseteq \bigcup_{e_i \in \mathcal{E}}\left\|e_i\right\|$.

Hence, $\rho = \left\| \sum_{f\in \mathcal{F}^+} f\right\|= \bigcup_{e_i \in \mathcal{E}}\left\|e_i\right\|$
\hfill
$\square$
\end{itemize}

This theorem means that these three parameters $\iota$, $\mu$ and $\rho$ can be computed with the help one finite generating set and furthermore that their values are independent of the chosen generating set (over $\mathbb{S}$ for the first and third item, but only over $\mathbb{N}$ or $\mathbb{Q}^+$ for the second item).
This is moderately astonishing considering that according to theorem \ref{th: over N}, minimal semiflows belong to any generating set over $\mathbb{N}$.
The third part of this theorem means that if $\mathcal{E} = \{e_1,...e_N\}$ is any generating set of a given Petri Net PN then $\bigcup_{e_i \in \mathcal{E}}\left\|e_i\right\|$ is also the largest support of PN (it would useless to look for other semiflow than $\mathcal{E}$).

\subsection{Canonical semiflows}
\label{sec: canonical}

A semiflow is \textit{canonical} (\cite{ColomS89}, \cite{ColomTS2003} p. 68) if and only if the gcd of its non-null coordinates is equal to one. In \cite{CMPW09}, such a semiflow is said to be \textit{scaled back}. 

\subsubsection{Canonical and minimal semiflows}
Minimal semiflows and canonical semiflows are two different notions. The following lemma and theorem help to compare them.

\begin{lemma}
\label{lem: canonical-minimal}
If a semiflow is minimal then it is canonical.

If a semiflow is canonical and its support is minimal then it is minimal.
\end{lemma}

The first point is quite evident: if $f$ is not canonical its gcd $k$ is such that $k>1$ so $\exists g \in \mathcal{F^{+}}$ such that $f= kg$ and $f$ would not be minimal.

The second point is a direct application of theorem \ref{th: min support}.
\hfill
$\square$

However, canonical semiflows are not necessarily minimal semiflows; minimal semiflows do not necessarily have a minimal support. 
For example, Figure \ref{STC3}, $f_1$ and $f_2$ are canonical and minimal, but their support is not minimal.
Any semiflow $f= ag_1 + g_2 + bg_3$ where $a, b \in \mathbb{N}$ and $a + b > 0$ is canonical and clearly not minimal. As $a$ and $b$ can be arbitrarily large, this shows that the number of canonical semiflows can be infinite. 

\subsubsection{About the number of canonical semiflows}
\label{subsec: nb of canonical}
We can observe in the previous example that the infinite sequence of canonical semiflows is constituted of semiflows with non-minimal support. This is hinted on the following theorem.

\begin{theorem}
\label{th: canonical}
Given a support $I$, $c_I\ = |\{f\  canonical\ semiflow \ | \ \left \| f \right \| = I \}|$,

If $I$ is a minimal support, then $c_I = 1$, else $c_I$ is infinite.

If $c_I = 1$, then $I$ is minimal.
\end{theorem}

From theorem \ref{th: min support}, there is a unique minimal semiflow having a given minimal support. From lemma \ref{lem: canonical-minimal}, a minimal semiflow is canonical. Hence, if $I$ is minimal, then $c_I = 1$.

If $I$ is not minimal, then $\exists\ e,\ f \in \mathcal{F}^+$ such that $ \ \left \| e \right \| \subset \ \left \| f \right \| = I$.
We can build an infinite sequence of semiflows $f_i,\ i \in \mathbb{N}$, such that 
$f_i = \alpha_i (f + k_ie)$, where $k_i \in \mathbb{N}$ and $1/\alpha_i$ is the gcd of the non-null coordinates of $f+k_ie$. 
$\forall i \in \mathbb{N}. f_i$ is canonical by construction. 
Let's consider $i,j$ such that $f_i=f_j$; then $\alpha_i (f + k_ie) = \alpha_j (f + k_je)$. 
This leads to: $ (\alpha_i - \alpha_j)f = (k_i - k_j)e$. 
However, since $ \ \left \| e \right \| \subset \ \left \| f \right \|$, we must have $\alpha_i = \alpha_j$ and $k_i = k_j$. 
Hence, we built an infinite sequence of canonical semiflows based upon an infinite sequence of non-negative integers.

If $c_I = 1$ then let $f$ be the unique canonical semiflow of support $I$. 
Let's consider $g \in \mathcal{F}^+$ such that  
$ \ \left \| g \right \| \subseteq I$. 
With the same construction as before, we can build a canonical semiflow $h= \alpha (f + kg)$ where $1/\alpha$ is the gcd of the non-null coordinates of $f+kg$, and $k \in \mathbb{N}$. We have $ \ \left \| h \right \| = I$ and $c_I = 1$; therefore, $h=\alpha (f + kg)=f$. Then, $g=((1-\alpha)/k\alpha)f$ which means that any semiflow of support included in $I$ is a multiple of $f$. Hence, $I$ is minimal.
\hfill
$\square$

The fact that the number of canonical semiflows can be infinite was already pointed out in \cite{Kruck86,ColomS89,CMPW09}. The fact that this number is infinite only when the considered support is non minimal as described in theorem \ref{th: canonical} is new to the best of our knowledge.
 
\subsection{Minimal generating sets, least generating sets, and fundamental sets}
\label{subsec: mgs}

Minimal generating sets have been defined over $\mathbb{N}$ in \cite{M78} and over $\mathbb{Q^+}$ in \cite{M83} p.39; few years later, the notion of least generating sets over $\mathbb{Q}$ was introduced in \cite{ColomS89} p.82 and can be found in \cite{GV03} p.68.
Similarly to the notion of generating set defined in Section \ref{subsec: gs-basic-results}, we slightly extend their definition to hold over a set $\mathbb{S} \in \{\mathbb{N}, \ \mathbb{Q^+}, \mathbb{Q}\}$. 

\begin{definition}
A \textit{minimal generating set} over $\mathbb{S}$ is a generating set that does not strictly include any generating set.

A \textit{least generating set of semiflows} ``is made up of the least number of elements to generate any semiflow" over $\mathbb{S}$ \footnote{More precisely, the least generating set is defined over $\mathbb{Q}$ in \cite{ColomS89,ColomTS2003} and over $\mathbb{N}$ in \cite{CMPW09}.}
In other words, $\mathcal{G}$ is a least generating set if and only if it does not exist a generating set $\mathcal{H}$ such that $\left| \mathcal{H} \right| \leq \left| \mathcal{G} \right|$. 
\end{definition}

\subsubsection{Coincidence between minimal and least generating sets}
A minimal generating set is defined with respect to set inclusion while a least generating set is defined with respect to its cardinality. In general, these two notions are different, however, in the case of generating sets of semiflows, theorem \ref{th: lgs=mgs} hereunder is a new result stating that these two notions are in fact equivalent over $\mathbb{S}$.

\begin{lemma}
\label{lemma: lgs=mgs}
If $\mathcal{G}$ is a generating set over $\mathbb{Q^+}$ or $\mathbb{N}$ and $I$ a minimal support, then $\exists g \in  \mathcal{G}$ such that $I= \left\|g \right\|$.
\end{lemma}
We consider $e$ a semiflow of minimal support, $\mathcal{G} = \{g_1,...g_k\}$, a generating set over $\mathbb{Q^{+}}$ or $\mathbb{N}$. Then, $e = \sum_{i=1}^{i=k}\alpha_ig_i$. 
All the coefficients are non-negative and $e \neq 0$, then $\exists j \leq k$ such that $\alpha_j > 0$ and $e \geq \alpha_jg_j$. Since $\left \| e \right \|$ is minimal, $\left \| e \right \| = \left \| g_j \right \|$. 
\hfill
$\square$

This lemma states that any generating set over $\mathbb{Q^+}$ or $\mathbb{N}$ contains at least one semiflow per minimal support. Indeed, this property is not true over $\mathbb{Q}$. In Figure \ref{STC2}, $\mathcal{G}_2$ is a minimal generating set over $\mathbb{Q}^+$ and $\{g_2, g_3, g_4\} \subset \mathcal{G}_2$ is a generating set over $\mathbb{Q}$ since $g_1= g_2 + g_3 -g_4$ is of minimal support but generated over $\mathbb{Q}$ (since one coefficient is negative) by the other minimal semiflows of minimal support.

\begin{theorem}
\label{th: lgs=mgs}

If $\mathcal{G}$ is a generating set over $\mathbb{S}$, where $\mathbb{S} \in \{\mathbb{N}, \mathbb{Q^{+}}, \mathbb{Q}\}$ then the two following properties are equivalent:

$\mathcal{G}$ is a minimal generating set,

$\mathcal{G}$ is a least generating set. 
\end{theorem}

First, the fact that a least generating set is a minimal generating set is straightforward.

Let's consider $\mathcal{G}$, a minimal generating set over $\mathbb{N}$. 
By applying theorem \ref{th: over N}, we conclude that $\mathcal{G}$ is the set of minimal semiflows and a least generating set.

Next, let's consider $\mathcal{G}$, a minimal generating set over $\mathbb{Q}^{+}$. Then, lemma \ref{lemma: lgs=mgs} can apply stating that $\mathcal{G}$ includes $\mathcal{G'}$ a family of exactly one semiflow for each minimal support. From theorem \ref{th: decomp}, we draw that $\mathcal{G'}$ is a generating set. $\mathcal{G}$ is minimal then $\mathcal{G} = \mathcal{G'}$. This being true for any minimal generating set, $\mathcal{G}$ is also a least generating set.

Finally, let's consider $\mathcal{G}$ a generating set over $\mathbb{Q}$. From $\mathcal{G}$ we can extract a subset $\mathcal{B}$ of linearly independent semiflows (see \cite{Lan02} p. 85 for basic results on vector spaces). Then, $\mathcal{B}$ is a least and minimal generating set over $\mathbb{Q}$. 
\hfill 
$\square$

\subsubsection{About fundamental sets}
Theorem \ref{th: canonical} states that there is exactly one canonical semiflow for each minimal support. This particularity characterizes the notion of a fundamental set introduced in \cite{STC1998} p. 319,. 
 
\begin{definition}
The set of all canonical semiflows of minimal support is called a  \textit{fundamental set}. 
\end{definition}   

\begin{corollary} (fundamental set) 
A fundamental set is a generating set over $\mathbb{Q^{+}}$ or $\mathbb{Q}$ but not necessarily over $\mathbb{N}$.
A fundamental set over $\mathbb{Q^{+}}$ is a minimal generating set but not necessarily over $\mathbb{Q}$.
\label{cor: fundamental set}
\end{corollary}

The first point of this corollary is a direct consequence of theorem \ref{th: decomp} and lemma \ref{lem: canonical-minimal}: we conclude that a fundamental set is one possible generating set over $\mathbb{Q^{+}}$ and therefore over $\mathbb{Q}$.

The second point is directly deduced from the first point and lemma \ref{lemma: lgs=mgs}.

The last parts of the two points of the corollary are illustrated by the two following counterexamples.

In Figure \ref{STC3}, $\mathcal{G}_1=\{g_1, g_2, g_3\}$ is a fundamental set that is not a generating set over $\mathbb{N}$ since $f_1$ or $f_2$ are minimal and cannot be decomposed as a linear combination of elements of $\mathcal{G}_1$ over $\mathbb{N}$ 
\footnote{In this regard, the statement p.143-147 of \cite{CMPW09} should be rewritten.}.

A fundamental set over $\mathbb{Q}$ is not necessarily a minimal generating set; in Figure \ref{STC2}, $\mathcal{G}_2$ is the fundamental set and is not a minimal generating set.
\hfill
$\square$
  
Belonging to a least generating set, hereunder denoted by \textit{lgs}, does not equip a semiflow $f$ with specific properties  
\footnote{The properties 2.2 p. 82 of \cite{ColomS89} and 5.2.5 p.68 of \cite{Colom2003} should be rewritten by taking the following statements into account.}:

\textbf{-} if $f \in lgs$ over $\mathbb{N}$, then $f$ is minimal but not necessarily canonical or of minimal support;

\textbf{-} if $f \in lgs$ over $\mathbb{Q}^+$, then $f$ has a minimal support but is not necessarily canonical or minimal;

\textbf{-} if $f \in lgs$ over $\mathbb{Q}$, then $f$ is not necessarily minimal and not necessarily canonical or of minimal support.

\subsubsection{About uniqueness}
\label{subsec: uniqueness}
\begin{corollary}(uniqueness)
The set of minimal semiflows is the unique minimal generating set over $\mathbb{N}$.

If $\mathcal{G}$ is a least generating set over $\mathbb{Q}^+$ then for any minimal support $I$ of semiflows, $\exists g \in \mathcal{G}$ unique such that $I = \left \| g \right \|$.  

The fundamental set is the unique minimal generating set of minimal semiflows over $\mathbb{Q}^{+}$. 
\label{cor: uniqueness}
\end{corollary}
The first point can be directly deduced from theorem \ref{th: over N}, the second and third  are directly deduced from theorems \ref{th: min support} and \ref{th: decomp} and lemma \ref{lem: canonical-minimal}.
\hfill
$\square$

The third point of this corollary can be considered as a variation of a statement in \cite{STC1998}.

However, a minimal generating set over $\mathbb{Q}^+$ or $\mathbb{Q}$ is not unique even among minimal semiflows of minimal support. 
In the example of Figure \ref{STC2}, 

$\{k_1g_1, k_2g_2, k_3g_3, k_4g_4\}$ where $k_i \in \mathbb{N}$ constitutes a family of minimal generating sets over $\mathbb{Q}^+$. 
Moreover, $ \mathcal{G}_3 = \{g_1,g_2,g_3\}$ and $\{g_2,g_3,g_4\}$ are two minimal generating sets over $\mathbb{Q}$.

\subsection{Results summary}
\label{sec: summary}
\begin{figure}[ht]
\centering
\includegraphics[width=1\textwidth]{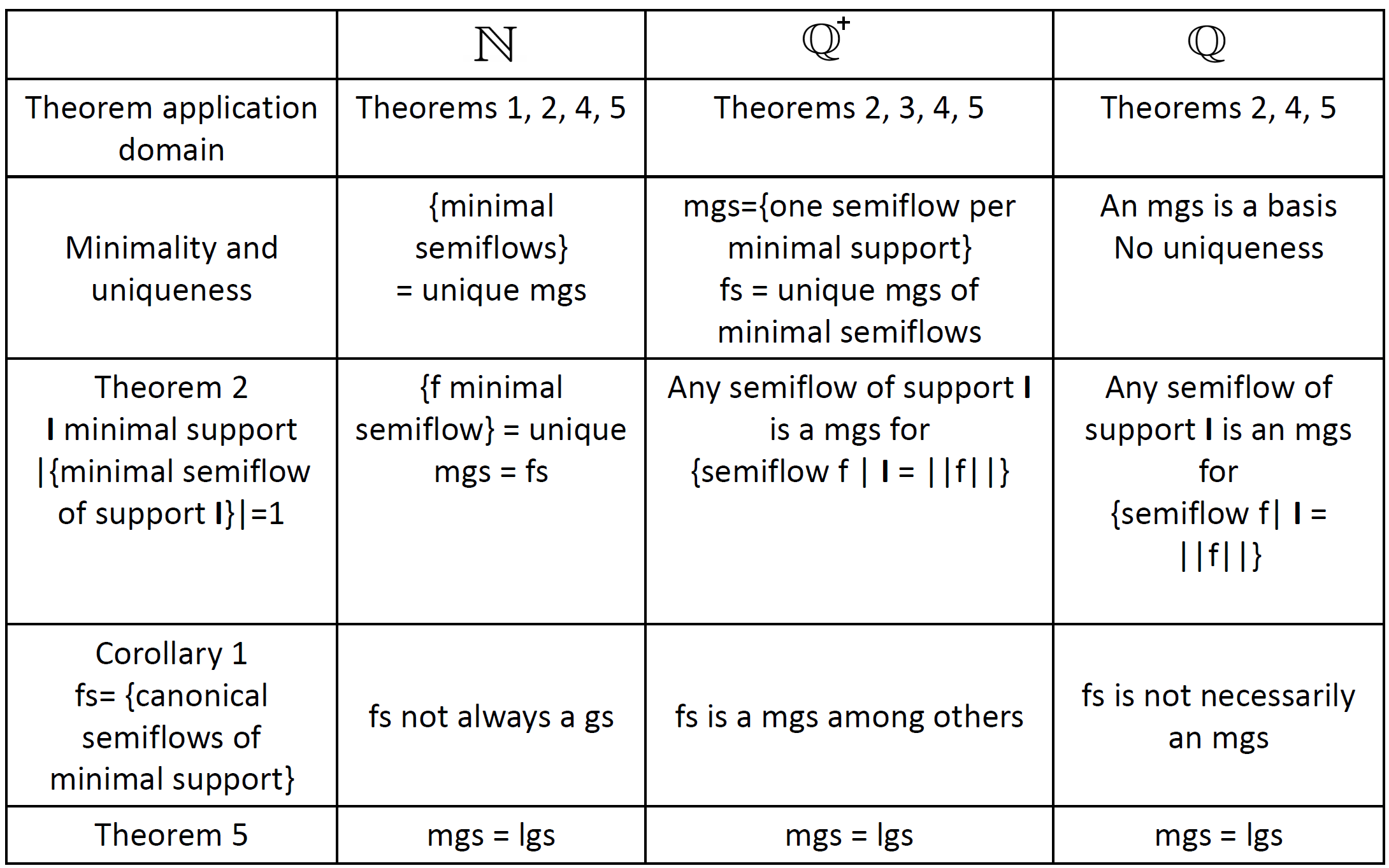}
\caption{gs, fs, mgs, lgs denote the generating set, the fundamental set, the minimal generating set, and the least generating set respectively.}
\label{table-gs}
\end{figure}

We have seen through several counterexamples that no result must be taken for granted and that any proposition must be carefully addressed.
Figure \ref{table-gs} summarizes the main results presented in this note. 

Our contribution in terms of new results consists of the following:

\textbf{-} the three theorems of decomposition in Section \ref{sec: 3theorems}, which have been slightly extended with a new proof for theorem \ref{th: over N} and could be used for infinite Petri Nets \cite{PETRI96};

\textbf{-} the theorem \ref{th: bounds} making possible to compute bounds directly deduced from semiflow generating sets, in terms of home spaces, boundedness, or supports;

\textbf{-} the lemma \ref{lem: canonical-minimal} and theorem \ref{th: canonical} of Section \ref{sec: canonical} about comparing minimal and canonical semiflows;

\textbf{-} lemma \ref{lemma: lgs=mgs} and theorem \ref{th: lgs=mgs} about the equivalence between minimal and least generating sets;

\textbf{-} corollaries \ref{cor: fundamental set} and \ref{cor: uniqueness} about fundamental sets and uniqueness. 

By considering $\mathbb{N}$, then $\mathbb{Q^+}$, then $\mathbb{Q}$, the size of a minimal generating set decreases as expected since an increasing amount of possibilities to combine semiflows are provided. More precisely, if $c$ is the number of minimal supports in a Petri Net then, we have the following corollary:

\begin{corollary}
\label{Corollary: c-size}
If $c$ is the number of minimal support of $\mathcal{F^{+}}$:

A minimal generating set over $\mathbb{N}$ is finite and has at least $c$ elements,

A minimal generating set over $\mathbb{Q^+}$ has exactly $c$ elements,

A minimal generating set over $\mathbb{Q}$ has, at most, $c$ elements.  
\end{corollary}

Together with the mathematical definitions associated with the notion of generating set in Section \ref{subsec: gs-basic-results}, this result contribute in differentiating the properties of generating sets over $\mathbb{N},\ \mathbb{Q^+}$, or $\mathbb{Q}$. A generating set over $\mathbb{N}$ is often too large; a generating set over $\mathbb{Q^+}$ allows deducing strong properties (boundedness in particular) but is complex to compute while a generating set over $\mathbb{Q}$ is easy to compute but looses the ability to deduce boundedness for the state variables in the union of its supports which is at the source of proving liveness as illustrated through the few following examples. 


\newpage
\section{Reasoning with invariants, semiflows, and home spaces}
\label{sec: ex}

Invariants, semiflows, and home spaces can be used to prove a rich array of behavioral properties of various conceptual models such as LTS or Petri Net even within different settings, in particular when using parameters. 

Analysis can be performed with incomplete information on the initial marking as shown in the first example below or even on a subsystem, exhibiting some compositionality ability. 
The model can be described with parameters which will make the invariant calculus of Section \ref{subsec: inv calculus} more complex but still tractable as shown in the subsequent examples. 
Most of the time, especially with actual system models, semiflows calculus can easily be performed and interesting invariants can be deduced since after all, most of the time specifications express invariants.
It will then be possible to conduct an analysis avoiding a painstaking symbolic model checking or a parameterized and complex development of a reachability graph (see for instance, \cite{DRvB01}).  

In this section, we present four parameterized examples with their algebraic or arithmetic analysis to illustrate how the properties on semiflows and home states can be articulated.
Some invariants are deduced from the structure of the Petri Nets in addition to the invariants directly deduced from semiflows. This is their combination which allows to prove behavioral properties (not only in terms of boundedness but also in terms of liveness).
The two first ones show how to use invariants, semiflows, and elementary arithmetic reasoning. 
The third one is more elaborated. Coming from the telecommunication industry, it utilizes a proof by \textit{home spaces refinement}. 
The last one illustrates the difficult question of fairness and starvation.

\subsection{A tiny example}
\label{subsec: tiny}
\begin{figure}[ht]
\centering
\includegraphics[width=0.6\textwidth]{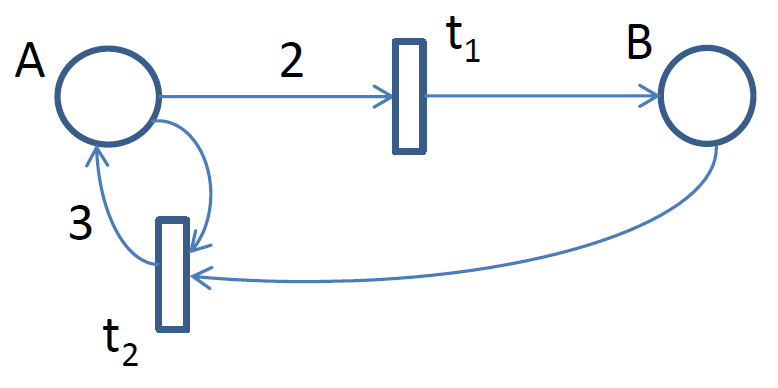}
\caption{Any semiflow $f^\top = (a,b)$ must verify the system of equation (\ref{eq: inv-semiflow}). 
In this example, the equations associated with transitions $t_1$ and $t_2$ are identical and we get $2a=b$ where $a$ and $b$ the variables associated to $A$ and $B$ respectively.
$f^\top  = (1, 2)$ is an obvious solution.
\newline
This tiny Petri Net is live if and only if $f^\top q_0$ is an odd number greater than one whatever is the initial marking of $B$.}
\label{fig: tiny}
\end{figure}

The Petri Net $TN = \langle \{A,B\}, \{t_1,t_2\}, Pre, Post \rangle$ in Figure~\ref{fig: tiny} is defined by: 

$Pre(\cdot,t_1)^\top =(2,0); Pre(\cdot,t_2)^\top =(1,1)$;

$Post(\cdot,t_1)^\top =(0,1); Post(\cdot,t_2)^\top =(3,0).$

This example can be first found in \cite{BR82} or in \cite{M83} without proof then in \cite{M19} with the main elements of a proof of liveness. 
Here, the analysis is revisited, completed, and enriched with new examples stemming from it. 
From parameters on the marking, the example is made more complex by adding a parameter on some edges before adding a last parameter expending the very structure of the model. 
Each time, a similar liveness analysis can be conducted. 

\subsubsection{A first level of analysis}
\label{subsubsec: TN}

$f^\top  = (1, 2)$ is a semiflow: the Kirchhoff{'}s law is easily verified for the nodes (i.e. transitions or equations) $t_1$ and $t_2$ of $TN$. 
Moreover, it is obvious that $f$ is a minimal semiflow  in $\mathbb{N}$ and that its support is minimal as well.

$TN$ models a formal machine that stops only when a given natural number $n$ is even or null. 
In our context, it means that it will always be possible to find a sequence of transitions such that $TN$ stops (i.e. $t_{1}$ and $t_{2}$ are not enabled anymore) only when $n$ as the initial marking of $A$ is even or null, and $TN$ is live otherwise.
This also means that there exists at least one sink (see definition \ref{def: source & sink}) reduced to a singleton in $RG(TN))$ given the initial marking $q_0$ such that $q_{0}(A)=n,\ q_0(B)=x$ where $n$ is even or null and $x \in \mathbb{N}$. 
Actually, we can do slightly better and also prove that $\left\langle TN,q_0 \right\rangle$ is live if and only if $f^\top q_0$ is odd and that if $f^\top q_0$ is even or null then there exists exactly one sink reduced to a single marking in $RG(TN)$.     

The scalar product $f^\top q$ does not vary, we have the following invariant: for any marking $q$ reachable from an initial marking $q_0$, $f^\top q = q(A)+2q(B) =  f^\top q_0$. 
This can be rewritten as $q(A) = f^\top q_0-2q(B)$ which means that $q(A)$ and $f^\top q_0$ have always the same parity and this will never change \footnote{Let's point out that we just used the phrases \textit{always} and \textit{never will} as mentioned in Section \ref{subsec: invariant-definition}: we expressed an invariant about the parity of the marking of the place $A$. This is not the usual formulation for an invariant generated by a semiflow.} (if we admit that zero is even). 

If $f^\top q_0 > 2$ and is odd, then $q(A)$ also is odd. 
If $q(A)=1$ and $f^\top q_0 > 2$ then $q(B)>1$ and $t_2$ is enabled in $q$; it can easily be deduced that $\left\langle TN,q_0 \right\rangle$ is live since $q(A)$ remains odd and strictly positive. 
If $q(A)>1$ then $t_1$ is enabled in $q$; again, the liveness of $TN$ can be easily deduced. 
Reversely, if $\left\langle TN,q_0 \right\rangle$ is live then $f^\top q_0 > 2$ since the enabling threshold of $t_2$ is $f^\top Pre(\cdot,t_2)=3$ (see property \ref{prop: enabling-threshold}).

Moreover, if $f^\top q_0$ is even, then $q(A)$ is even and the Petri Net is not live ($t_1$ can always be executed until $A$ has no token which means that it is not possible to enable $t_1$ or $t_2$ anymore). 
The marking $q_s$ such that $q_s(A)=0, \ q_s(B)=f^\top q_0 / 2$ is the only possible marking such that $A$ has no token therefore, it is the only sink of the reachability graph of $\left\langle TN,q_0 \right\rangle$ and at the same time, the only home state of the Petri Net since for ant reachable marking $q$ ($q$ being even) we can execute $t_1 \ q(A)/2$ times and reach $q_s$ (see property \ref{prop: sink & home spaces})
\hfill 
$\square$ 

Let us make few comments here.

First, let us point out that the loop between $A$ and $t_2$ is critical and its disappearance from the incidence matrix prevents to draw this kind of result with the sole usage of the state equation (\ref{eq: state}).

Since $f>0$, we have $A$ and $B$ bounded by $f^\top q_0$ for any initial marking $q_0$. $\{A,B\}$ is therefore structurally bounded (property \ref{prop: boundedness-alg}.

What is worth noticing about the analysis of this tiny example is that it was not necessary to develop the reachability graph in order to decide whether or not the Petri Net is live or bounded. 
We could analyze $TN$ considering $q_0(A)=n$ as a parameter and without even considering the value $x$ taken by $q_0(B)$.

An interesting lesson to be drawn from this tiny example has to do with liveness: once live, adding more tokens to some places of a Petri Net is far from guaranteeing that liveness will persist. It can only increase the size of the reachability set.
As a matter of fact, while a minimum of tokens are required (refer for instance to the notion of enabling threshold in \ref{subsubsec: semi}) for $TN$ to be live, adding one token to the place A will make $TN$ not live while adding a second token will make it live again and so on.

Even if adding more resources to the system under study can bring confidence that some actions will eventually be performed, this  does not necessarily result in solving a deadlock issue; this could even create a deadlock situation! 
In this regard, adding more resources can be a ``false good idea” that is encountered with many students and engineers certainly taking its origin with the notion of monotonicity seen section \ref{sec: behav}. 

Last but not least, it is possible to use the same $TN$ to semi-decide whether $n$ is odd simply by considering $q_0(A)=n+1$.
This remark will be used Section \ref{subsubsec: euclidean} to build $TNED(i)$, a parameterized Petri Net that recognizes the remainder of the euclidean division of a natural number $n$ by a given number $i$.

\subsubsection{Adding a parameter to the model}
\label{subsubsec: TN(i)}

\begin{figure}[ht]
\centering
\includegraphics[width=0.6\textwidth]{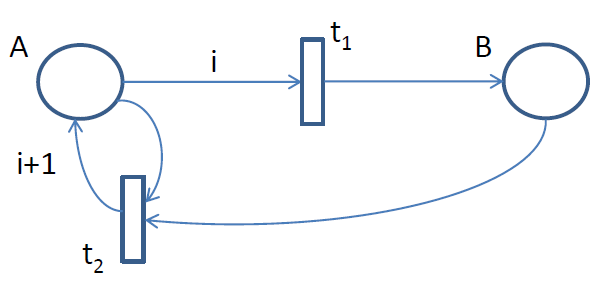}
\caption{
Semiflows must verify the equation: $i \times a = b$ for which $g^\top  = (1,i)$ is an obvious solution.
\newline
This parameterized version $TN(i)$ of $TN$ is live if and only if $g^\top q_0 > i$ and is not a multiple of $i$, whatever is the initial marking of $B$. 
For $i=1$, $TN(i)$ has no live transition whatever is the initial marking.
\newline
We have $TN(2) = TN$ of Figure~\ref{fig: tiny}.}
\label{fig: tinyi}
\end{figure}

We can expand the tiny example $TN$ of Figure \ref{fig: tiny} by adding  a parameter $i$ such that $i>1$, only by changing the labeling of the edges $(A,t_1)$ and $(t_2,A)$ by $Pre(A,t_1)=i$ and $Post(A,t_2)=i+1$ respectively as in Figure~\ref{fig: tinyi}. 
Calling $TN(i)$ this Petri Net, we have $TN(2) = TN$ as expected. 

This time, we have the following minimal semiflow of minimal support: $g^\top  = (1, i)$ and we can prove that $\left\langle TN(i),q_0 \right\rangle$ is not live if and only if $g^\top q_0 \leq i$ or $g^\top q_0 = n \times i$ where $n \in \mathbb{N}$, independently of the value of $q_0(B)$.
In other words, $TN(i)$ recognizes whether a given number $n$ is a multiple of $i$ the same way as $TN$ recognizes whether a number is even.

First, if $g^\top q_0 < i$ then the enabling threshold of $t_1$ can never be reached (property \ref{prop: enabling-threshold}) and neither $t_1$ nor $t_2$ can be executed (since $q_0(B)$ is necessarily null to satisfy the inequality). 
Second, if $g^\top q_0 \geq i$ then we consider the Euclidean division of $g^\top q_0$ by $i$ giving: $g^\top q_0 = n \times i + r$ where $r < i$ then since $g$ is a semiflow, $g^\top q = q(A)+iq(B) \equiv r\ (mod \ i)$ therefore $q(A) \equiv r\ \forall q \in RS(TN(i),q_0)$. If $r=0$  then we have $q(A)= n \times i - i \times q(B)$ and $t_1$ can be executed $n - q(B)$ times to reach a marking with zero token in $A$. 

If $r \neq 0$ and $g^\top q_0 > i$ then $q(A) \neq 0$ and either $q(A) > i$ or $q(B) \neq 0 \ \forall q \in RS(TN(i),q_0)$. 
In the first case, $t_1$ is enabled; in the second case $t_2$ is enabled. 
It is easy to conclude that the Petri Net $TN(i)$ is live if and only if $g^\top q_0 > i$ and is not a multiple of $i$ whatever is the initial marking of $B$.
\hfill 
$\square$ 

\subsubsection{euclidean division}
\label{subsubsec: euclidean}
\begin{figure}[ht]
\centering
\includegraphics[width=0.6\textwidth]{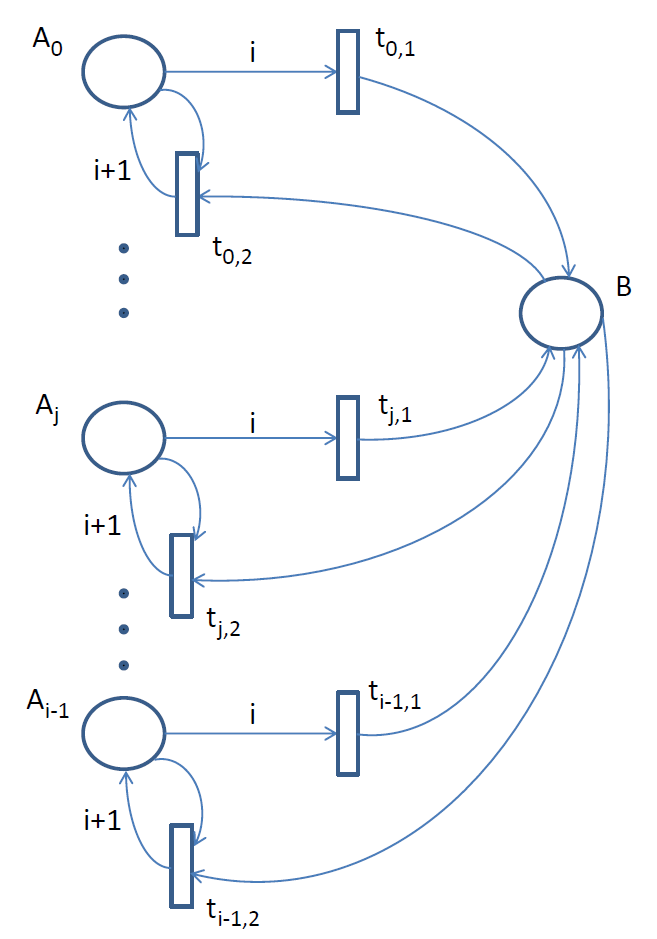}
\caption{
This time, we have a system of $i$ equations: 
$i \times a_j = b$ with $j \in \left[0,i-1\right]$
\newline
for which $g$ such that $g(A_j) = 1$ for $j \in \left[0,i-1\right]$ and $g(B)=i$ is the minimal semiflow of minimal support in $\mathbb{N}$.
\newline
This parameterized Petri Net allows to know the remainder of the euclidean division of a natural number $n$ by $i$.
}
\label{fig: euclidean}
\end{figure}

From the properties of $TN$ and $TN(i)$, it is natural to progress by one more step and propose to design a Petri Net with the ability not only to recognize whether a natural number $n$ is a multiple of a given natural number $i$, but more generally to recognize the remainder of the euclidean division of $n$ such that $n>0$ by $i$ such that $i>0$.
To this effect, we consider the Petri Net $TNED(i)$ of Figure \ref{fig: euclidean}, and the parameter $i \geq 2$. We have:
$P = \{\{A_j |\ j \in \left[0 , i-1\right]\}, B\}$, 
$T=\{t_{j,1} ,t_{j,2} |\ j \in \left[0 , i-1\right]\}$, where Pre and Post are defined by :
\newline
$Pre(A_j,t_{j,1})=i,\ Pre(B,t_{j,1})=\ Pre(A_j,t_{j,2})=1$,
\newline
$Post(A_j,t_{j,2})=i+1,\  Post(B,t_{j,1})=1$ where $j \in  \left[0 , i-1\right]\ $.
\footnote{$TNED(i)$ can be more succinctly represented by an equivalent Colored Petri Net with $i$ different colors and a graph isomorphic to $TN(i)$ of figure \ref{fig: tinyi}.}

The initial marking is such that $q_0(A_j) = n+j$ where $j \in \left[0 , i-1\right]$, and $q_0(B) = x$ where $n>0$ and $x$ are natural numbers.


$g_i^T = (1, \cdots 1,i)$ such that $g(A_j) = 1$ for $j \in \left[0,i-1\right]$ and $g(B)=i$ is the minimal semiflow of minimal support in $\mathbb{N}$.
We have a first invariant $I_1$: $\forall q \in RS(TNED(i),q_0)$,

$\  g^{\top}q_0 = g^{\top}q
= \sum_{j=0}^{j=i-1}q_0(A_j) + i q_0(B)
= i \times (x + n + \frac{i-1}{2})$.

Then, we need to notice that any place $A_j$ is connected to only two transitions $t_{j,1}$ and $t_{j,2}$ such that:

$Post(A_j,t_{j,1})-Pre(A_j,t_{j,1}) = -i$,

$Post(A_j,t_{j,2})-Pre(A_j,t_{j,2}) = i$.

Hence, $\forall q \in RS(TNED(i),q_0),\ \forall j \in \left[0,i-1\right]$, $q(A_j)$ can only vary by $\pm i$. We then have a family of invariants $I(j)$: $\forall q \in RS(TNED(i),q_0)$,

$\  q(A_j) \equiv q_0(A_j) \pmod i$.

Let's perform the euclidean division of $q_0(A_j)$ by $i$, we have: $q_0(A_j) = n + j = a_j \times i + \alpha_j$ where $\alpha_j < i$

$\forall j \in \left[0,i-1\right]$, a new family of invariants $I'(j)$ can be directly deduced from $I(j)$:

$I'(j): \forall q \in RS(TNED(i),q_0),\  q(A_j) \geq \alpha_j$.

Furthermore, it must be pointed out that $\{\alpha_0, \cdots \alpha_{i-1}\}$ is a permutation of $\{0, \cdots i-1\}$. Indeed, if $\exists j < i , j'<i$ such that $\alpha_j = \alpha_{j'}$, then $n+j - a_j \times i = n+j' - a_{j'}\times i $ and $|j-j'|=|(a_j-a_{j'}|\times i$. Since $|j-j'|<i$, we have $a_j=a_{j'}$ and $j=j'$. Therefore:
\begin{itemize}
    \item[(a)] $\sum_{j=0}^{j=i-1}q_(A_j) \geq \frac{i(i-1)}{2}$,
    \item[(b)] there is a unique $k \in \left[0,i-1\right]$ such that $\alpha_k = 0$.
\end{itemize}

From (a) and $I_1$, we deduce: $\forall q \in RS(TNED(i),q_0),\ q(B) \leq x+n$ (which is a better bound that the one that can be deduced from proposition \ref{prop: boundedness-alg}).

From $I(j)$ we can define any reachable marking $q$ such that:

$\forall q \in RS(TNED(i),q_0),\  q(A_j)=y_j \times i + \alpha_j$.

We can then execute the sequence $\sigma_q = t_{0,1}^{y_{0}} \cdots t_{i-1,1}^{y_{i-1}}$ and reach the marking $q_h$ such that:

$\forall j \in \left[0,i-1\right],\ q_h(A_j)=\alpha_j$ and $q(B)=x+n$.

$q_h$ is a home state since $\sigma_q$ is defined for any reachable marking (Note that $q_0$ is not a home state in general). From property \ref{prop: home-state-liveness} we deduce that since any transition $t_{j,2}$ where $j \neq k$ is executable (since $q_h(B) > 1$ and $q_h(A_j) = \alpha_j > 0$) then $t_{j,2}$ is live and therefore the corresponding transitions $t_{j,1}$ are also live.

From (b), we have $q_h(A_k) = 0$ from which we deduce that $t_{k,1}$ and $t_{k,2}$ are not live \footnote{Actually, it suffices to notice that $A_k$ is an empty deadlock that remains empty \cite{BR82}.}
Finally, we have $n+k = a_k \times i$ and the remainder of the euclidean division of $n$ by $i$ is $i-k$.

$TNED(i)$ provides the ability to recognize this result since $(t_{k,1},t_{k,2})$ is the only couple of transitions not live
\hfill 
$\square$ 

\subsection{Revisiting the mutual exclusion example}
\label{subsec: mutex-param}

One could argue that the Petri Nets of Figures \ref{mutex}, \ref{fig: tiny}, \ref{fig: tinyi}, or even \ref{fig: euclidean} are relatively simple and that it would be easy to obtain similar results without reasoning with semiflows. This may be true (for instance, using the LRGs of the Petri Nets of Figure \ref{mutex} or \ref{fig: tiny} is particularly trivial), however, it must be stressed out that semiflows allow to elegantly develop reasoning in the presence of parameters. 

In this regard, let's revisit and generalize the mutual exclusion example of Section \ref{subsubsec: modelization in a PN} by introducing few parameters. 
This time, we allow $k$ instances of Program 1 and $l$ instances of Program 2 to run concurrently. The two edges from $\textbf{S}$ to ${Semp_1}$ and from ${Semv_1}$ to $\textbf{S}$ are labeled by $x$; the two edges from $\textbf{S}$ to ${Semp_2}$ and from ${Semv_2}$ to $\textbf{S}$ are labeled by $y$ (by construction, we have $x>0$ and $y>0$ ). 
The initial state $q_0$ is being defined with three non- negative parameters: $k$ tokens in $A$, $l$ tokens in $D$, and $z$ tokens in $\textbf{S}$; and, $q_0(B) = q_0(E)= 0$.

This time, we authorize several instances of the same program to be simultaneously in their \texttt{critical section} however, we still want the mutual exclusion to hold between instances of Program 1 and Program 2. 
That is to say that we must have either $q(B) > 0$ or $q(E) > 0$ for any reachable marking. 
From the four versions of invariant representing mutual exclusion in section \ref{subsubsec: modelization in a PN} only (ME) and (ME)' still make sense in this parameterized context. Therefore, we want to determine under which conditions (i.e. for which values of the parameters) the Petri Net is live and that: 

(ME): $q(B)\times q(E)=0 $ for any marking reachable from $q_0$ is an invariant.

\begin{figure}[ht]
\centering
\includegraphics[width=0.8\textwidth]{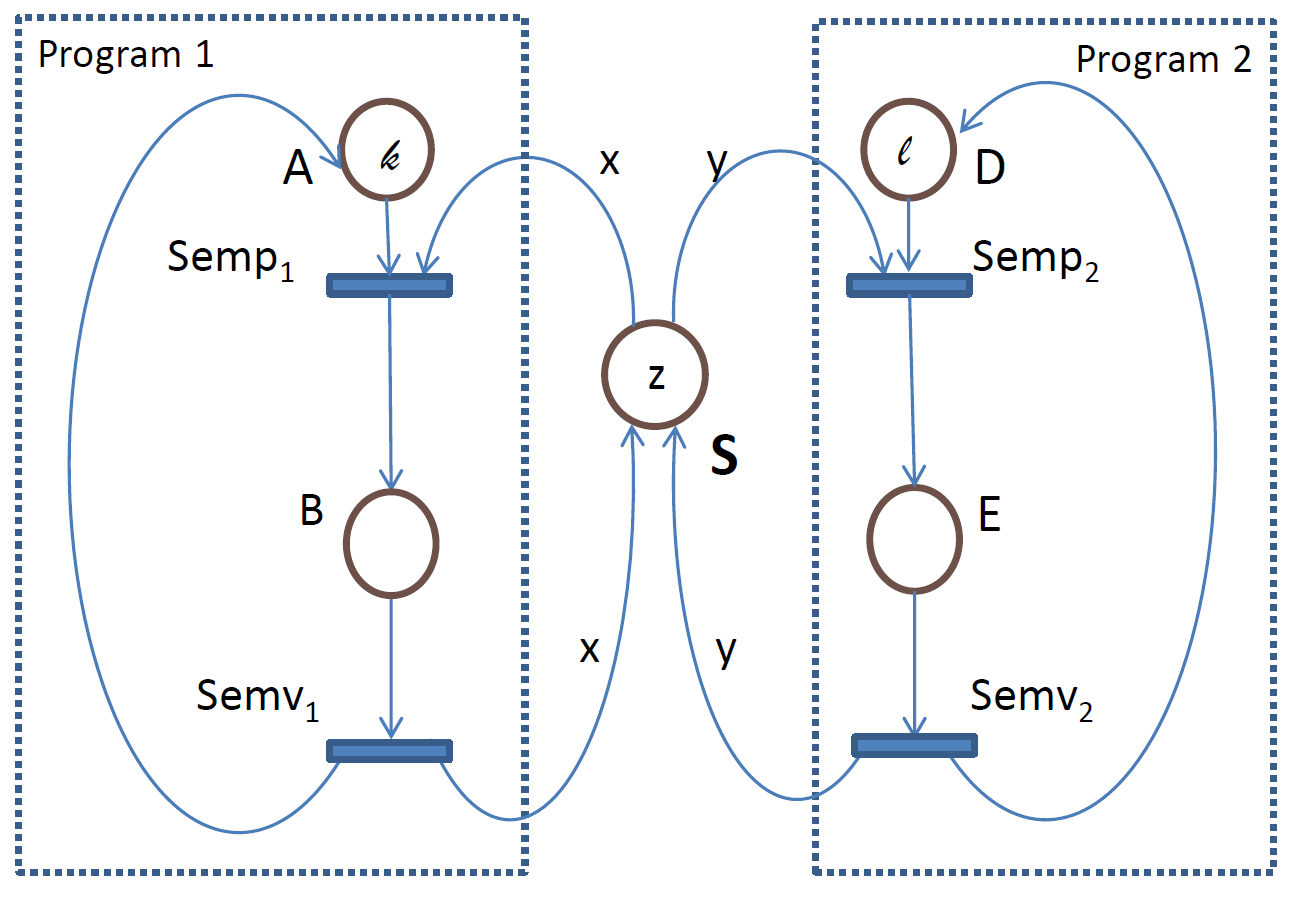}
\caption{\textit{k} instances of Program1 and \textit{l} instances of Program 2 can be executed concurrently. 
This time, unlike the Petri Net in Figure~\ref{mutex}, the mutual exclusion invariant depends on the various values of the three parameters $x, y, z$ independently of the parameters $k$ and $l$. This cannot be straightforwardly deduced from any semiflow.}

\label{mutex-param}
\end{figure}

The Petri Net Figure~\ref{mutex-param} has still 3 invariants directly deduced from 3 semiflows. The semiflows $f_1$ and $f_2$ are unchanged from Figure~\ref{mutex}; the third semiflow $sem_2$ is different and such that:
$sem_2(B)=x$, $sem_2(E)=y$, $sem_2($\textbf{S}$)=1$, $sem_2(p)=0$ for any other place $p$.

For any reachable marking $q$ from $q_0$,we have the following invariants:

$I : x\times q(B)+ y\times q(E)+ q($\textbf{S}$)= z$

$F_1: q(B)+q(A) = k$

$F_2: q(D) + q(E) = l$

We propose to proceed in three steps.

First, it is easy to prove that $q_0$ is a home state: 
let's assume $u \in  \mathbb{N}$,  
$v \in \mathbb{N}$ and $q \in RS$ such that $q(B)= u$ and $q(E) = v$.
Therefore, we can execute the sequence $Semv_1^{u}Semv_2^{v}$ for any value of $u$ and $v$. 
We then reach a marking $q'$ such that $q'(B)=q'(E)=0$. 
From $I, F_1, F_2$, we easily conclude that $q'= q_0$ without any assumption on the values of the parameters $k, l, z, x, y$.

Second, let us see that the Petri Net is live if and only if:

$k>0, l>0$ and $z \geq max(x,y)$.

If the Petri Net is live then $F_1$ (saying that there is no more than $k$ tokens in $A$) involves $k = q_0(A)>0$, $F_2$ involves $l = q_0(D)>0$ and $I$ involves that $q(\textbf{S})$ is bounded by $z$, therefore we must have $z \geq max(x,y)$ to have a chance to reach the enabling threshold of $Semp_1$ or $Semp_2$ and execute them. Reversely, if the 3 inequalities hold, we use the fact that $q_0$ is a home state to easily deduce that $Semp_i$ are live then $Semv_i$ are also live (since they need only $B$ or $E$ to contain one token that $Semp_i$ would have produced) and the Petri Net is live.

Third, we can now prove that (ME) holds if and only if $z<x+y$ or $k=0$ or $l=0$. 
The condition is obviously necessary otherwise if $(z \geq x+y)\ and\ (k \neq 0)\ and\ (l \neq 0)$ then we could execute the sequence $Semp_1Semp_2$ from $q_0$ and (ME) does not hold. Reversely, if $k=0$ or $l=0$ (ME) is obviously verified since at least one of the two program would have no instance. Now, from $z<x+y$ and $I$ we can write:

$q(\textbf{S}) = z -  x \times q(B) - y \times q(E) < x+y - x \times q(B) - y \times q(E)$

If $q(B)$ and $q(E)$ were both strictly positive then we would have:
$q(\textbf{S})<0$ which is impossible. Therefore, $q(B)$ and $q(E)$ cannot be both strictly positive for the same marking and the invariant is satisfied independently of the values of $k$ and $l$.

We can conclude that this PN paired with $q_0$ is live and (ME) is an invariant if and only if $k>0$, $l>0$, and $z \in [max(x,y),x+y-1]$.
\hfill 
$\square$

In addition, let us assume $z<x+y$ and $x<y$.
We have just seen that: 
$q($\textbf{S}$) < x\times (1-q(B))+ y\times (1-q(E))$. 
If $x<y$ then  $0 \leq q($\textbf{S}$) < -x\times q(B) + y\times (2-q(E))$ which can hold ony if $q(E)<2$ which means that:
(ME2): if $x<y$ and $z<x+y$ then there cannot be more than one token in $E$ for any reachable marking $q$ from $q_0$ is a second invariant for this Petri Net.

We can prove similarly that if $x=y$ and $z<x+y$ then we have: $0<x \times (2-q(E)-q(B))$. therefore, (ME3): if $x=y$ and $z<x+y$ then $q(B)+q(E) \leq 1$ for any reachable marking $q$ from $q_0$ is an invariant.

\subsection{An example from the telecommunication industry}
\label{subsec: example-summary}
\begin{figure}[ht]
\centering
\includegraphics[width=0.9\textwidth]{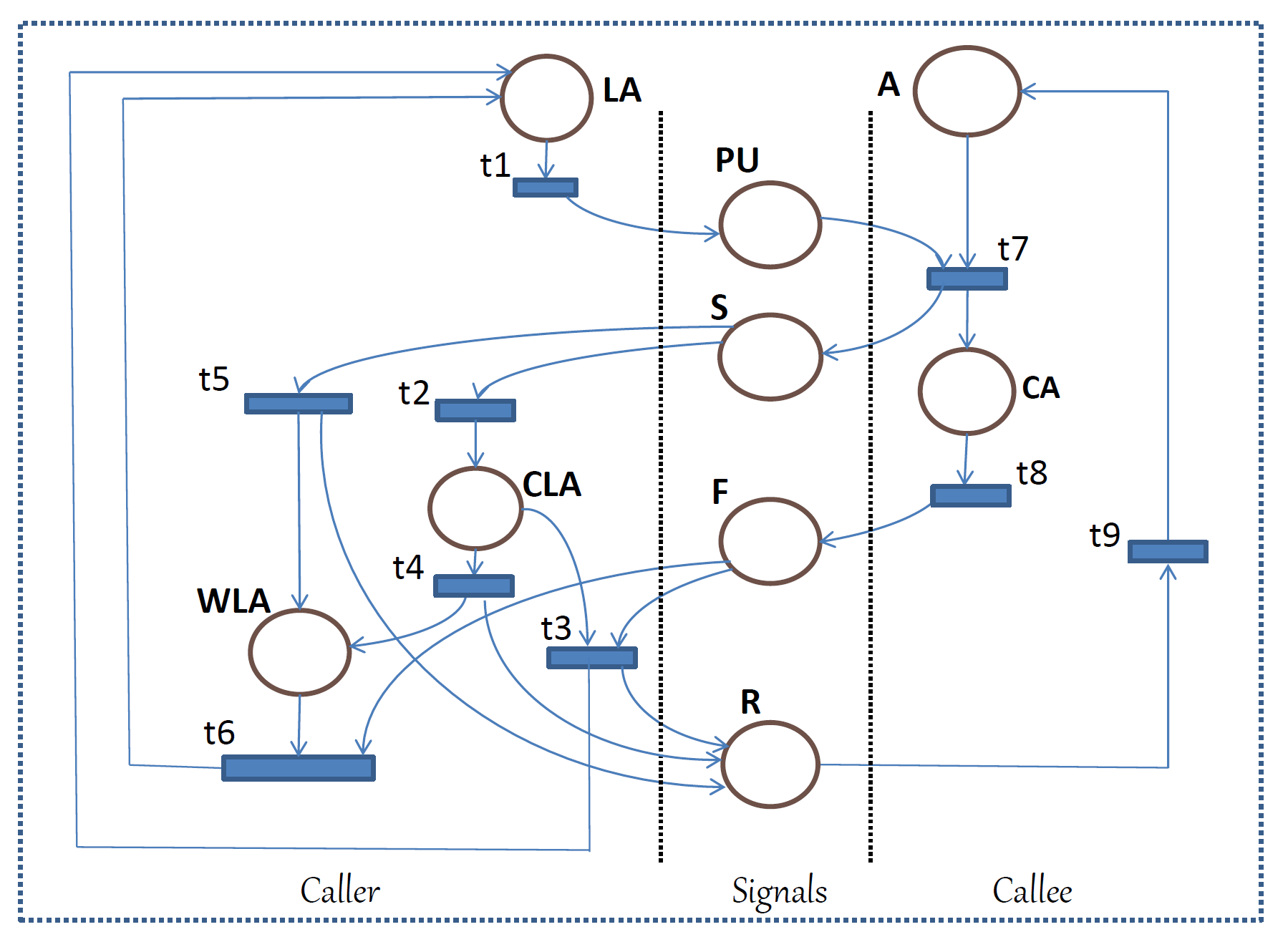}
\caption{This Petri Net TEL has exactly three minimal semiflows of minimal support, constituting a minimal generating set over $\mathbb{Q}^+$: $\mathcal{GB}_1 = \{f_1,\ f_2,\ f_3\}$ such that: 
\newline
$f_1(LA)=f_1(CLA)=f_1(WLA)=f_1(PU)=f_1(S)=\ 1$ and $f_1(p)=\ 0$ for any other place,
\newline
$f_2(LA)=f_2(PU)=f_2(F)=f_2(CA)=\ 1$ and $f_2(p)=\ 0$ for any other place,
\newline
$f_3(CLA)=f_3(S)=f_3(R)=f_3(A)= \ 1$ and  $f_3(p)=\ 0$ for any other place.
}
\label{Mame}
\end{figure}

The Petri Net $TEL= \langle P_{TEL},T_{TEL},Pre,Post \rangle$ is described Figure \ref{Mame}

where $P_{TEL} = \{LA, CLA, WLA, A, CA, PU, S, F, R\}$ 

and $T_{TEL}= \{t_1, t_2, t_3, t_4, t_5, t_6, t_7, t_8, t_9\}$.

TEL is a reduced  Petri Net version of a Fifo Net published in \cite{MemFink85, MeMa81}. 
It is well known that reduction rules preserve liveness, however, it is also well known that some rules do not preserve boundedness \footnote{See \cite{BR82,Colom2003} for Petri Nets reduction rules.}.
A sketchy version of the analysis hereunder is published in \cite{M23}.

This example is representing two subscribers, ``a caller" and a ``callee," having a conversation (places $CLA$ and $CA$ respectively).
We simplify the model by taking into account only the actions related to calling for the caller and the actions related to being called for the callee.
Initially, they are in an idle state with places $LA$ and $A$ marked with one token.

Signals $PU$ and $R$ are sent from the caller to the callee and signals  $S$ and $F$ from the callee to the caller. 
The overall desired behavior is that caller and callee cannot go back to their idle state as long they have not received all the signals sent to them despite the fact that they both can hang up at any time making the order in which signals $F$ and $R$ are sent and received undetermined.

From their idle state (place $LA$), the caller can pick up their phone (transitions $t_1$) sending the signal $PU$ to the callee. 
From their idle state (place $A$), the callee, upon receiving the signal $PU$, can pick up their phone (transition $t_7$), send the signal $S$ and go
to conversation (place $CA$) from where they can hang up (transition $t_8$) at any time sending the signal $F$ to the caller. 
Receiving the signal $S$, the caller can go (transition $t_2$) to the conversation (place $CLA$).
They can also hang up at any time (transitions $ t_3
, t_4, t_5$) sending the signal $R$ to the callee.
After hanging up via $t_4$ or $t_5$, the caller will have to wait (place $W$) until they receive the signal $F$ from the callee before going back (transition $t_6$) to their initial idle state $LA$.
The callee can go back to their idle state $A$ only upon receiving the signal $R$ (transition $t_9$).

The initial state $q_0$ is such that $q_0(LA)=q_0(A)=1$ and $q_0(p)=0$ for any other place: the modeled system is in its idle state.
\subsubsection{Optimizing Sperner's bound}
First, let us notice that inequality (\ref{eq: Sperner}) gives us the following bound for $c$, the number of minimal supports:
$ c \leq \binom{9}{\left \lfloor 9/2 \right \rfloor}=126 $. 
However, it is easy to notice that transitions $t_1, t_2, t_8, \text{and}\  t_9$ have only one input and one output, which means that any support including such an input also includes the corresponding output to satisfy the equations associated to $t_1, t_2, t_8, \text{and}\ t_9$. 
The bound can be optimized to: $c \leq \binom{5}{\left \lfloor 5/2 \right \rfloor}=10$. It can further be improved by reasoning on transitions $t_5\ \text{and}\ t_6$ and reach $c \leq 3$.
\subsubsection{A proof scheme using minimal semiflows of minimal support}
We want to verify the following property (known as safeness \cite{GV03} p.489):

$\mathcal{P} = \forall q \in RS(TEL, q_0), \forall p \in P_{TEL}, q(p) \leq 1 $.

We also want to prove that the initial marking $q_0$ is a home state (see definition \ref{def: homespace} and \ref{def: home state}) 
from which it is easy to deduce that the Petri Net is live.

These two important properties are proven hereunder, starting from 
$\mathcal{GB}_1=\{f_1,f_2,f_3\}$ the set of minimal semiflows of minimal support defined Figure \ref{Mame} without considering every reachable marking. 

Similarly as in equation (\ref{eq: invariant}), three invariants can directly be drawn from $\mathcal{GB}_1$: for any reachable marking $q$ from $q_0$,
$f_1^\top q=f_1^\top q_0=1$, 
$f_2^\top q=f_2^\top q_0=1$, 
$f_3^\top q=f_3^\top q_0=1$. 
\footnote{In the following, we will omit the phrase ``for any reachable marking $q$ from $q_0$".}

In other words, there is exactly one token in the support of any semiflow of $\mathcal{GB}_1$. Hence, $\mathcal{P}$ is true since $P = \bigcup_{i=1}^{i=3}\left\|f_i \right\|$.

In order to prove that $q_0$ is a home state, we can start to prove that
$\mathcal{HS}=\{q\ |\ q(LA)=1\}$ is a home space (i.e., a set of markings such that whatever is the evolution of the Petri Net, it is always possible to reach one element of the set). 
If we consider the invariant deduced from $f_2$, we know that any reachable marking $q$ is necessarily in one of the 4 following cases.

i) $q(LA)=1$, then $q \in \mathcal{HS}$,

ii) $q(F)=1$, since $f_2^\top q=1$, we have $q(LA)=q(PU)=q(CA)=0$.
$f_1^\top q=f_1(CLA)q(CLA)+f_1(WLA)q(WLA)+f_1(S)q(S)=1$ and it then remains only 3 sub-cases to explore:

ii-1) $q(CLA)=1$, $t_3$ can occur from $q$ and we can reach $q' \in \mathcal{HS}$

ii-2) $q(WLA)=1$, $t_6$ can occur from $q$ and we can reach $q' \in \mathcal{HS}$

ii-3) $q(S)=1$, $t_2$ can occur from $q$ then we are in the sub-case ii-1).

iii) $q(CA)=1$, $t_8$ can occur from $q$ then we are in the case ii).

iv) $q(PU)=1$, then considering the invariants deduced from $f_1$ and $f_2$, we have: 
$q(LA)=q(CLA)=q(WLA)=q(F)=q(S)=q(CA)=0$

From $f_3^\top q=1$, we have two sub-cases:

iv-1) $q(A)=1$, $t_7$ can occur from $q$ then we are in the case iii).

iv-2) $q(R)=1$, $t_9$ can occur then we are in the case iv-1) again. 

From these simple 4 cases and 5 sub-cases, we directly deduce that $\mathcal{HS}$ is a home space from where it is easy to conclude that $q_0$ is a home state and the Petri Net is live.

We let the interested reader develop the same proof scheme from $\mathcal{GB}_2=\{f_1,\  g=f_2+f_3,\ h=f_1+f_3\}$ or from $\mathcal{GB}_3 = \{g,\ h,\ l=f_1+f_2\}$ which are minimal generating sets 
over $\mathbb{Q}$ and find out that the analysis becomes more complex even with small additional changes.
We conclude that the smaller the support, the more effective the analysis will be since the number of cases and sub cases of the proof scheme depends on the number of elements of each considered support. 

\subsubsection{Proving TEL(x,y)}
\begin{figure}[ht]
\centering
\includegraphics[width=0.8\textwidth]{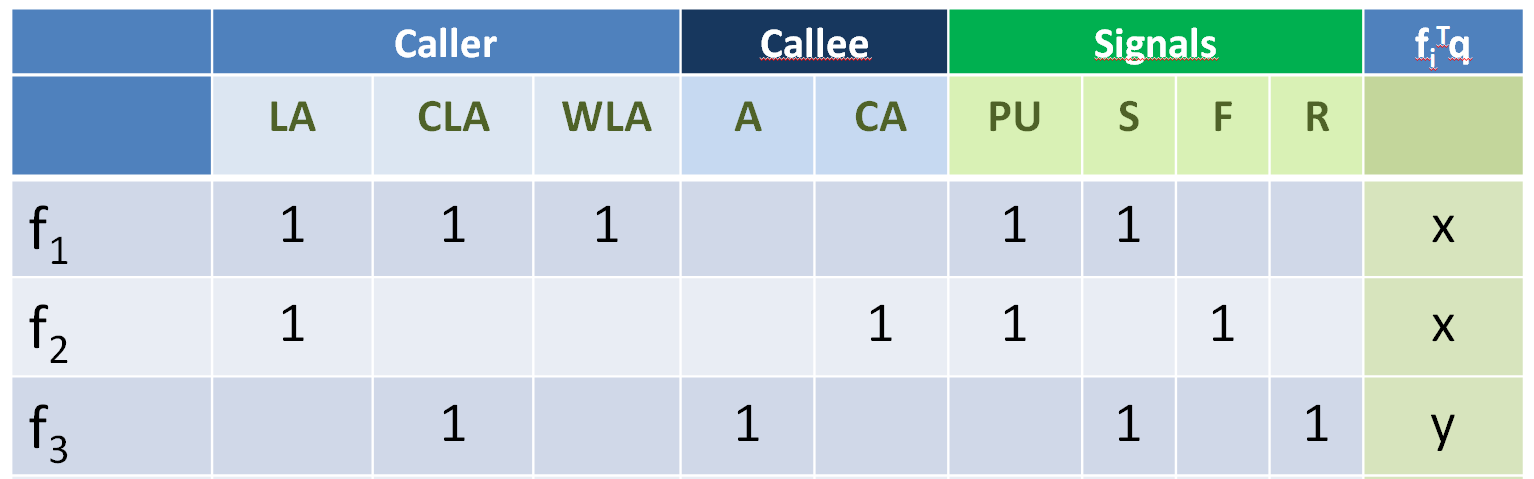}
\caption{This tableau is concise way to represent a generating set and corresponding invariants of $\mathcal{GB}_1$. It reads for instance: $f_1(LA) = 1$ or $f_1(A)=0$. The three semiflows $f_1,\ f_2,\ f_3$ do not depends on the initial state (even enriched with parameters); only the evaluation functions of the three corresponding invariants are depending on the initial state, therefore are defined with parameters.}
\label{fig: tableau-parameterrized}
\end{figure}

A similar proof scheme could be conducted with a colored Petri Net or the Petri Net of Figure \ref{Mame} enriched with two parameters to model $x$ callers and $y$ callees, where $x>0$ and $y>0$. 
Considering the parameterized Petri Net TEL(x,y) of Figure \ref{Mame}, 
the initial state $q_0$ becomes: 
$q_0(LA)=x, \ q_0(A)=y$ and $q_0(p)=0$ for any other place $p$. 
The generating set $\mathcal{GB}_1$ is unchanged, but its three associated invariants become:

$ \forall q \in RS(TEL, q_0)$:

$f_1^\top q=f_1^\top q_0=x$, 
$f_2^\top q=f_2^\top q_0=x$, 
$f_3^\top q=f_3^\top q_0=y$,

as concisely described Figure \ref{fig: tableau-parameterrized}.

This time, we want to prove that TEL(x,y) is live and satisfies the following properties:

$\mathcal{P}_1(x,y) : \mathcal{HS}(z) = \{q|\ q(CLA)=q(CA)=z \ \text{and}\ z \leq min(x,y)\} $ is a home space,

$\mathcal{P}_2(x,y) : \forall q \in RS(TEL(x,y), q_0)$,

$q(CLA) \leq min(x,y)$, 
$q(CA) \leq min(x,y)$,
$Q(LA)+q(CLA)+q(WLA) \leq x$,
$q(A)+q(CA) \leq y$.

In other words, we want at most $min(x,y)$ simultaneous conversations, no more than one token per subscriber ever.

First, from property \ref{prop: inter-home-spaces} that $\mathcal{HS}_0 = \{q \in Q\ |\ f_1^\top q=f_2^\top q=x, \text{and} 
f_3^\top q=y$ \} is a home space.

From any state $q \in \mathcal{HS}_0$, it is always possible to execute the sequence:

$\sigma_1=(t_5t_9)^{q(S)}(t_4t_9)^{q(CLA)}t_9^{q(R)}$.
We then reach a state $q_1$ such that $q_1(S)=q_1(CLA)=q_1(R)=0\ $ defining a second home space $\mathcal{HS}_1= \{q \in \mathcal{HS}_0 \ |\ q(S)=q(CLA)=q(R)=0\}$.
From the invariant associated with $f_3$, it can be directly deduced that $\forall q_1 \in HS_1$, we have $q_1(A)=y$.
Similarly, it is always possible to empty $CA$ from its tokens: $\forall q_1 \in HS_1$, it is always possible to execute
$\sigma_2=t_8^{q(CA)}$ reaching a state $q_2$ such that $q_2(CA)=0$ defining a third home space $\mathcal{HS}_2= \{q \in \mathcal{HS}_1 \ |\ q(CA)=0\}$.

Since $y>0$, from any state $q_2$ in $HS_2$, it is always possible to execute the sequence: 

$\sigma_3=(t_7t_5t_9t_8)^{q_2(PU)}$.
We then reach a state $q_3$ still in $HS_2$ (each time $t_7$ puts a token in $S$ and $CA$, we can execute $t_5t_9$ and $t_8$ respectively to return to $HS_2$) such that $q_3(PU)=0$ defining a fourth home space $HS_3= \{q \in \mathcal{HS}_2 \ |\ q(PU)=0\}$.
From the two invariants associated with $f_1$ and $f_2$, it can be deduced that $\forall q \in HS_3$, we have $q(LA)+q(WLA)=x$ and $q(LA)+q(F)=x$ respectively, hence necessarily, $q(WLA)=q(F)$. 
Therefore, from any state $q_3$ in $HS_3$, it is always possible to execute the sequence: $\sigma_4=t_6^{q_3(F)}$ reaching a state $q_4$ still in $HS_3$ such that $q_4(WLA)=q_4(F)=0$ therefore, $q_4(LA)=x$. 
The only such state is $q_0$ which being always reachable via the sequence $\sigma_1\sigma_2\sigma_3\sigma_4$, is a home state.
From there, $t_1$ is live (from property \ref{prop: home-state-liveness}) and it becomes easy to prove that the Petri Net is live.
\begin{figure}[ht]
\centering
\includegraphics[width=0.8\textwidth]{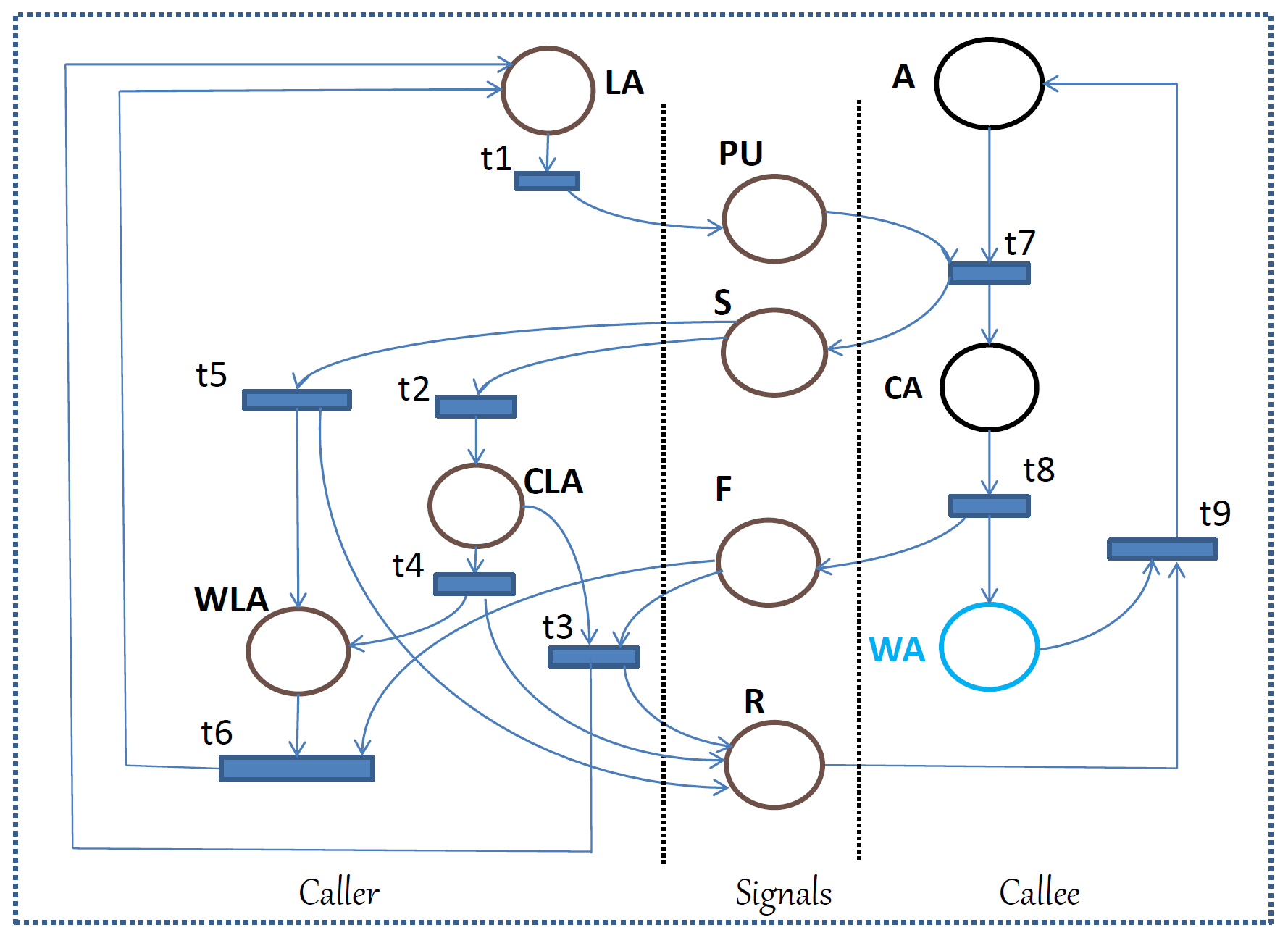}
\caption{TEL2(x,y): By adding the place $WA$ and connecting it to the transitions $t_8$ and $t_9$, we generate a fourth minimal semiflow without changing the three first ones.
\newline
$f_4(A)=f_4(CA)=f_4(WA)=\ 1$ and $f_1(p)=\ 0$ for any other place}
\label{fig: mame2}
\end{figure}

From $q_0$ home state, we can always execute the sequence $\sigma_z = (t_1t_7t_2)^z$ as long as $z \leq min(x,y)$; therefore
$\mathcal{HS}(z) = \{q|\ q(CLA)=q(CA)=z \ \text{and}\ z \leq min(x,y)\} $ is a home space meaning that it is always possible to have $z$ pairs of subscribers in a conversation at the same time which satisfies $\mathcal{P}_1(x,y)$.

Furthermore, by applying theorem \ref{th: bounds} using $\mathcal{GB}_1$, we directly have:

$\forall q \in RS(TEL(x,y),q_0), q(CLA) \leq \mu(CLA,q_0)= min(x,y)$.

This is the first half of property $\mathcal{P}_2(x,y)$ and it means that we can only have up to $min(x,y)$ simultaneous conversations.

However, the same theorem allows deducing not only that:

$\forall q \in RS(TEL(x,y),q_0), q(CA) \leq \mu(CA,q_0)= x$,
but also that there is not hope to find a better bound for $CA$ with the only help of semiflows.
Worse, it can be shown that $\mu(CA,q_0)= x$ can be reached and that $\mathcal{P}_2(x,y)$ is not satisfied. In the case where $x > y$, this would model the fact that a callee can be in a simultaneous conversation with several callers (which is a possibility today, but was not expected here). The last part of property $\mathcal{P}_2(x,y)$ is not satisfied either.

This situation is due to an oversimplification of TEL. By restoring one reduction back and remembering that in particular, the callee process was modeled by a state machine, we construct the Petri Net $TEL2(x,y)$ of Figure \ref{fig: mame2}.
$TEL2(x,y)$ has an augmented generating set $\mathcal{GB}_2 = \mathcal{GB}_1 \cup f_4$ where $f_4$ is a new minimal semiflow such that $f_4(A)=f_4(CA)=f_4(WA)=\ 1$ and $f_1(p)=\ 0$ for any other place and is directly associated with the following invariant: $\forall q \in RS(TEL2(x,y),q_0),\ f_4^\top q=f_4^\top q_0=y$.  
We can apply theorem \ref{th: bounds} again to $CA$ and this time obtain 
$\forall q \in RS(TEL2(x,y),q_0),\ q(CA) \leq \mu(CA,q_0) = min(x,y)$ and finish to prove $\mathcal{P}_2(x,y)$. Indeed, it can be shown that with similar methods of proof, $TEL2(x,y)$ satisfies $\mathcal{P}_1(x,y)$ and is live.

\subsection{A brief illustration of fairness and starvation}

The Petri Net Figure~\ref{mutex-param} has a \textit{fairness} issue despite the fact that it is live for $k>0$, $l>0$, and $z \in [max(x,y),x+y-1]$. As soon as $k>1$ and $z\geq 2x$, Program1 can \textit{starve} Program2 by having constantly at least one instance of program1 in its \texttt{critical section}: the marking of $B$ remains strictly positive and an infinite loop between $Semp_1$ and $Semv_1$ lets no instance of program2 to ever have the opportunity to enter in its \texttt{critical section}. Similarly, as soon as $l>1$  and $z \geq 2y$, Program2 can \textit{starve} Program1: there exists an infinite computation such that $Semp_1$ can never have the opportunity to be executed from any marking of this sequence. 

\begin{figure}[ht]
\centering
\includegraphics[width=0.8\textwidth]{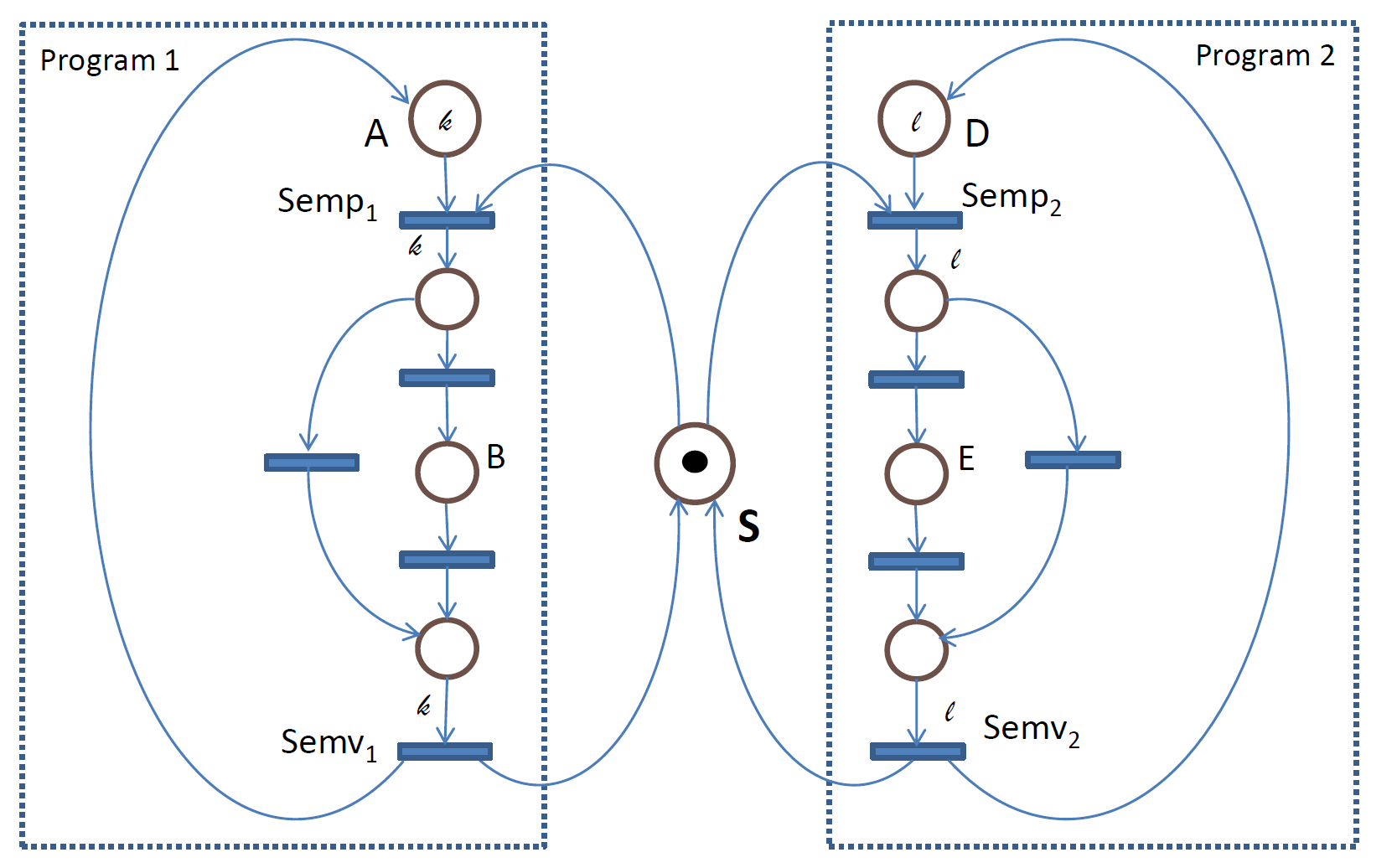}
\caption{There does not exist an infinite progression such that a Program $i$ can prevent the other one to have the opportunity to execute its \texttt{critical section}, in other words, to never have the possibility for $Semp_j$ (where $j=3-i$) of being enabled.}
\label{mutex3}
\end{figure}

The Petri Net Figure~\ref{mutex3} addresses this issue since the transition $Semv_1$ is enabled only when the marking of $B$ is null. This solution can be compared to another example page 111 of \cite{BR82} where a privilege is given to one Program to starve the other one. This time, it can be proven that any infinite sequence offers to $Semp_i$ an infinite number of opportunities to be executed.

\newpage
\section{Conclusion}
\label{sec: concl}
As soon as we can associate the behavior of a system under study with a set of state variables and follow its evolution through sequences of states forming a reachability graph (or a structure) where each edge is associated with a transition, we can assume the existence of an underlying Transition System. Doing so, invariants and home spaces can be checked or proven along all possible sequences and help understanding and verifying the functioning of such a system.

About Petri Nets, it can never be sufficiently stressed how the combination of their two key concepts puts Petri Nets apart from many other conceptual models. First, a bipartite graph (a topology) captures and models so well the relationship and dependencies between actions and resources. 
Second, the \textit{“token game”} allows for simulations and construction of a reachability graph given an initial marking (or state). This combination is clearly able to depict concurrency and is supporting somehow a notion of parallelism. 
It has been seen how semiflows create a link from the static topology (the bipartite graph) to the dynamic evolution (the variation of the number of tokens) of the Petri Net. 
They support constraints over all possible markings which greatly help analyzing and discovering behavioral properties (even some unspecified ones). 
Semiflows infer a class of invariants with a constant evaluation function that can be computed from the initial marking of the Petri Net under consideration. 
Last but not least, they can be characterized by a generating set that can be used in order to support some elegant level of behavioral analysis even with the presence of some level of parameterization.

We were able to regroup and clarify some key algebraic results scattered in the literature by considering only the two notions of minimality in terms of semiflow and support respectively. 

The example of Section \ref{subsec: example-summary} showed how loose the Sperner's bound can be, as well as a first pathway to improve upon it by considering the connections imposed by the homogeneous system of equations (\ref{eq: inv-semiflow}). 
The same example provides reasons to consider non-negative semiflows of minimal support. 
With that said, a comprehensive complexity analysis of the complete process of computing a generating set along with analyzing a Petri Net properties is required in order to quantitatively evaluate the trade-off between manipulating a generating set over $\mathbb{Q^+}$ or over $\mathbb{Q}$.  
Let us also point out that in the literature, there exist additional reasons to consider minimal supports described, for instance in \cite{Colom2003}.


We claimed that results in \cite{CMPW09} or in \cite{ColomTS2003} p. 68 need to be rephrased in considering the formulation by the same authors in
one of their previous publications  \cite{STC1998}.


We believe that these results may be enriched along two different alleys. 
From a mathematical point of view, the relation with integer linear programming or convex geometry has been investigated many times, particularly in \cite{ColomS89}, however, we believe it could be fruitful to look at the notion of toric varieties and saturated semigroups \cite{Oda12}. 
From a Petri Net and, even more broadly, from a transition system theory point of view, applying these new results to a variety of models (for example, colored Petri Nets or any transition system that can be associated with a system of equations such as (\ref{eq: inv-semiflow})) remains to be done. 

Finally, we would like to stress the possibility to automate the proof scheme exhibited throughout the example of Section \ref{subsec: example-summary}. Even though some proof steps can be shortened, we chose to develop our proofs in a systematic way to provide a clear road map on how an algorithm could proceed for solving such properties.



\bibliography{mybibfile}

\end{document}